\documentclass[a4paper,11pt]{report}
\usepackage[utf8]{inputenc}
\usepackage[english]{babel}

\usepackage{bbm}
\usepackage{xcolor}
\usepackage{amsbsy}
\usepackage{breqn}
\usepackage{amssymb}
\usepackage{subfig}
\usepackage{enumerate}
\usepackage{amsfonts}
\usepackage{float}
\usepackage{graphicx}
\usepackage{slashed}
\usepackage{setspace}
\usepackage{bbm}
\usepackage{breqn}
\graphicspath{{.\images}}
\usepackage{float}
\usepackage{amsmath}
\usepackage{ragged2e}
\usepackage{amssymb}
\usepackage{float}
\usepackage{mathtools}
\usepackage{pifont}
\usepackage{multirow}

\usepackage{bm}
\usepackage{physics}
\usepackage{listings}
\usepackage{algorithm2e}
\usepackage{tikz}
\usepackage{abstract}
\addto\captionsenglish{\renewcommand{\abstractname}{}}   

\usepackage[compat=1.1.0]{tikz-feynman}
\usetikzlibrary{backgrounds}
\usepackage{mathrsfs}
\usepackage[bottom]{footmisc} 
\usepackage[nottoc]{tocbibind}
\usepackage{wallpaper}
\usepackage[width=150mm,top=25mm,bottom=25mm,bindingoffset=6mm]{geometry}
\usepackage[colorinlistoftodos]{todonotes}
\usepackage[colorlinks=true, pdfborder={0 0 0},linkcolor=black, citecolor=red]{hyperref}
\usepackage{array}
\newcolumntype{P}[1]{>{\centering\arraybackslash}p{#1}}
\usepackage{authblk}
\usepackage[square,numbers]{natbib}
 \usepackage{fancyhdr}
\renewcommand{\headrulewidth}{1.5pt}
 
\newlength\FHoffset
 \usepackage{etoolbox}
\newbox\FHline
\setbox\FHline=\hbox{\hsize=\paperwidth%
}

\title{\vspace{-5cm}\textbf{\huge THE CASIMIR EFFECT FOR LATTICE FERMIONS}\\
\vspace{1cm}{\Large \textbf{\Large A THESIS}\\\textit{\Large submitted in partial fulfillment of the requirements\\for the award of the dual degree of}\\\vspace{1.5cm}\textbf{\Large{Bachelor of Science - Master of Science}}\\\textit{in}\\\textbf{\Large{PHYSICS}}}}
\vspace{-2cm}
\author{\Large \textit{by}\\\textbf{\Large{YASH VIKAS MANDLECHA}}\\\textbf{\Large(17327)}}
 \date{}
 \def\maketitle{
\begin{titlepage}
\begin{center}
\begin{doublespace}

\vspace{1cm}

\textbf{\doublespacing \huge THE CASIMIR EFFECT FOR\\[0.4cm] LATTICE FERMIONS}

\vspace{1cm}

{\Large
\textbf{\fontsize{14pt}{0.7cm}\selectfont A THESIS}\\
\textit{\fontsize{14pt}{0.9cm}\selectfont submitted in partial fulfillment of the requirements\\
for the award of the dual degree of}\\[1.5cm]
\textbf{\Large Bachelor of Science - Master of Science}\\[0.2cm]
\textit{in}\\
\textbf{\Large PHYSICS}
}

\vspace{0.4cm}

{\Large \textit{by}}\\[0.2cm]
\textbf{\Large YASH VIKAS MANDLECHA}\\
\textbf{\Large (17327)}

\end{doublespace}

\vfill

\includegraphics[width=4.5cm]{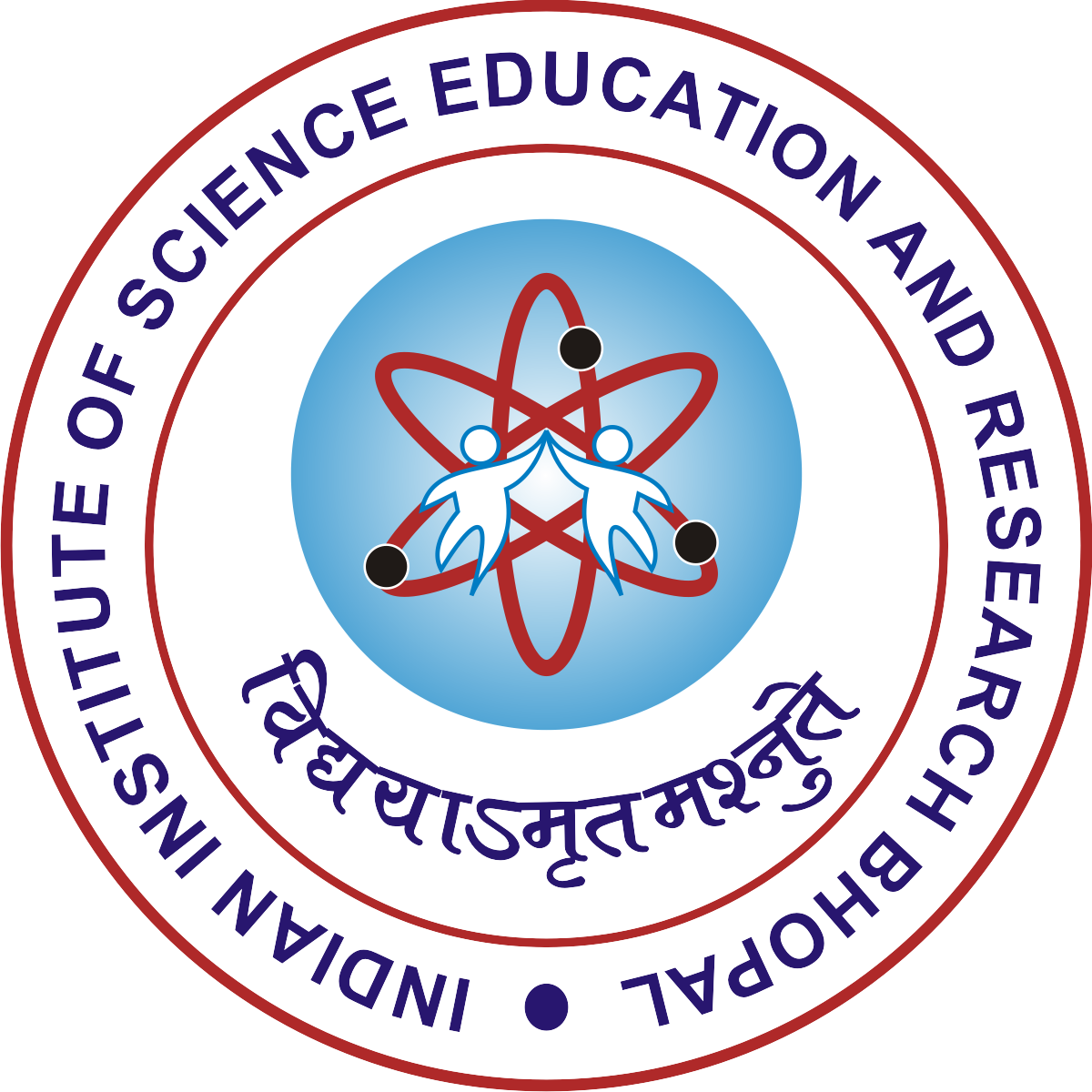}

\vspace{0.5cm}

\textbf{\large
DEPARTMENT OF PHYSICS,\\
INDIAN INSTITUTE OF SCIENCE EDUCATION AND RESEARCH\\
BHOPAL, INDIA-462066\\[5mm]
April 2022
}

\end{center}
\end{titlepage}
}

\begin{document}
\renewcommand{\abstractname}{\vspace{-\baselineskip}}

\maketitle
\thispagestyle{empty}

\newpage
\patchcmd{\tableofcontents}{\contentsname}{\centering\MakeUppercase\contentsname}{}{}
\newpage 
\fancypagestyle{certificate}{
    \fancyhf{}
    \cfoot{}
    \chead{}
    \rhead{}
    \renewcommand{\headrulewidth}{0pt}
}
\thispagestyle{empty}
\ThisURCornerWallPaper{0.96}{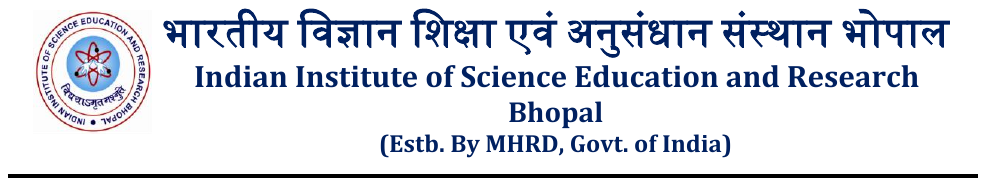}
\vspace*{1cm}
\begin{center}
\fontsize{20pt}{16pt}\selectfont{\textbf{CERTIFICATE}}
\end{center}
\phantomsection
\addcontentsline{toc}{chapter}{Certificate}
\vspace{1cm}

  This is to certify that \textbf{Yash Vikas Mandlecha}, BS-MS (Physics), has worked on the project entitled {\bf `The Casimir effect for lattice fermions'} under my supervision and guidance. The content of this report is original and has not been submitted elsewhere for the award of any academic or professional degree.
\vspace{3cm}\\
\textbf{April 2022\\IISER Bhopal}
\begin{flushright}
\textbf{Prof. Rajiv V. Gavai }
\end{flushright}
\vspace{4cm}
\begin{center}
\begin{tabular}{ccc}
\textbf{Committee Member} & \textbf{Signature} & \textbf{Date} \\
\\
\rule{17em}{0.5pt} & \rule{12em}{0.5pt} & \rule{8em}{0.5pt} \\
\\
\rule{17em}{0.5pt} & \rule{12em}{0.5pt} & \rule{8em}{0.5pt} \\
\\
\rule{17em}{0.5pt} & \rule{12em}{0.5pt} & \rule{8em}{0.5pt} \\
\end{tabular}
\end{center}
 \newgeometry{left=1.2in,right=0.8in,top=1in,bottom=1in}
   \newpage 
 \begin{center}
  \addcontentsline{toc}{chapter}{Academic Integrity and
Copyright Disclaimer}
\vspace{1cm}
  \centering{\textbf{\huge ACADEMIC INTEGRITY AND\\\vspace{0.3cm}
COPYRIGHT DISCLAIMER}}
  \end{center}
  \vspace{1.6cm}
  
I hereby declare that this project is my own work and, to the best of my knowledge, it contains no materials previously published or written by another person, or substantial proportions of material which have been accepted for the award of any other degree or diploma at IISER Bhopal or any other educational institution, except where due acknowledgement is made in the document.\\

I certify that all copyrighted material incorporated into this document is in compliance with the Indian Copyright Act (1957) and that I have received written permission from the copyright owners for my use of their work, which is beyond the scope of the law. I agree to indemnify and save harmless IISER Bhopal from any and all claims that may be asserted or that may arise from any copyright violation.\\
\vspace{7cm}\\
\textbf{April 2022}\\
\textbf{IISER Bhopal}
\begin{flushright}
\textbf{Yash Vikas Mandlecha}
\end{flushright}
   \newpage 
   \begin{center}
  \addcontentsline{toc}{chapter}{Acknowledgement}
  \centering{\textbf{\huge ACKNOWLEDGEMENT}}
  \end{center}
  \vspace{1.5cm}
  
  First of all, I offer my sincere gratitude to Professor Rajiv V. Gavai for his persistent support and guidance throughout this project. I would also like to thank him for offering an introductory course on Lattice Gauge theory in the summer of 2021, which served as a great introduction to the topic and a vital resource throughout the project. He always found time to address my queries and have constructive and exhilarating discussions, which helped me deepen my understanding. I consider myself very fortunate to have got an opportunity to work with him on this project. I am grateful for his reproach as much
as his generous advice and appreciation that helped me step forward towards being a better researcher.\\

I am thankful to Dr. Suhas Gangadharaiah for his valuable insights into my project and for allowing me to waive my courses in the final semester after completing the course load in his capacity as the DUGC, Department of Physics. Due to this, I was able to pay more attention to this project. I thank INSPIRE and KVPY, Department of Science and Technology,
Govt. of India for their generous funding throughout my graduation. I would also like to thank the Office of the Department of Physics, who enabled my access to an efficient desktop computer and a professional workplace, without which this work would not have been possible.\\

Last but not the least, I am indebted to my parents, Dr. Vikas and Dr. Vaishali Mandlecha, my elder sister CA Shraddha Mandlecha and friends for their constant motivational and emotional support towards my studies and projects.\\
\vspace{1cm}
\begin{flushright}
\textbf{Yash Vikas Mandlecha\footnote{Email address: yashmandlecha@gmail.com, mandlec1@msu.edu}}
\end{flushright}
 \newpage
 \addcontentsline{toc}{chapter}{Abstract}
 \begin{center}
  \centering{\textbf{\huge ABSTRACT}}
  \vspace{0.7cm}
\end{center}

The Casimir effect for photons and Dirac fermion fields and its generalization to ($D+1$)-dimensional spacetime in the continuum,  was studied. 
We proposed applying the MIT Bag model, and corresponding boundary conditions to treat the slab with perfectly conducting parallel plates as a bag with confined fermion fields \cite{Mandlecha:2022cll, Gavai:2024vdu}. In this thesis, these boundary conditions are realized on the lattice. The formalism developed in Ref. \cite{Ishikawa:2020ezm} is used to calculate and study the Casimir energy for free lattice fermions, where the Naive and Wilson lattice fermions are treated analytically using the Abel-Plana formulae for finite range in ($1+1$)-dimensions and numerically for higher dimensions. Casimir energy for Overlap fermions with Möbius domain wall (MDW) kernel is also calculated numerically. The results obtained from the MIT Bag boundary conditions were encouraging and matched precisely with the continuum expressions for all three lattice fermions in the zero limit of the lattice spacing. As expected, doubling was observed in the results for the Naive fermion. No oscillation of Casimir energy was observed from the MIT Bag boundary conditions for massive and massless lattice fermions.
 These studies were later also conducted for the periodic and antiperiodic boundary conditions. The Casimir energy for the Wilson and Overlap fermion, in this case, exactly matches the continuum expressions 
 in the zero limit of the lattice spacing. However, it is observed from existing literature that the result for the Naive fermion rapidly oscillates with the odd and even lattice sizes for the periodic and anti-periodic boundary conditions. It also approaches two different expressions in the continuum limit. Contrary to the claim in Ref. \cite{Ishikawa:2020ezm} that the Naive fermion cannot be used to compute the Casimir energy for the Dirac fermion in continuum, which seemed to be an apparent violation of the universality property, we have numerically shown that the Casimir effect for Dirac fermions can also be computed from the Naive fermion in the zero limit of the Casimir spacing. Casimir effect for negative mass Wilson fermions and Overlap fermions has exciting applications in condensed matter systems. The negative mass Wilson fermions and Overlap fermions with MDW kernel correspond to the bulk and surface fermions of the topological insulator, respectively. Our work \cite{Mandlecha:2022cll, Gavai:2024vdu} already published is based on this thesis.
\tableofcontents
\newpage
\pagenumbering{arabic}

\chapter{Introduction to lattice fermions}
\label{intro}
\section{Fundamentals of Lattice Quantum Chromodynamics}
Quantum Chromo Dynamics (QCD) is the Yang-Mills theory of the strongly interacting particles and ﬁelds, namely the quarks and gluons. This chapter shall briefly discuss the theory of QCD from the perspective of adopting the lattice quantization and introducing various lattice fermions to the reader.

Lattice QCD aims to study the regime where the coupling gets large, and the physics describing the interactions is non-perturbative. It is a systematic regularization approach to discretize the Lagrangian on a hypercubic spacetime lattice with lattice spacing $a$ to numerically evaluate the Green’s functions and extrapolate the resulting observables in the continuum limit. A lattice calculation proceeds with discretizing the classical Euclidean action by introducing a UV regulator, which is the finite constant lattice spacing $a$, and the construction of the measure for the integration over all conﬁgurations of the classical ﬁelds. The final procedure includes the removal of the regulator by taking the continuum limit $a \rightarrow 0$ in order to obtain the continuum result.
Note that taking the continuum limit is only possible in theories where the coupling does not diverge in the UV regime and respects asymptotic freedom. We calculate the expectation values of observables numerically using techniques like Monte-Carlo simulations. Thus, a lattice formulation of the fermionic Casimir effect will allow us to study its consequences without losing the non-perturbative effects in interacting fermionic systems, like colour confinement, chiral symmetry breaking Etc.

 Let spacetime position be denoted by $x$. 
Also, Einstein summation convention is followed by all of the indices in expressions.
In the Euclidean path integral convention, one uses the quark field $\psi(x)$ and the antiquark field $\overline{\psi}(x)$ as independent integration variables, unlike the Minkowskian operators where the two ﬁelds are related as $\overline{\psi} = \psi^\dagger \gamma_0$. Note that we omit colour and Dirac indices for simplicity if not necessary.
We now consider the gauge ﬁelds, represented as $A_\mu(x)$, and describe the gluons which act as the exchange particles for the strong interaction between quarks. In addition to the Euclidean spacetime index $x$, the gauge ﬁelds constitute a vector ﬁeld carrying the Lorentz index $\mu$ in the Euclidean action. Note that the ﬁeld $A_\mu(x)$ is a traceless, hermitian $3 \times 3$ matrix for a given $x$ and $\mu$. 
The QCD action is split into a fermionic part and a gluonic part, which we shall briefly discuss below.
 This couples the quarks to the gluons. 
We first consider the Euclidean action for a single ﬂavor:
 \begin{equation}
 \label{ferac}
       S_E[\psi. \overline{\psi},A] = \int d^4x\left\{ \overline{\psi}(x)\left(\gamma^\mu D_\mu(x)+m\right)\psi(x)+  \frac{1}{4} F^{(i)}_{\mu\nu}(x)F^{\mu\nu, {(i)}}(x) \right\}
 \end{equation}
where $D_\mu(x) = \partial_\mu +igA_\mu(x)$ and $g$ is the coupling strength of the gauge fields to the quarks. \\
 In QCD, we require invariance of the fermionic and  gluonic part of the action under local rotations among the color indices of the quarks of the special unitary group $SU(3)$.  
Consequently, we encounter the cubic and quartic interaction terms in addition to the quadratic terms in the gauge action. These additional terms give rise to the self-interactions of the gluons, which make QCD so complex and nontrivial. These self-interactions are presumably responsible for the mysterious phenomenon of colour conﬁnement.

Let us introduce a $4D$ space-time lattice `$\Lambda$’ consisting of the vector label points $n$ on the space-time lattice separated by a constant lattice spacing $a$. 
\begin{equation}
    \Lambda = \{(n_i,n_4)\in \mathbb{N}\;| n_i = 0, 1,\dots, N_{\sigma}-1 ; n_4 = 0, 1,\dots, N_{\tau}-1\} \;\;\;\text{here}\;\;i = 1, 2, 3.\end{equation}
The theory is modelled such that the spinors will be placed on the lattice points.
We shall now introduce the gauge fields on the lattice. We have to ensure the gauge invariance under the $SU(3)$ gauge group for interactions between different quark fields.
Thus, we introduce $U_\mu(n)$, the matrix-valued variables that link the lattice points and thus are referred to as \textit{link variables}. Here, $\hat{\mu}$ and $-\hat{\mu}$ are the forward and backward directional indices on lattice respectively.
  $U_\mu(n)$ transform such that they form the group elements of the gauge group $SU(3)$. 
\begin{figure}[h]
    \centering
    \includegraphics[width = 13cm]{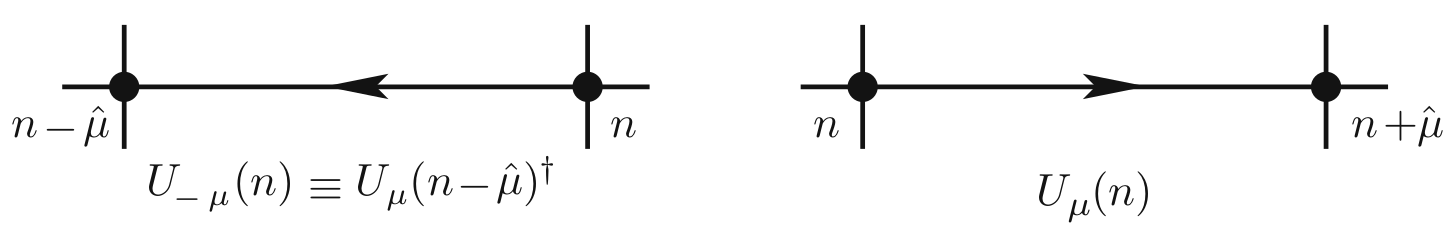}\\
    \caption{\footnotesize{The link variables  $U_\mu(n)$ and  $U_{-\mu}(n)$}}
    \label{link}
\end{figure}
If we consider $n$ and $n + \hat{\mu}$ as endpoints of a path, as shown in Fig. \ref{link}, the transformation properties of the gauge transporter in continuum are the same as for our link variables $U_\mu(n)$. These can be interpreted as the lattice counterpart of the gauge transporter connecting the points $n$ and $n + \hat{\mu}$. From this argument, one proposes the representation of the link variables in terms of the lattice gauge ﬁelds $A_\mu(n)$ as:
\begin{equation}
\label{UasA}
U_\mu(n) = \exp(i gaA_\mu(n))   
\end{equation}
The $SU(3)$ group valued link variables $U_\mu(n)$ in the lattice formulation, are considered as the fundamental variables, and not just an auxiliary construction from the $su(3)$ Lie algebra-valued ﬁelds $A_\mu(x)$ in continuum. 
\begin{figure}[h]
\centering    
\includegraphics[width = 7cm]{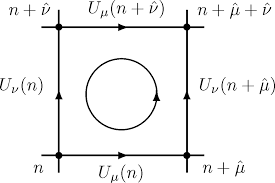}
    \caption{\footnotesize{The four link variables which build up the plaquette $P_{\mu\nu}(n)$ . The circle
indicates the order in which the links are run through in the plaquette.}
}\label{plaquette}
\end{figure}
For constructing the gluon action, it is suﬃcient to use the shortest, nontrivial closed loop on the lattice, called the \textit{plaquette}. The \textit{plaquette variable} $P_{\mu\nu}(n)$ on the lattice is a product of only four-link variables deﬁned as:
\begin{align}
  P_{\mu\nu}(n) 
  &=  U_{\mu}(n) U_{\nu}(n + \hat{\mu}) U_{\mu}(n  +\hat{\nu})^\dagger U_{\nu}(n)^\dagger
\end{align}
 and represented in Fig. \ref{plaquette}. Kenneth G. Wilson \cite{Wilson} was the first to formulate the gauge action on a space-time  lattice. The Wilson gauge action is obtained by taking a sum over all lattice points $n$ where plaquettes with an independent orientation are located, with an overall sum over the Lorentz indices $\mu, \nu \in [1,4]$.
The Wilson’s lattice gauge action is obtained as:
\begin{equation}
\label{gauactU}
    S_G[U] = \frac{2}{g^2}\sum_{n\in\Lambda}\sum_{\mu<\nu}  \text{Re}\left(\text{tr}[\mathbbm{1} - P_{\mu \nu}(n)]\right) 
\end{equation}
    After discretizing the gauge action and obtaining the Wilson gauge action on the lattice, we can prove that it indeed approaches the continuum form in the limit $a \rightarrow 0$.
    
    Thus,  $\lim_{a\rightarrow0} S_G[U] = S_G[A]$.
\section{Fermions on Lattice}
To study the Casimir effect for fermions on a lattice, we first need to take a closer look at the Nielsen-Ninomiya theorem, which plays an important role in the discretization of the fermion action. After this, we discuss different types of lattice fermion actions \cite{Susskind}.
\subsection{Nielsen-Ninomiya theorem}
 \label{NNthm}
In $2k$ Euclidean spacetime dimensions, apart from going over to the right action in the continuum limit, one can not have a lattice Dirac operator which satisfies all of the following conditions simultaneously:
\begin{enumerate}
    \item ultra-locality, meaning that only nearest neighbours enter the Dirac operator,
    \item hermiticity,
    \item translational invariance and
    \item chirality
\end{enumerate}   
of the massless Dirac operator, and then describe the right number of fermions from the Dirac operator. The following sections discuss different types of lattice fermion actions and how they confirm the theorem. Later, we shall study the Casimir effect for these lattice fermions in Chapter (\ref{ch3}).
\subsection{Naive discretization of fermions}
This process begins with the naive discretization of the fermionic part in the Lagrangian, followed by understanding the phenomenon of naive fermion doubling and rectifying it using Wilson’s action.
The spinor fields $\psi(n)$ and $\overline{\psi}(n)$ on lattice carry the same colour and flavour indices as their counterparts in the continuum case. The free fermionic action $S^0_F[\psi,\overline{\psi}]$ in continuum, once we set $A_\mu = 0$ is given by:
\begin{equation}
    S^0_F[\psi, \overline{\psi}] = \int d^4x \overline{\psi}(x)\left(\gamma_\mu \partial_\mu +m\right)\psi(x)
\end{equation}
The partial derivative is discretized by replacing the integral with the sum over the lattice points. One obtains the free fermionic action on the lattice as:
\begin{equation}
    S^0_F[\psi, \overline{\psi}] = a^4 \sum_{n \in \Lambda} \overline{\psi}(n) \left(\sum_{\mu = 1}^{4} \gamma_\mu \frac{\psi(n+\hat{\mu}) - \psi(n-\hat{\mu})}{2a}  +m_f\psi(n)\right)
\end{equation}
We can generalize and make the free fermionic action in an external gauge field $U$, gauge invariant and write it as:
\begin{equation}
\label{feractU}
    S_F[\psi, \overline{\psi},U] = a^4 \sum_{n \in \Lambda} \overline{\psi}(n) \left(\sum_{\mu = 1}^{4} \gamma_\mu \frac{ U_\mu(n)\psi(n+\hat{\mu}) - U_{-\mu}\psi(n-\hat{\mu})}{2a}  +m_f \psi(n)\right)
\end{equation}
This fermionic action on lattice is now gauge invariant.
Naive discretization is not necessarily the best way to perform the discretization. Therefore, we discuss other possibilities of discretizing fermion actions and the properties one wants to achieve of lattice fermions, with an overall sum over the Lorentz indices $\mu, \nu \in [1,4]$.
\subsection{Fermion doubling and Wilson’s fermion action}
\label{Doublingwilson}
 Fermions are defined as particles that they obey Fermi statistics. The vacuum expectation value must gain a minus sign and be antisymmetric if the quantum numbers of any two fermions are swapped.
All fermionic degrees in the lattice theory anti-commute with each other. In terms of anti-commuting Grassman numbers, the fermionic path integral may be derived from the canonical anti-commutation relations for fermions by introducing coherent states. The fermionic action (\ref{feractU}), being bilinear in $\overline{\psi}$ and $\psi$, can be written in a convenient form
\begin{equation}
   S_F[\psi, \overline{\psi},U] = a^4 \sum_{n,m \in \Lambda}  \overline{\psi}(n) D(n|m) {\psi}(m)  
\end{equation}
where the naive Dirac operator $ D(n|m)$ on the lattice is then given by :
\begin{equation}
\label{Dnm}
   D(n|m)  =  \sum_{\mu = 1}^{4} (\gamma_\mu) \frac{ U_\mu(n) \delta_{n+\hat{\mu},m} - U_{-\mu}(n) \delta_{n-\hat{\mu},m}}{2a}  +m_f\delta_{n,m}
\end{equation}
Let us now compute the Fourier transform of the lattice Dirac operator $D(n|m)$ for the case of free lattice
fermions when the gauge ﬁelds are  trivial, $U_\mu(n) = \mathbbm{1}$. Fourier transformation is applied independently to the two space–time arguments $n$ and $m$. The color indices are omitted for notational convenience in the Dirac space. The Fourier transform of the Dirac operator in (\ref{Dnm}) for trivial gauge ﬁeld reads :
\begin{align}
    \tilde{D}(p|q) &= \frac{1}{|\Lambda|}\sum_{n,m\in\Lambda} e^{-ip\cdot na}D(n|m) e^{-iq\cdot ma}\\
    &= \frac{1}{|\Lambda|}\sum_{n\in\Lambda} e^{-i(p-q)\cdot na}\left(\sum_{\mu = 1}^{4} \gamma_\mu\frac{e^{+iq_\mu a}-e^{-iq_\mu a}}{2a}  +m\mathbbm{1}\right)\\
    & = \delta(p-q)\tilde{D}(p)
\end{align}
Here, $|\Lambda|$ represents the total number of lattice points. The Fourier transform of the lattice Dirac operator is obtained by substituting $D(n|m)$. One calculates the inverse of the Dirac operator $D^{-1}(n|m)$  in real  space, by simply computing the inverse $\tilde{D}(p)^{-1}$ and inverting the Fourier transformation. Thus, obtaining:
\begin{align}
    \tilde{D}_{nf}(p) &= m_f\mathbbm{1} +\frac{i}{a}\sum_{\mu=1}^4 \gamma_\mu\sin(p_\mu a) \label{naive}\\
     \Rightarrow \tilde{D}_{nf}(p)^{-1} &= \frac{m_f\mathbbm{1}-ia^{-1}\sum_\mu \gamma_\mu\sin(p_\mu a)}{{m_f}^2 +a^{-2}\sum_\mu \sin(p_\mu a)^2}\\
     \Rightarrow D_{nf}^{-1}(n|m) &= \frac{1}{|\Lambda|}\sum_{p \in \overline{\Lambda}}\tilde{D}_{nf}(p)^{-1}e^{ip\cdot(n-m)a}\label{qo}
\end{align}
 \begin{figure}
     \centering
     \includegraphics[width=10cm]{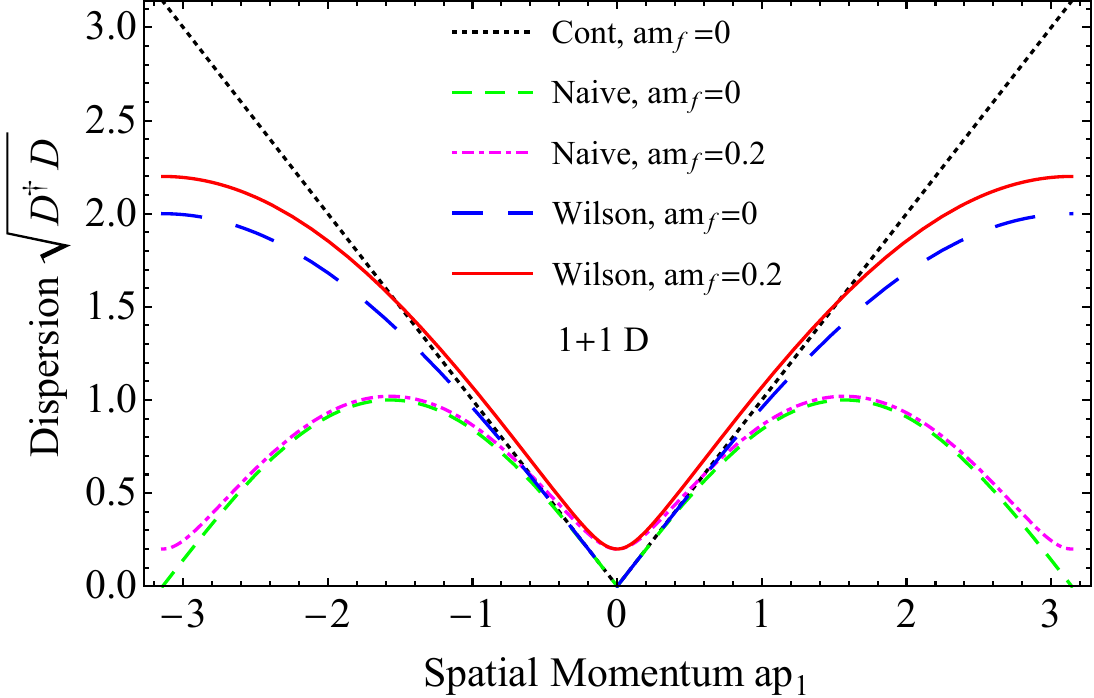}
     \caption{\footnotesize{Dispersion Relations for massless and massive, naive and Wilson lattice fermions in ($1+1$)- dimensional spacetime.}}
     \label{dispersionrelations}
 \end{figure}
The inverse of the Dirac Operator calculated for the free fermions in \ref{qo} is referred to as the \textit{quark propagator}. The quark propagator governs the behaviour of $n$-point functions and therefore is an important expression to analyze for free fermions in momentum space:
\begin{equation}
\label{uni}
    \tilde{D}_{nf}(p)^{-1} |_{m=0}= \frac{ -ia^{-1}\sum_\mu \gamma_\mu\sin(p_\mu a)}{a^{-2}\sum_\mu \sin(p_\mu a)^2} \xrightarrow{a\rightarrow0}\frac{-i\sum_\mu \gamma_\mu p_\mu}{p^2}
\end{equation}
Note that infinitely many lattice actions are possible, which all reproduce the continuum
action in the continuum limit $a \rightarrow 0$. Universality of lattice fermions demands that they all lead to the same result for observables in a quantum field theory. This also enables us to produce specific improved actions keeping the Nielsen Ninomiya theorem in mind. The universality property will play a significant role in the problem we will be addressing in the following chapters for the calculated Casimir energy of the Naive fermion. 

Note that the continuum expression for momentum space propagator for massless fermions to the far right in (\ref{uni}), has a pole at $p = (0, 0, 0, 0)$. It is this pole that actually corresponds to the single fermion described by the continuum Dirac operator. But, notice that the situation on the lattice, as represented by the central term in (\ref{uni}), is diﬀerent. The propagator for free fermions on the lattice has poles not only at $p_\mu = 0$, but also at $p_\mu = \pi/a$. The momentum space contains all momenta  within the range $p_\mu \in (-\pi/a, \pi/a]$. The fifteen additional unphysical poles from the lattice propagator are enlisted here: 
\begin{equation}
    p=(\pi/a,0,0,0),(0,\pi/a,0,0),\dots,(\pi/a,\pi/a,\pi/a,\pi/a).
\end{equation}
These additional unphysical degrees of freedom lead to the problem of \textit{doubling}. In order to remove the doublers, we need to distinguish between the proper pole $p_\mu = 0$ and the doublers that contain momentum components $p_\mu = \pi/a$. Kenneth Wilson proposed to add an extra term, such that compared to expression (\ref{naive}), the momentum space Dirac operator now reads
\begin{equation}
\label{Dp}
    \tilde{D}_{\text{W}}(p) = \tilde{D}_{nf}(p) +\mathbbm{1}\frac{1}{a}\sum_{\mu=1}^4 (1-\cos(p_\mu a))
\end{equation}
where $\tilde{D}_{nf}(p)$ is given by (\ref{naive}). This Wilson term acts as an additional mass term. But, it breaks all chiral symmetries on the lattice and hence, escapes the fermion doubling in accordance with the Nielsen Ninomiya theorem. For components with $p_\mu = 0$ the Wilson term simply vanishes. For each component with $p_\mu = \pi/a$ it provides an extra contribution $\frac{2}{a}$. This is exactly what we needed to eliminate doubling.  
This is the expression for Wilson’s fermion action on the lattice. The dispersion relations of the Naive and Wilson fermions (massive and massless) in ($1+1$)-dimensional spacetime have been plotted in Fig. \ref{dispersionrelations} to compare their dispersion relations to the continuum Dirac fermion. For additional references, refer to standard texts on the topic like Gattringer and Lang \cite{lang}, Jan Smit \cite{smit} or Montvay and Munster \cite{montvay} for theoretical aspects.
\subsection{Overlap fermions with Möbius domain wall kernel}
\label{Overlap}
We primarily want to achieve a Dirac operator which does not break chiral symmetry in this section. One has to find a massless Dirac operator which satisfies the Ginsparg-Wilson equation on the lattice to do so
\begin{equation}
    \{D,\gamma_5\} = 2aD\gamma_5 D
\end{equation}
which reduces to the continuum equation zero lattice spacing limit. Neuberger found such an operator, called it the overlap fermion operator, to serve this purpose:
\begin{equation}
\label{DOV}
    D_{\text{OV}}= \frac{1}{2}[1+\gamma_5 \text{sign}(H)], \;\;\;H=\gamma_5A
\end{equation}
where $A$ is a suitable $\gamma_5$-hermitian kernel Dirac operator and operator $H$ is hermitian. 
However, the computation of the signum function in (\ref{DOV}) is very expensive. 
One method to approximate the signum function is the domain wall fermions (DWF), first proposed by David Kaplan \cite{Kaplan}. 
The idea of domain wall fermions is to define a five-dimensional Euclidean lattice action with four-dimensional left- and right-handed fermions located on the different boundaries of the fifth dimension.
For simplicity, the extra dimension is considered to be of infinite length. 
The M\"{o}bius domain wall (MDW) kernel operator is:
\begin{equation}
\label{MDW}
    aD_{\text{MDW}}\equiv \frac{b(aD_{\text{W}})}{2+c(aD_{\text{W}})}
\end{equation}
where, the $b$ and $c$ are called M\"{o}bius parameters, and $D_{\text{W}}$ is the previously defined Wilson Dirac operator with $r=1$. The operator (\ref{MDW}) at values $b=c=1$ corresponds to the conventional Shamir type, whereas at $b=2$ and $c=0$ is the Boriçi-type formulation. The original fermion mass ($m_f$) in (\ref{Dp}) is now replaced by the domain wall height as $am_f \rightarrow -M_0$. We define the Overlap Dirac operator in terms of the MDW kernel operator $D_{\text{MDW}}$ as:
\begin{equation}
    aD_{\text{OV}}\equiv(2-cM_0)M_0am_{\text{PV}}\times\frac{(1+am_f)+(1-am_f)V}{(1+am_{\text{PV}})+(1-am_{\text{PV}})V}
\end{equation}
where $m_f$ is the fermion mass, $m_{\text{PV}}$ is the Pauli-Villars mass and expression $V$ is defined as:
\begin{equation}
\label{V}
    V\equiv\gamma_5\;\text{sign}(\gamma_5 a D_{\text{MDW}}) = \frac{D_{\text{MDW}}}{\sqrt{D_{\text{MDW}}^\dagger D_{\text{MDW}}}}
\end{equation}
Note that $V^\dagger V=1$. 
Using the definition of the energy momentum dispersion relation we can calculate the dispersion relation for Overlap fermion to be:
\begin{equation}
\label{EOV}
    a^2E^2_{\text{OV}} = [(2-cM_0)M_0 m_{\text{PV}}]^2\times\frac{2[1+(am_f)^2]+[1-(am_f)^2](V^\dagger+V)}{2[1+(m_{\text{PV}})^2]+[1-(m_{\text{PV}})^2](V^\dagger+V)}
\end{equation}
Using the commutation relation of $(V^\dagger +V)$ and $V$, we obtain the following expression in terms of the Wilson fermion Dirac operator $D_{\text{W}}$:
\begin{align}
    V^\dagger + V &=\left(D^\dagger_{\text{MDW}}+D_{\text{MDW}}\right)\frac{1}{\sqrt{D_{\text{MDW}}^\dagger D_{\text{MDW}}}}\nonumber\\
    &=2(D_{\text{W}}+D^\dagger_{\text{W}}+cD^\dagger_{\text{W}}D_{\text{W}})\frac{1}{\sqrt{D^\dagger_{\text{W}}D_{\text{W}}}}\times\frac{1}{\sqrt{4+2c(D^\dagger_{\text{W}}+D_{\text{W}})+c^2D^\dagger_{\text{W}}D_{\text{W}}}}
\end{align}
Additionally, note that this Overlap fermion operator and the dispersion relation are independent of the M\"{o}bius parameter $b$. It can also be explicitly shown that an overlap formulation with Wilson fermion kernel is doubler free. 
 
 \chapter{The Casimir effect: A brief review}
 
The Casimir effect is regarded as one of the few and most interesting directly observable consequences of the vacuum of quantum fields. The earliest theoretical derivation of this phenomenon by H.B.G Casimir in 1948 \cite{Casimir:1948dh} was based on calculating the attractive force between two neutral parallel conducting plates caused by the electromagnetic field's quantum fluctuations in vacuum. The quantum vacuum is filled with continuously fluctuating fields and exhibits properties very unlike the corresponding classical vacuum. The virtual particles are a transient quantum fluctuation that can be created and annihilated back to the vacuum for a time interval allowed by the uncertainty relation
\begin{equation}
\Delta E \cdot \Delta t \geq \frac{\hbar}{2}.
\end{equation}
The presence of macroscopic bodies like the perfectly conducting parallel plates in vacuum introduce non-trivial boundary conditions in the system. This modifies the spectrum of zero-point fluctuations in the vacuum state and restricts the normally incident component of the allowed wave modes between the parallel plates to discrete values. Its manifestation is seen as an attractive force due to a finite lowering of the vacuum energy, inversely proportional to the distance between parallel plates. The setup of parallel plates and explanation through quantum vacuum fluctuations is the simplest of the many possible configurations and explanations of the Casimir effect. This rather counterintuitive and remarkable phenomenon has now been generalized to various physical situations and geometrical setups \cite{Milton:2001yy, Mostepanenko_1988}. A repulsive Casimir effect  in the case of a perfectly conducting sphere was theoretically predicted by Boyer \cite{Boyer}, where the quantum fluctuations of the electromagnetic field produce a repulsive force on the wall of the sphere.

\begin{figure}[t!]
    \centering
    \includegraphics[width=7.5cm]{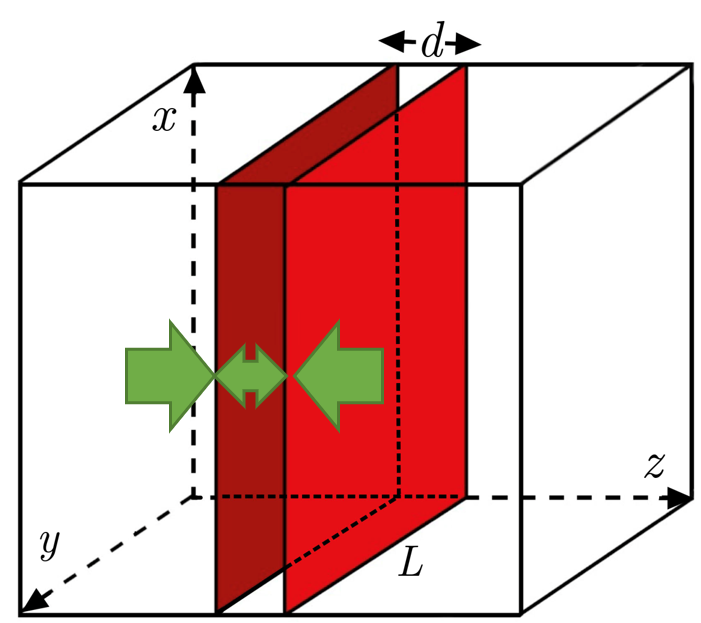}
    \caption{\footnotesize{Here, $L$ is the transverse dimension of the plate. The distance between the plates is $d$, and the area of the plates $(A) = L^2$.}}
    \label{casimirsetup}
\end{figure}

  In an arrangement of the parallel conductor plates, as shown in Fig. \ref{casimirsetup}, the components of the wave vector parallel to the plate surfaces remain unaffected and continue to form a continuous spectrum. Also, note that introducing perfectly conducting parallel plates in vacuum with continuous-wave modes affects both the contribution of modes allowed between the plates and the surrounding region.
The perfectly conducting surface imposes three conditions on the electromagnetic field as it restricts the transverse components of the electric field  $E_\parallel$ in two orthogonal directions and the normal component of the magnetic field $B_\perp$. Assuming the parallel plates lie in the $x$-$y$ plane, as shown in Fig. \ref{casimirsetup}, the standing waves are :
\begin{equation}
    \psi({x,y,z,t}) = e^{-i\omega_n t}e^{ik_xx +i k_yy} \sin(k_zz)
\end{equation}
Here, $k_x$ and $k_y$ are the wave vector components in directions parallel to the plates.

In the same year, Casimir, along with Dirk Polder, described a similar effect experienced by a neutral atom in the vicinity of a macroscopic interface referred to as Casimir–Polder force.\cite{Polder} 
This established a duality between the Casimir effect and van der walls force between molecules, thus showing that the structure of gross matter is intimately tied to the Casimir effect.

  These forces have been measured, and the phenomenon was confirmed to a high degree of precision in several experiments \cite{Lamoreaux, Mohideen} recently. One might conclude quantum phenomena like these to be esoteric, with a limited practical consequence, but their effects are very relevant and of increasing significance in nanotechnology, like the silicon integrated circuit technology based on micro-and nano-electromechanical systems, as characteristic distances get smaller. The fact that energy density in certain regions of space is negative relative to the ordinary vacuum energy was established by the Casimir effect. This is exciting, as such effects contribute to the stability of hadrons and model colour-confinement \cite{Johnson:1975zp}, might make it possible to stabilize a traversable wormhole \cite{Morris}, and provide significant insights into the cosmological constant problem \cite{MAHAJAN20066}.

\section{Massless Bosonic Casimir effect}
\label{photons}
We now introduce boundary conditions in the direction parallel to surface normals for calculating the Casimir energy and the force per unit area between two infinitely large ($L\rightarrow\infty$), perfectly conducting parallel plates a distance `$d$' apart in $(3+1)$-dimensions. We have,  
\begin{equation}
k_x L = l \pi\;\;;\;\;  
k_y L = m \pi\;\;;\;\;   
k_z d = n \pi\;\;\;\;\;\;(l,m,n \in \mathbb{N}) 
\end{equation}
Thus, the frequency is expressed in terms of the momentum componenets as:
\begin{equation}
\omega_{lmn} = k_{lmn}c = \pi c\left[\frac{l^2}{L^2} + \frac{m^2}{L^2} +\frac{n^2}{d^2}\right]^{\frac{1}{2}} 
\end{equation}
The zero-point energy of the field inside the cavity is therefore :
\begin{equation}
\label{discrete}
{\sum_{l,m,n}}^\prime (2)\left(\frac{1}{2}\right)\hbar \omega_{lmn} = \sum_{l,m,n}\pi \hbar c \left[\frac{l^2}{L^2} + \frac{m^2}{L^2} +\frac{n^2}{d^2}\right]^{\frac{1}{2}}
\end{equation}
Here, a factor of (2) arises from two independent polarizations of modes ($l,m,n \neq 0$). The prime on summation indicates that if one of  $l,m,n$ is zero, this factor of two must be removed.
Assuming $L$ to be infinitely large, as compared to $d$, we replace the sums over `$l$' and `$m$' in (\ref{discrete}) by appropriate integrals as:
\begin{equation}
\label{cont}
\sum_{l,m,n} \rightarrow {\sum_n}^\prime\frac{L^2}{\pi^2}\int dk_x \int dk_y  
\end{equation}
{Therefore, the vacuum energy density (per unit area $A$)} of the configuration with distance $d$ between the plates, denoted by $\langle E(d)  \rangle$ is :
\begin{equation}
\label{Edexp}
 {\langle E(d)  \rangle}  = {\sum_{l,m,n}}^\prime (2)\left(\frac{1}{2}\right)\hbar \omega_{lmn} \rightarrow \frac{\hbar c}{\pi^2} {\sum_n}^\prime\int_0^\infty dk_x \int_0^\infty dk_y \left(k_x^2 + k_y^2 + \frac{n^2\pi^2}{d^2}\right)^{\frac{1}{2}} 
\end{equation}
We note that this quantity diverges to infinity. In the case when the parallel plates are very far apart and`$d$' is taken to be arbitrarily large, the sum over `$n$' in (\ref{Edexp}) will be replaced by :
\begin{equation}
{\langle E(d\rightarrow\infty)  \rangle}  = \frac{\hbar c}{\pi^2} \frac{d}{\pi} \int_0^\infty dk_x \int_0^\infty dk_y \int_0^\infty dk_z\left(k_x^2 + k_y^2 + k_z^2\right)^{\frac{1}{2}} 
\end{equation}
which also diverges to infinity.\\
The potential energy of the system with a separation of $d$ between plates for the photon field ($p$) is $E^{3+1\text{D},\text{cont},p}_{\text{Cas}}(d)$, given as the difference between the Energy at separation $d$ ($\langle E(d)  \rangle $) and the energy at infinite separation ($\langle E(d\rightarrow\infty)  \rangle $).
\begin{equation}
E^{3+1\text{D},\text{cont},p}_{\text{Cas}}(d) =\langle E(d)  \rangle - \langle E(d\rightarrow\infty)  \rangle
\end{equation}
 which is the energy needed to bring the plates from an infinite separation to a separation `$d$' :
\begin{align}
{E^{3+1\text{D},\text{cont},p}_{\text{Cas}}(d)} &=\frac{\hbar c}{\pi^2} \Big[ {\sum_n}^\prime \int_0^\infty dk_x \int_0^\infty dk_y \left(k_x^2 + k_y^2 + \frac{n^2\pi^2}{d^2}\right)^{\frac{1}{2}} \\&- \frac{d}{\pi} \int_0^\infty dk_x \int_0^\infty dk_y \int_0^\infty dk_z\left(k_x^2 + k_y^2 + k_z^2\right)^{\frac{1}{2}} \Big] 
\end{align}

To simplify the expression, we rewrite the above equation using polar co-ordinates `$ u $' and `$ \theta $' in the $ k_x k_y $ plane ($ dk_x dk_y = u \;du \;d\theta $). Here, $ k_x^2 + k_y^2 = u^2 $. We have  
\begin{equation}
\label{polar}
{E^{3+1\text{D},\text{cont},p}_{\text{Cas}}(d)} = \frac{\hbar c}{\pi^2} \frac{\pi}{2} \Big[ {\sum_n}^\prime\int_0^\infty du u \left( u^2 + \frac{n^2\pi^2}{d^2}\right)^{\frac{1}{2}} - \frac{d}{\pi} \int_0^\infty dk_z \int_0^\infty du u \left(u^2 + k_z^2\right)^{\frac{1}{2}} \Big] 
\end{equation}
where, $ \theta \in \left(0,\frac{\pi}{2}\right)$ for $k_x, k_y>0$.\\
Introducing a cutoff function $ f(k) = f([u^2 + k_z^2]^{\frac{1}{2}}) $ such that:
\begin{equation} 
f(k) = \Big\{ \begin{array}{cc}
                1 & \text{if $k<<k_m$}\\ 
                0 & \text{if $k>>k_m$}\\
               \end{array}  
\end{equation}
Our assumption of a perfectly conducting wall breaks down at small wavelengths, especially for wavelengths small compared to an atomic dimension.
  The quantity $k_m$ is assumed to be on the order of plasma frequency of the metals comprising the plates, therefore, $  k_m \approx \frac{\omega_p}{c} \approx \frac{1}{a_0}$. The primary assumption here is that the Casimir effect is a low-frequency phenomenon.

Rewriting (\ref{polar}) by taking $ x = \frac{u^2d^2}{\pi^2}$ and $\kappa = \frac{k_z d}{\pi}$, we have:
\begin{align}
E^{3+1\text{D},\text{cont},p}_{\text{Cas}}(d) &= \frac{\hbar c}{4\pi}\frac{\pi^3}{d^3} \Bigg[ {\sum_n}^\prime \int_0^\infty dx ( x +  n^2)^{\frac{1}{2}} f\left(\frac{\pi}{d}[x+n^2]^{\frac{1}{2}}\right) \\&- \frac{d}{\pi} \int_0^\infty dk_z \int_0^\infty dx (x + \kappa^2)^{\frac{1}{2}}  f\left(\frac{\pi}{d}[x+\kappa^2]^{\frac{1}{2}}\right) \Bigg] 
\end{align}
Note that here, we have introduced the cut off function. Further expressed as :
\begin{equation}
\label{subs}
{E^{3+1\text{D},\text{cont},p}_{\text{Cas}}(d)}  =   \frac{\pi^2 \hbar c}{4a^3}\left[ \frac{1}{2}F(0) + \sum_{n=1}^\infty F(n) - \int_0^\infty d\kappa F(\kappa)\right] 
\end{equation}

where, $F(\kappa) = \int_0^\infty dx\left(x+\kappa^2\right)^{\frac{1}{2}} f\left(\frac{\pi}{d}[x+\kappa^2]^{\frac{1}{2}}\right)$.\\
Let the $k^{th}$ derivative be denoted as $ F^{(k)}(0)$. We use the Euler-Lagrange summation formula :
\begin{equation}
\sum_{n=1}^\infty F(n) - \int_0^\infty d\kappa F(\kappa) = -\frac{1}{2}F(0) -\frac{1}{12}F^{(1)}(0)+ \frac{1}{720}F^{(3)}(0) ... \;\;\;
\end{equation}
We substitute $u=x+\kappa^2$ in $F(\kappa)$ and obtain its higher derivatives using the Leibniz Integral rule:
\begin{align}
                   F(\kappa) &= \int_{\kappa^2}^\infty du\sqrt{u}f\left(\frac{\pi}{d}\sqrt{u}\right)\\                    \Rightarrow F^{(1)}\left(\kappa\right) &= -2{\kappa^2} f\left(\frac{\pi}{d}\kappa\right)
\end{align}
All the higher derivatives vanish if we assume that all derivatives of the cutoff function $`f$' vanish at $\kappa = 0$. We have:
\begin{equation}
\label{result}
\sum_{n=1}^\infty F(n) - \int_0^\infty d\kappa F(\kappa) = -\frac{1}{2}F(0) -\frac{4}{720};
\end{equation}
Substituting (\ref{result}) in (\ref{subs}), we get: 
\begin{equation}
\label{Ecas3+1}
{E^{3+1\text{D},\text{cont},p}_{\text{Cas}}}  = \frac{\pi^2\hbar c}{4d^3}\Big(\frac{-4}{720}\Big) = - \frac{\pi^2\hbar c}{720d^3} \\ 
\end{equation}
 {The attractive force per unit area is calculated as: }
\begin{align}
{F^{3+1\text{D},\text{cont},p}_{\text{Cas}}} &= -\frac{\partial E^{3+1\text{D},\text{cont},p}_{\text{Cas}}}{\partial d} \nonumber\\
             \Rightarrow {F^{\text{3+1,cont}}_{\text{Cas}}}  &= -\frac{\pi^2\hbar c}{240d^4}  \label{CForce}
\end{align}
Note that the Casimir energy per unit area in (\ref{Ecas3+1}) is a finite and purely quantum quantity. It is also independent of the cut-off function  $f(k)$. Thus, changes in the infinite zero-point energy of the quantum electrodynamic vacuum can be finite and observable.\\
Alternatively, the Abel-Plana formulae can also be employed to obtain the exact same result. A generalized expression for the Casimir energy in $(D+1)$-dimension has been obtained by both the cut-off method \cite{Svaiter:1989gz} and the zeta-function regularization \cite{Milton:2001yy}. The expression is 
   \begin{equation}
   \label{Fp}
         F^{\text{D+1},\text{cont},p}_{\text{Cas}} = -2D(4\pi)^{\frac{D+1}{2}}\Gamma\left(\frac{D+1}{2}\right)\zeta(D+1)\frac{1}{d^{D+1}} .
   \end{equation}
  This can be used to obtain the expression for Casimir energy in any arbitrary dimensional spacetime. If there exists a cut-off length $d_c \approx \frac{1}{k_c} $ in nature, acting as a UV regulator, this suppresses the modes with wavelengths $\lambda \lesssim d_c$ leading to modification in the Casimir force between the plates. 
  The maximum wavelength allowed between plates separated by distance `$d$', is called $\lambda_{max} \approx 2d$. So the modes with wavelengths lying between the range $d_c \lesssim \lambda \lesssim 2d \label{2d}$ only contribute to our expression of the Casimir effect. For wavelengths much smaller than the $d_c$, it is as if the plates become transparent (they penetrate) for the electromagnetic waves. As a result, the boundary conditions, which were essential for the discreteness of the spectrum, can no longer be obeyed. Now, we shall briefly discuss an interesting approach to this result.  
   
\subsection{Approach by Radiation Pressure}
\label{radpress}
 There exists an alternate interpretation to approach this problem found by Milloni \cite{MILONNI1994253}. The zero point virtual photons of the vacuum carry linear momentum $\frac{\hbar k}{2}$. Their reflections off the plates effectively act to push the plates together.  
Consider the radiation pressure exerted by a plane wave incident normally on a plate.   
 One cosine component gives the normal component of the imparted linear momentum and the increase factor in the incident elemental area.
 A mode of frequency `$\omega$' is formed by reflections off the plates contribute pressure:
\begin{align}
P &= 2\left(\frac{1}{2}\right)\left(\frac{1}{2}\hbar \omega\right)\cdot\frac{1}{V}\cdot {\cos}^2(\theta)
= \frac{\hbar\omega}{2V}\cdot\frac{k_z^2}{ k^2}\label{RP}
\end{align}
 where $\theta$ is the angle of incidence. The zero-point energy of each mode is divided equally between waves propagating toward or away from each plate.
  For large plates, $k_x$ and $ k_y$ take continuum values while $k_z = \frac{n\pi}{d}, \;(n\in \mathbb{N})$.
  Therefore, adding contributions from all modes of space between the plates, we get the total outward pressure $P_{out}$ from (\ref{RP}) using similar techniques as the previous section (\ref{photons}).
  A continuum of allowed frequencies contribute to the inward pressure $P_{in}$ to push the plates together.
 Both these quantities diverge. We calculate the difference between the two pressures, $P_{\text{out}}-P_{\text{in}}$. 
 It is rather insightful to find that, from detailed computation given in \cite{MILONNI1994253}, one obtains the exact same expression for the Casimir force (per unit area $A$) in (\ref{CForce}) for the above quantity. This interpretation may help one to understand the role Casimir effect plays in the MIT Bag model better. 
 

 \section{MIT Bag Model and Boundary Conditions}
In the theory of elementary particles, the Casimir effect most directly affects the physical properties of hadrons in the MIT Bag model \cite{Johnson:1975zp} and, in turn, affects the physics around us. The MIT Bag model is a relativistic, gauge invariant, and heuristic model of quark confinement, which attempts to study the substructure of strongly correlated systems called hadrons as a bag, which can include confined fluctuating quark and anti-quark in a meson or three quark fields in a baryon to the interior of a spherical cavity of radius $R$.
 
One of the first attempts at this was Nikolay Bogoliubov's Bag model, where quarks were given an enormous mass $m$, thereby confining them by making them unable to propagate. The quarks were confined to a spherical cavity bound with a strongly attracting field of strength proportional to the quark mass, $m$. This was done to model the asymptotic freedom observed at a very close range. The MIT Bag model has evolved to include an additional term called the ``bag constant", $B$, in the Lagrangian density (\ref{Lag}). This inward vacuum pressure $B$ on the surface of the bag balances the outward pressure of the quarks confined in the interior. Several studies have shown a relationship between this constant and the Casimir energy. 

From QED, we know that the screening effect for photons, even in the absence of a polarizable material, yields permittivity $\epsilon>1$ in vacuum leading to a decreasing Coulomb force with increasing distance between the particles, $r$. Similarly, the colour confinement of quarks in QCD overpowers the screening and yields $\epsilon_c<1$ for sufficiently large $r$, leading to increasing attractive force with distance. However, as $r$ decreases, the color permittivity $\epsilon_c\rightarrow1$, as shown in the Fig. \ref{MITbag}(a). If $\mu_c$ denotes the colour permeability of the medium, the asymptotic freedom is implemented by modelling the interior of the cavity as a chromomagnetic vacuum ($\epsilon_c=\mu_c=1$), while the exterior as a perfect chromomagnetic conductor ($\epsilon_c=0,\;\mu_c=\infty$). If one exchanges $\epsilon_c$ with  $\mu_c$, the ideal bag representation would be analogous to a perfect conductor in classical electrodynamics, thus providing a direct analogy. With the boundary condition of a perfect conductor i.e $\hat{n}\times \overrightarrow{\text{E}}=0$ and $\hat{n}\cdot \overrightarrow{\text{B}}=0$ in mind, the boundary condition for the color fields as evident from Fig. \ref{MITbag}(b) becomes $\hat{n}\times \overrightarrow{\text{B}}=0$ and $\hat{n}\cdot \overrightarrow{\text{E}}=0$. \cite{Bhaduri:1988gc}

The equations of motion are found by minimizing the action $S$, given by integral of the Lagrangian density $\mathcal{L}$ over spacetime.
     \label{MITbag}
  \begin{figure}
\centering
    \subfloat\footnotesize{(a)}{{\includegraphics[width=9.9cm]{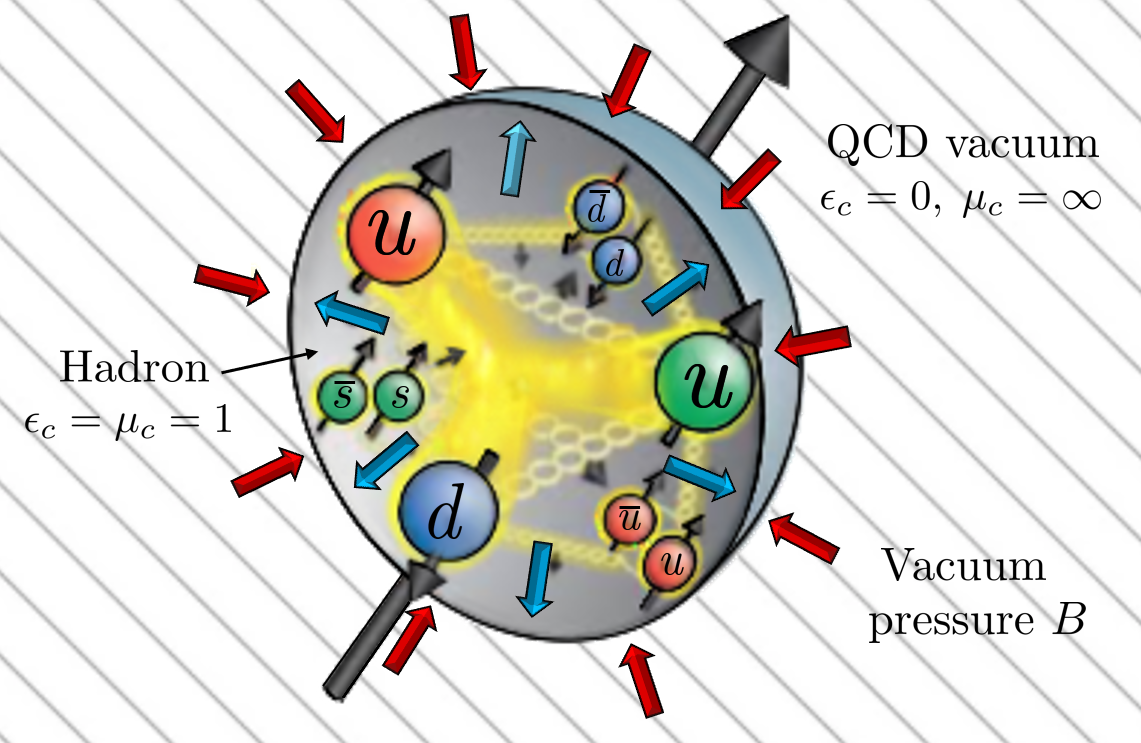}}} 
    \subfloat\footnotesize{(b)}{{\includegraphics[width=4.9cm]{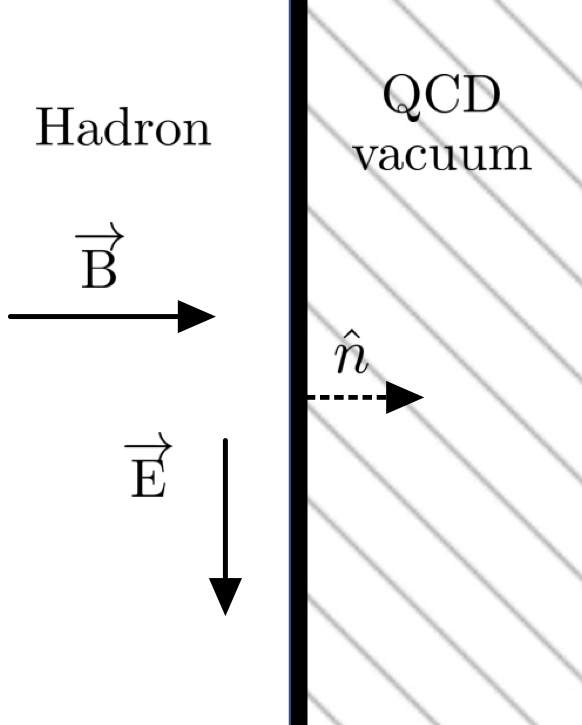} }}%
\caption{\footnotesize{(a) Color permeability and
permittivity in the bag-model. The outward pressure from the quark fields is balanced by the inward pressure on the spherical cavity modelled as a hadron. (b) Color fields at the surface of the bag.}}%
\label{MITbag}
\end{figure}
 The field variables correspond to only the subset of points inside an extended particle modelled as a bag. 
 The Lagrangian density for the MIT Bag Model is :
 \begin{equation}
 \label{Lag}
     \mathcal{L} = \left[\frac{i}{2}(\overline{\psi}\gamma^\mu\partial_\mu\psi-(\partial_\mu\overline{\psi})\gamma^\mu\psi)-B)\right]\theta_v(x)-\frac{1}{2}\overline{\psi}\psi\Delta_s
 \end{equation}
 where, ${\psi}$ and $\overline{\psi}$ are the quark and  anti-quark fields respectively, associated only with points in the interior of the hadron bag. Note that their value is zero outside, by definition. $\theta_v$ is the step function valued one inside, and zero outside the bag. $\Delta_s$ is the derivative of $\theta_v$:
 \begin{equation}
     \theta_v = \theta(R-r),\;\;\; \frac{\partial \theta_v}{\partial x^\mu} = n_\mu\Delta_s,\;\;\; \Delta_s = \delta(R-r)
 \end{equation}

 Here, $\delta$ is the Kronecker delta, $n_\mu$ is a unit outward normal vector to the surface, and $R$ is the radius of the bag. We minimize the action, $S$, by the principle of least action. Correspondingly setting $\delta S=0 $, we obtain the boundary condition for $\psi$ by inserting $\overline{\psi}$ as the field in Euler-Lagrange equations of motion for all independent fields. 
 \begin{equation}
 \label{PEL}
     \frac{\partial\mathcal{L}}{\partial\overline{\psi}}-\partial_\mu\left(\frac{\partial\mathcal{L}}{\partial(\partial_\mu\overline{\psi})}\right) = 0
 \end{equation}
 Substituting the Lagrangian density $\mathcal{L}$ in (\ref{Lag}) and calculating the two terms separately yields:
 \begin{align}
       \frac{\partial\mathcal{L}}{\partial\overline{\psi}}&= \frac{\partial}{\partial\overline{\psi}}\left(\left[\frac{i}{2}(\overline{\psi}\gamma^\mu\partial_\mu\psi-(\partial_\mu\overline{\psi})\gamma^\mu\psi)-B)\right]\theta_v(x)-\frac{1}{2}\overline{\psi}\psi\Delta_s\right)\\  \label{first}
       &= \left(\frac{i}{2}\gamma^\mu\partial_\mu\psi\right)\theta_v(x)-\frac{1}{2}\psi\Delta_s
 \end{align}
 
 And the second term is calculated as :
 \begin{equation}
 \label{second}
     \partial_\mu\left(\frac{\partial\mathcal{L}}{\partial(\partial_\mu\overline{\psi})}\right) =  \partial_\mu\left[\left(-\frac{i}{2}\gamma^\mu\psi\right)\theta_v(x)\right] = \left(-\frac{i}{2}\gamma^\mu\partial_\mu\psi\right)\theta_v(x)-\frac{i}{2}\gamma^\mu n_\mu\psi\Delta_s
 \end{equation}
 Substituting back (\ref{first}) and(\ref{second}) in (\ref{PEL}):
 \begin{align}
     \left(\frac{i}{2}\gamma^\mu\partial_\mu\psi\right)\theta_v(x)&-\frac{1}{2}\psi\Delta_s - \left[\left(-\frac{i}{2}\gamma^\mu\partial_\mu\psi\right)\theta_v(x)-\frac{i}{2}\gamma^\mu n_\mu\psi\Delta_s\right] = 0\\
     &\Rightarrow \left(i\gamma^\mu\partial_\mu\psi\right)\theta_v(x) + \frac{1}{2}\left(i\gamma^\mu n_\mu\psi-\psi\right)\Delta_s = 0
 \end{align}
 From the above-simplified expression, two different boundary conditions are derived:
 \begin{enumerate}[(I)]
     \item Inside the bag, where $\Delta_s = 0$ and $\theta_v = 1$ we have:
     \begin{equation}
         i\gamma^\mu\partial_\mu\psi = 0
     \end{equation}
     \item On the surface of the bag, where $\Delta_s = \infty$ and $\theta_v = 0$ we have:
     \begin{equation}
         \frac{1}{2}\left(i\gamma^\mu n_\mu\psi-\psi\right)\Delta_s = 0
     \end{equation}
     But as $\Delta_s = \infty $, the expression in parenthesis must be zero on the surface of the bag. Thus, we get the boundary condition for the quark field $\psi$ to be followed on the surface of the bag:
     \begin{equation}
     \label{boundpsi1}
         i\gamma^\mu n_\mu\psi = \psi
     \end{equation}
     Following a similar procedure of inserting the $\psi$ field in the equation (\ref{PEL}), the boundary condition for the corresponding anti-quark field $\overline{\psi}$ on the surface of the bag is:
     \begin{equation}
     \label{boundpsi2}
         -i\gamma^\mu n_\mu\overline{\psi} = \overline{\psi}
     \end{equation}
     \end{enumerate}
     The local expectation value of the field on the bag surface is given by calculating the $\langle{\overline{\psi}}{\psi}\rangle$. Substituting (\ref{boundpsi1}, \ref{boundpsi2}) in this expression yields:
     \begin{equation}
         \langle{\overline{\psi}}{\psi}\rangle = \braket{\overline{\psi}}{(i\gamma^\mu n_\mu)\psi} = \braket{-i(\overline{\psi}\gamma^\mu n_\mu)}{\psi}
     \end{equation}
     We find that the local expectation value $\langle{\overline{\psi}}{\psi}\rangle = 0$, and consequently, the probability density of the quark field on the bag surface is zero. Thus, this boundary condition ensures that there is no quark current on the surface of the bag and that the quarks are confined within the bag modelled as a hadron. Although, it is interesting to note that $\langle{\overline{\psi}}{\psi}\rangle$ is nonzero and rapidly varying between the plates \cite{Lutken}. From the Lagrangian density of the model, it is found that $B$ must be negative. In accordance with the radiation pressure interpretation (\ref{radpress}), this suggests that there exists an outward pressure exerted by the confined particles within the bag, countering the inward vacuum pressure $B$ shown in Fig \ref{MITbag}(a).  
  The mass and magnetic moment of hadrons is predicted using certain parameters, principally the bag constant $B$, the ``zero-point energy" dependence, radius of the hadron $R$, and strong coupling constant $\alpha_s$. \cite{Bernotas}
\section{Fermionic Casimir effect}

We described the Casimir effect for the photon field using the quantum electrodynamical vacuum in section (\ref{photons}). We shall now investigate and extend this phenomenon on parallel conducting surfaces for fermions and other gauge fields. The free gluon fields behave exactly like the electromagnetic field and thus must lead to the same result. The zero-point energy of a bosonic oscillator is  $+\frac{1}{2}\hbar\omega$, while for a fermionic oscillator it is  $-\frac{1}{2}\hbar\omega$. Thus, we expect the quarks which obey the Fermi-Dirac statistics to behave differently. Surprisingly, this is not the case. The force between two parallel confining plates for a fermionic field is attractive and has the exact same dependence on the distance between parallel plates `$d$' as the Casimir energy for photons. In order to prevent the crossing of parallel conducting plates by the fermionic current, we adopt the MIT bag model boundary conditions \cite{Johnson:1975zp} discussed and obtained in (\ref{boundpsi1}) appropriate for the Dirac equation.
  
  Let us consider the Dirac fermionic field $ \psi $ on a general ($D+1$)-dimensional spacetime. We solve for a case where one spatial dimension, $x^1$ is compact. The field $\psi$ has $N_D$ components given by $2^{\frac{D+1}{2}}$ if $D$ is odd and by $2^{\frac{D}{2}}$ if $D$ is even, and follows the Dirac equation given by :
  \begin{equation}
  \label{diraceqn}
      (i\gamma^\mu\partial_\mu-m)\psi  = 0
  \end{equation}
  
The field $\psi$ is confined between the two parallel plates placed at $x^1 = 0$ and $x^1 = d$ and thus, satisfies the MIT Boundary conditions on these plates. The boundary conditions are :
\begin{equation}
\label{bound}
(1+i\gamma^\mu n_\mu)\psi\Big\vert_{x^1 =0,d}= 0
\end{equation}
 Here, $ n_\mu $ represents the unit outward normal vector to the boundary surfaces, and $\gamma^\mu$ are the Dirac matrices. For our configuration setup, the parallel plates placed at $x^1=0$ and $d$ we have unit outward normals as $\left.n_\mu\right\vert_{x^1=0}=-\delta^{x^1}_\mu$ and $\left.n_\mu\right\vert_{x^1=d}=\delta^{x^1}_\mu$, and thus the boundary conditions are respectively:
\begin{equation}
(1 \mp i \gamma^1)\psi = 0
\end{equation}
These boundary conditions (\ref{bound}) satisfy the natural requirement that the current of particles across the boundary must vanish. Using the chiral representation of the Dirac matrices,
\begin{equation}
\label{matrix}
    \gamma^0 = \left( \begin{array}{cc}
1 & 0 \\
0 & -1
\end{array} \right), \;\;\;\;
\gamma^j = \left( \begin{array}{cc}
0 & \sigma_j\\
-\sigma_j & 0
\end{array} \right)
\end{equation}
 where $\sigma_j$ are the Pauli matrices and satisfy $\{\sigma_j,\sigma_l\} = 2\delta_{jl}$ with $j = 1,\dots, D$.\\ The standard positive and negative frequency solutions of the Dirac equation in (\ref{diraceqn}), with time dependence of the form $e^{\mp i\omega t}$ are respectively: \cite{tong}
 \begin{equation}
  \label{varphi1}
 \psi^{(+)} = e^{-i\omega t}\left( \begin{array}{cc}
\varphi^{(+)} \\
\dfrac{-i\bm{\sigma\cdot\nabla}\varphi^{(+)}}{\omega +m}
\end{array} \right), \;\;\text{and  } 
 \psi^{(-)} = e^{i\omega t}\left( \begin{array}{cc}
\dfrac{i\bm{\sigma\cdot\nabla}\varphi^{(-)}}{\omega +m}\\
\varphi^{(-)} 
\end{array} \right),
 \end{equation}
 where $\bm{\sigma} = (\sigma ^1,...,\sigma ^{D})$ and the spinors $\varphi^{(+)}$ and $\varphi^{(-)}$ correspond to the particle and antiparticle fields respectively. They can be stated in the following explicit form: \cite{Burdman, Henrik}
 \begin{equation}
 \label{varphi2}
     \varphi^{(\pm)} = ( \varphi_+^{(\pm)} e^{ik_1x^1} + \varphi_-^{(\pm)} e^{-ik_1x^1}) \exp{\pm i\sum_{j=2}^D k_jx^j}
     \end{equation}
 and the the energy-momentum dispersion relation is given by $\omega^2 = \sum_{j=1}^D {k_j}^2 + m^2$.\\
    On applying the massless boundary conditions on the plate located at $x^1 = 0$, we have:  
 \begin{align}
 \label{x=0}
 &\Rightarrow (1 - i \gamma^1)\psi^{(+)} = 0
    \end{align}
    Note that the outward unit normal for this plate would be in the negative $x^1$ direction. The detailed derivation after substitution of (\ref{matrix}) and (\ref{varphi1}, \ref{varphi2}) in (\ref{x=0}) is done in Appendix A (\ref{derx}). We obtain the following equation from both the elements of the matrix, since $\sigma_1^2 = \mathbbm{1}$. Similarly, we can as well solve for the relation between $\varphi_{+}^{(-)}$ and $\varphi_{-}^{(-)}$. Combining both the results at $x^1=0$ we get:
 \begin{equation}
 \label{rel1}
      \varphi_{+}^{(\pm)}= -\frac{m(m+\omega) + {k_1}^2 - \sigma_1 k_1  \bm{\sigma}_\parallel \cdot \bm{k}_\parallel}{(m\mp ik_1)(m+\omega)}\varphi_{-}^{(\pm)}
 \end{equation}
 Now, we similarly compute a relation between $ \varphi_{+}^{(\pm)}$ and $\varphi_{-}^{(\pm)}$ using the boundary condition at $x^1 = d$ with the corresponding outward unit normal along the positive $x^1$ direction, we get:
 \begin{align}
 &\Rightarrow (1 + i \gamma^1)\psi^{(+)} = 0
\end{align}
    We obtain the following equation from both matrix elements after calculating the matrix. Then, substituting (\ref{varphi2}) at $x^1=d$ in (\ref{phiexp2}) leads us to the final expression.
Similarly,  getting the relation for $\varphi_{+}^{(-)} $ and $\varphi_{-}^{(-)}$, and combining the two results, we get:
\begin{equation}
\label{rel2}
    \varphi_{+}^{(\pm)}= -\frac{m(m+\omega) + {k_1}^2 - \sigma_1 k_1 \bm{\sigma}_\parallel \cdot \bm{k}_\parallel}{(m \pm ik_1)(m+\omega)}\varphi_{-}^{(\pm)}e^{-2ik_1 d}
\end{equation}
 In order to have non-trivial consistent solutions for $(\varphi_{+}^{(\pm)},\varphi_{-}^{(\pm)})$, on comparing equations (\ref{rel1}) and (\ref{rel2}), one gets the following relation to be satisfied by any compact dimension $k_1$ \cite{Khoo:2011ux}:
\begin{equation}
\label{trans}
    md\sin(k_1d)+ k_1d\cos(k_1d) = 0
\end{equation}
For the massless fermion case, from the above transcendental equation in the $m\rightarrow 0$ limit we obtain the massless boundary conditions:
\begin{align}
    &\Rightarrow \cos(k_1d) = 0\\
    \text{Therefore,   }& k_1d = \left(n+\frac{1}{2}\right)\pi \;\;\;\; \text{where} \;\; n\in \mathbb{Z^{+}} \label{allow}
\end{align}
These boundary conditions will be used to compute the Casimir energy of massless fermions in a general $(D+1)$-dimensional spacetime in section (\ref{mlfer}). On the the compact dimension $x^1$, let us assume that the field $\psi$ satisfies the general periodicity conditions:
  \begin{equation}
  \label{comp}
      \psi({l,\bm{x}+L_j e_j}) = e^{2\pi i\alpha_j}\psi{(l,\bm{x})}
  \end{equation}
  where, $\bm{x} = (x^1,\dots,x^D)$ and $e_j$ is the unit vector in the $x^j$ direction. Moreover, the boundary conditions on the compact dimension in (\ref{comp}), for massive fermionic case are of the form:
\begin{equation}
\label{massivebound}
  k_1d = {2\pi(n +\alpha_n)} = {\lambda_n} 
\end{equation}
 The cases when values of $\alpha_j = 0$ and when $\alpha_j = \frac{1}{2}$ correspond to the untwisted and twisted fields respectively. But, in general $0\leq \alpha_j <1$ for $j=1,\dots,D$. The Casimir energy of massive fermions in $(3+1)$-dimensional spacetime is calculated in section (\ref{mvfer}).
 \subsection{Massive fermions in $(3+1)$-dimensional spacetime}  
\label{mvfer}
   The boundary conditions on the compact dimensions in (\ref{comp}), for the massive fermionic case, are of the form in (\ref{massivebound}). Consider a three-dimensional space with the standard $x$, $y$ and $z$ dimensions as our orthogonal axes. 
 
 The compact $z$ dimension the quantity $k_z d = \lambda_n$ as described earlier satisfies the transcendental relation derived in (\ref{trans}). In general for the massive fermionic effect in $(3+1)$-dimensional spacetime, we have the vacuum energy density (per unit area) denoted as ${\langle E^0\rangle}$, is given by
\begin{equation}
    {\langle E^0\rangle} = -\frac{\hbar c}{2\pi^2} {\sum_n}^\prime\int_{-\infty}^\infty dk_x \int_{-\infty}^\infty dk_y (k_x^2 + k_y^2 + \frac{{\lambda_n}^2}{d^2}+ m^2)^{\frac{1}{2}} 
\end{equation}
One obtains a finite and well defined result for the Casimir energy on renormalization of this expression using the generalized Abel-Plana summation formula, given below:  \cite{Saharian:2007ph, BelussiSaharian}
\begin{equation}
\label{abelplana}
  \sum_{n=1}^\infty \frac{\pi F(n)}{1-\frac{\sin(2\lambda_n)}{2\lambda_n}} = -\frac{\pi d m F(0)}{2(md+1)} +\int_{0}^{\infty}F(z)dz - i\int_0^\infty \frac{\Big(F(it)-F(-it)\Big)dt}{\frac{t+md}{t-md}e^{2t}-1} 
\end{equation}
  
 Considering the denominator of the left hand side of (\ref{abelplana}), we find:
 \begin{align} 
     \frac{\sin(2\lambda_n)}{2\lambda_n} &= 1-\frac{2}{2\lambda_n}\left(\frac{\lambda_n}{\sqrt{{\lambda_n}^2 +(md)^2}}\right)\left( -\frac{md}{\sqrt{{\lambda_n}^2 +(md)^2}} \right)\\
     &= 1+ \frac{md}{{\lambda_n}^2 +(md)^2}
 \end{align}
 We find that the regularized vacuum energy is given by :
\begin{align}
\label{ap}
    {\langle E^0  \rangle}=-\frac{\hbar c}{2\pi^3 d}  \int_{-\infty}^\infty dk_x \int_{-\infty}^\infty dk_y \left(-\frac{\pi md F(0)}{2(md+1)} +\int_{0}^{\infty}F(z)dz - i\int_0^\infty \frac{\Big(F(it)-F(-it)\Big)dt}{\frac{t+md}{t-md}e^{2t}-1}\right) 
\end{align}
where $F(z)$ is defined as :
\begin{equation}
\label{Fz}
    F(z) = \sqrt{(k_x^2 + k_y^2)d^2 + z^2 + (md)^2}\Big(1+ \frac{md}{{z}^2 +(md)^2}\Big)
\end{equation}
 Therefore, making the change of variable $t = ud$, and substituting (\ref{Fz}) in  (\ref{ap}) we can write the Casimir energy for a massive fermionic field ($f$) as:  \cite{Cruz, Elizalde:2002wg}
\begin{dmath}
     {E^{3+1\text{D},\text{cont},f}_{\text{Cas}}}=  -\frac{id\hbar c}{2\pi^3}  \int_{-\infty}^\infty dk_x \int_{-\infty}^\infty dk_y \int_0^\infty du\frac{u-m}{(u+m)e^{2ud} +u-m}\Big(1+\frac{m}{d(m^2-u^2)}\Big)\times \Big( \sqrt{k_x^2 + k_y^2 + (iu)^2 + m^2} -  \sqrt{k_x^2 + k_y^2 + (-iu)^2 + m^2} \Big)
\end{dmath}

 We separate the above integral over the variable $u$ in two intervals. If $u^2 < k_x^2 + k_y^2 + m^2 $, the above integral vanishes due to the last bracket. Although, when $u^2 > k_x^2 + k_y^2 + m^2
 \newline\left(u\in[\sqrt{k_x^2 + k_y^2 + m^2},\infty]\right)$ we rewrite the expression for Casimir energy as:
 \begin{dmath}
 \label{uterm}
     {E^{3+1\text{D},\text{cont},f}_{\text{Cas}}} =  -\frac{id\hbar c}{2\pi^3} \int_{-\infty}^\infty dk_x \int_{-\infty}^\infty dk_y \int_{\sqrt{k_x^2 + k_y^2 + m^2}}^\infty du \sqrt{u^2 - k_x^2 - k_y^2 - m^2} \times \Big( \frac{d(u-m) -m/(u+m)}{(u+m)e^{2ud} + u-m}\Big)
\end{dmath}
We see that the last term of the above integrand can be written as :
\begin{equation}
    \left( \frac{d(u-m) -m/(u+m)}{(u+m)e^{2ud} + u-m}\right) = -\frac{1}{2} \frac{d}{du}\ln\left(1+\frac{u-m}{u+m}e^{-2ud}\right)
\end{equation}
Also, we have a suitable integral representation as :
\begin{equation}
    \int d\vec{k_p}\int_{\sqrt{\vec{k_p}^2 + m^2}}^\infty du (u^2 - \vec{k_p}^2 - m^2)^{\frac{s+1}{2}}f(u) = \frac{\pi^{p/2}\Gamma(\frac{s+3}{2})}{\Gamma(\frac{p+s+3}{2})}\int_m^\infty dx(u^2-m^2)^{\frac{p+s+1}{2}}f(u)
\end{equation}
Using integration by-parts, substituting the change of variable $ud = y + md$ and above results in (\ref{uterm}) the expression for Casimir energy is:
\begin{equation}
\label{yterm}
    {E^{3+1\text{D},\text{cont},f}_{\text{Cas}}} = -\frac{\hbar c}{\pi^2 d^3} \int_0^\infty dy(y+md)\sqrt{y(y+2md)}\ln\left(1+\frac{y}{y+2md}e^{-2(y+md)}\right)
\end{equation}
 The integral (\ref{yterm}) obtained for massive fermionic fields has no closed expression in terms of standard functions. Although we solve further for approximate solutions in the two limiting cases, namely $md \ll 1$ and $md \gg 1$. \cite{Elizalde:2002wg}
 \begin{enumerate}[(I)]
     \item 

 In the case for $md \ll 1$ we expand the integrand in (\ref{yterm}) upto first order in the powers of $(md)$ to get :
 \begin{equation}
      {E^{3+1\text{D},\text{cont},f}_{\text{Cas}}} = -\frac{\hbar c}{\pi^2 d^3} \int_0^\infty dx x^2 \ln(1+e^{-2x}) - \left(2\frac{xe^{-2x}(x+1)}{1+e^{-2x}}-2x\ln(1+e^{-2x})\right)md
 \end{equation}
Performing the above integral, we obtain the Casimir force per unit area in the limit $md \ll 1 $ is given by :
\begin{equation}
\label{massiveferless}
     {E^{3+1\text{D},\text{cont},f}_{\text{Cas}}} = -\frac{7}{4}\frac{\pi^2\hbar c}{720d^3}  \left(1-\frac{120md}{7\pi^2}\right)
\end{equation}

\item Now, approximating the integrand in (\ref{yterm}) in the limiting case $md \gg 1$, we see that only the exponential term accounts for maximum contribution. The approximated integral is:
\begin{equation}
     {E^{3+1\text{D},\text{cont},f}_{\text{Cas}}} = -\frac{\hbar c}{\pi^2 d^3} \int_0^\infty dx \sqrt{\frac{md}{2}}x^{3/2}e^{-2(x+md)}
\end{equation}
 By the above approximation, the Casimir energy in the limit $md \ll 1 $ is given by:
 \begin{equation}
 \label{massivefer}
     {E^{3+1\text{D},\text{cont},f}_{\text{Cas}}} = -\frac{3}{32}\sqrt{\frac{m}{\pi^3d^5}}e^{-2md}
 \end{equation}
 where the Casimir energy decays exponentially with $md$.
  \end{enumerate} 
  One can obtain the expression for the Casimir energy for massless fermions in ($3+1$)-dimensions from (\ref{massiveferless}) in the $m\rightarrow0$ limit. Let us now check if it matches with a general expression for the Casimir energy of massless fermions in $(D+1)$-dimensions.
\subsection{Massless fermions in $(D + 1)$-dimensional spacetime}
\label{mlfer}
In this section, a general expression for the massless fermionic Casimir energy in the $(D+1)$-dimensions is obtained. Quantum field submitted to classical boundary conditions gives rise to the Casimir effect, just as in the MIT Bag model, we expect the colour confinement to give rise to the Casimir energy of the gluon and quark fields. 

Let us say the boundary conditions are satisfied in the $x^1$ direction and are given by (\ref{allow}) for the massless fermionic field. As stated earlier, the Dirac fermionic field $\psi$ has $N_D$ components and be confined between the two parallel plates placed at $x^1 = 0$ and $x^1 = d$. Fermions also have a characteristic negative sign. The vacuum energy density (per unit area) of the Dirac fermionic field in the slab-bag configuration of parallel plates is :
\begin{align}
    E_D &= \frac{ \bra{0}\hat{H}\ket{0}}{\prod_{i=1}^{D-1}d_i}= -\frac{N_D}{(2\pi)^{D-1}}\sum_{n=0}^{\infty}\int_0^\infty d^{D-1}k\left(\left(n+\frac{1}{2}\right)^2\frac{\pi^2}{d^2} + k^2_2 +\dots +k^2_D\right)^{\frac{1}{2}} \label{E_D}
\end{align}
The expression above in the summation and integration variables is divergent. We will use dimensional regularization for continuous variables and extend the Hurwitz zeta-function appearing after dimensional regularization analytically. From dimensional regularization we know
\begin{equation}
\label{dimreg}
    \int \frac{d^Du}{(u^2+a^2)^s} = \frac{\pi^{\frac{D}{2}}}{\Gamma\left(s\right)}\Gamma\left(s-\frac{D}{2}\right)\frac{1}{(a^2)^{s-\frac{D}{2}}}
\end{equation}
Using (\ref{dimreg}) in (\ref{E_D}), the vacuum energy density is given by
\begin{equation}
\label{ED}
    E_D = -\frac{N_D \pi^{(D)/2}}{2^{D}d^D}\Gamma\left(-\frac{D}{2}\right)\sum_{n=0}^{\infty}\left(n+\frac{1}{2}\right)^D.
\end{equation}
The Hurwitz zeta-function, analytic for $Re(z)>1$ is given by:
\begin{equation}
\label{zeta}
    \zeta(z,q) = \sum_{n=0}^{\infty}\frac{1}{(n+q)^z}, \;\;\;q\neq 0, \;\mathbb{Z}^{(-)}
\end{equation}
Thus, we can rewrite the above expression (\ref{ED}) in this configuration using (\ref{zeta}) as:
\begin{equation}
\label{expb}
    E_D = -\frac{N_D \pi^{(D)/2}}{2^{D}d^D}\Gamma\left(-\frac{D}{2}\right)\zeta\left(-D,\frac{1}{2}\right)
\end{equation}
Now, the following reflection formula, valid  $\forall\; s$ is used:
\begin{equation}
    \Gamma\left(\frac{s}{2}\right)\zeta(s) = \pi^{s-1/2}\Gamma\left(\frac{1-s}{2}\right)\zeta(1-s)
\end{equation}
and the relation $\zeta(s,\frac{1}{2}) = (2^s-1)\zeta(s)$ in the above expression (\ref{expb}).
The vacuum energy density and the corresponding pressure exerted on the plates by the vacuum are:
\begin{align}
    E_D &= -\frac{N_D (1-2^{-D})}{2^{D}\pi^{\frac{D+1}{2}}d^D}\Gamma\left(\frac{D+1}{2}\right)\zeta\left(D+1\right)\\
    \Rightarrow F_D = -\frac{\partial}{\partial d}E_D &=- \frac{DN_D (1-2^{-D})}{2^{D}\pi^{\frac{D+1}{2}}d^{D+1}}\Gamma\left(\frac{D+1}{2}\right)\zeta\left(D+1\right)  \label{Fmf}
\end{align}
  Alternatively, the Schwinger proper time representation also yields us the same result \cite{Milton:2001yy}. Note that in the cut-off function method employed in section (\ref{photons}), regularized zero-point energy of fields both inside and outside the cavity contributed to the total energy. The zero-point energy was divergent, but we obtained a finite shift in the total energy due to changes in the configuration. Although in the bag configuration, a more natural approach is to evaluate the fermionic Casimir energy is the analytic extension of zeta function, where the field is confined to one region, and these divergent terms\footnote{Vacuum energy density as series in ($3+1$)-dim is ${E^0}  = -\Omega^4\frac{6}{\pi^2} - \frac{7}{4}\frac{\pi^2}{720}\frac{1}{d^4}$. The first term   diverges in the limit $\tau=\frac{1}{\Omega}\rightarrow0$. It is independent of $d$ and is also present when the plates are drifted far apart, $d\rightarrow\infty$. Thus, this is the free energy of the fermion field in the absence of any confining boundaries. The second term is the free finite energy present due to the plates. Here, $\tau$ is the damping constant of the cut-off function.} vanish. 
 
For future reference, the continuum massless fermionic ($mf$) Casimir energy in ($1+1$)-, ($2+1$)- and ($3+1$)-dimension yield the result :
  \begin{equation}
 \label{fercont}
E^{1+1\text{D},\text{cont},mf}_{\text{Cas}}(d) = -\frac{\pi\hbar c}{24 d}\;\;;\;\;E^{2+1\text{D},\text{cont},mf}_{\text{Cas}}(d) = -\frac{3}{4}\frac{\zeta({3})\hbar c}{8\pi d}\;\;;\;\;E^{3+1\text{D},\text{cont},mf}_{\text{Cas}}(d) = -\frac{7}{4}\frac{\pi^2\hbar c}{720 d^3}.
\end{equation}
The motive behind modelling the hadron as a bag is to study the dynamics of colourless states of quarks and gluons completely confined within the interior of a classical cavity moving freely inside. Yet, as we compare the expressions and properties for the massless fermionic Casimir energy (\ref{fercont}) to the electromagnetic case (\ref{Ecas3+1}), we realize that there is a striking similarity between them. The fermionic expression in ($3+1$)-dimension is exactly $\frac{7}{8}\times 2$ times the electromagnetic energy density for the photon field and matches with (\ref{massiveferless}) in $m\rightarrow0$ limit. 

In the next chapter, we shall present a novel method \cite{Ishikawa:2020ezm} to calculate the Casimir energy for Dirac fermions by calculating it for lattice fermions in the continuum limit. Apart from studying this phenomenon from the perspective of Quantum Chromodynamics and colour confinement in hadrons using the MIT Bag model, we also use these results to study the properties of some condensed matter systems like topological insulators. The periodic and antiperiodic boundary conditions are often used as a simple setup in lattice simulations for condensed matter physics. The procedure used in this section can be followed to calculate the Casimir energy for periodic and antiperiodic boundary conditions in ($D+1$)-dimensional spacetime. The continuum results for these boundary conditions have been enlisted below for future reference and comparing the results obtained for lattice fermions to the continuum results:
\begin{subequations}\label{PAPcont}
\begin{align}
E^{\text{1+1D,cont,P}}_{\text{Cas}} &= \frac{\pi}{3d},
&\qquad
E^{\text{1+1D,cont,AP}}_{\text{Cas}} &= -\frac{\pi}{6d},
\label{PAP1p1} \\
E^{\text{2+1D,cont,P}}_{\text{Cas}} &= \frac{\zeta(3)}{\pi d^2},
&\qquad
E^{\text{2+1D,cont,AP}}_{\text{Cas}} &= -\frac{3\zeta(3)}{4\pi d^2},
\label{PAP2p1} \\
E^{\text{3+1D,cont,P}}_{\text{Cas}} &= \frac{\pi^2}{45d^3},
&\qquad
E^{\text{3+1D,cont,AP}}_{\text{Cas}} &= -\frac{7\pi^2}{45d^3}.
\label{PAP3p1}
\end{align}
\end{subequations}
These will serve as a reference for comparing the results for lattice fermions to the continuum expressions in ($1+1$)- and higher dimensional spacetime.

\chapter{Definition of Casimir energy on Lattice} 
 
 \label{ch3}
   Since the beginnings of quantum field theory, handling divergent quantities and the necessity for renormalization have kept physicists occupied. Regularization schemes are employed to remove the divergences during the calculation of processes through a Feynmann diagram. The lattice regularization goes beyond the diagrammatic approach and provides a  non-perturbative cut-off for the removal of ultraviolet infinities in a quantum field theory. It directly eliminates all wavelengths less than twice the lattice spacing `$a$' \cite{Creutz:1984mg}, which is also the case when all frequencies below $d_c$ are cut off, as discussed regarding the wavelengths contributing to the Casimir effect in section (\ref{2d}). The continuum physics of a particular system can be extracted from the numerical calculations by sending the lattice spacing limit to zero.
   
   The theoretical construction of the lattice fermions closely relates to the Nielsen-Ninomiya no-go theorem, as discussed in section (\ref{NNthm}). Apart from preserving non-perturbative effects, another important motivation for studying the Casimir effect for various lattice fermions is that their theoretical form also appears in condensed matter systems, which are currently areas of active and exciting research. The dispersion relation of Wilson lattice fermion can be regarded and modelled as the low energy band structure in Dirac semi-metals. The mechanism of gapless surface modes induced from the gapped bulk fermions is also effectively found to be the same as chiral fermions realized in the domain wall fermion formulation \cite{Kaplan}. Similarly, the domain wall fermions also act as an analogy to the zero-mode Dirac fermions on the surface of the topological insulator. Such applications are also observed in the physics of ultracold atom systems. Solving this problem in generality can help us understand various condensed matter systems in a better fashion \cite{Ishikawa:2020icy}. 
   \begin{figure}
       \centering
       \includegraphics[width=11cm]{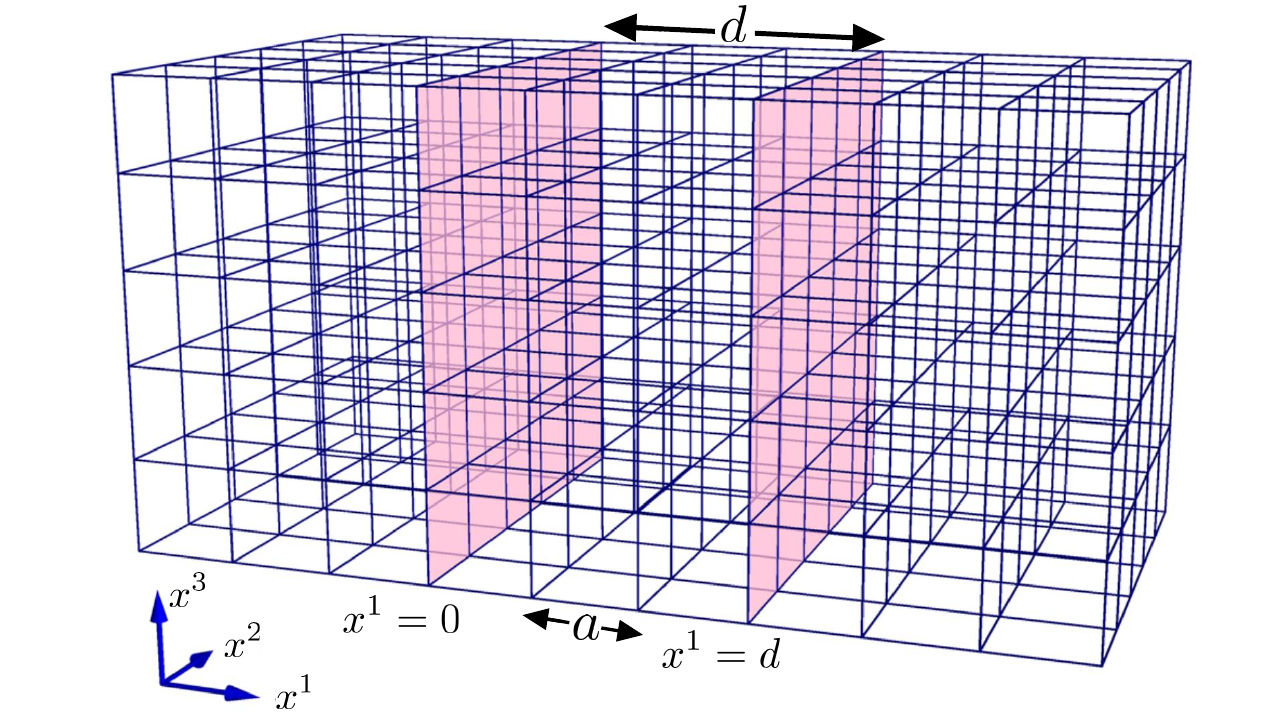}
       \caption{\footnotesize{The geometry of the Casimir effect on lattice between two infinitely large parallel plates, indicated by the shadowed planes for three spatial dimensional lattice. Here, suitable boundary conditions are implemented.}}
       \label{latticediagram}
   \end{figure}
  Note that in this formalism, one spatial dimension $x^1$ is compactified by the boundary condition. The corresponding spatial momentum component $p_1$ is then discretized, while the other momenta components remain continuous. Initially, the time component is kept continuous, and thus the temporal components of momentum are unaffected by latticization.
 Let the non-zero lattice spacing be denoted as `$a$', with an ultraviolet cut-off scale `$1/a$' in momentum space. From the Dirac operator we defined in Chapter (\ref{intro}), the energy-momentum dispersion relation for lattice fermions is defined as:
\begin{equation}
\label{Drel}
a E(ap) = a \sqrt{D^{\dagger} D}  
\end{equation}
Such that the zeroes of this expression are the position of poles in the momentum space fermion propagator. For our case, the Dirac operator, $D$, does not include the temporal momenta. Since the temporal component is not discretized, only the spatial momenta and mass are included.
The discretized momentum under periodic (P) and antiperiodic (AP) boundary conditions is:
\begin{equation}
\label{PAP}
ap_1 \rightarrow ap^{\text{P}}_1(n) = \frac{2n\pi}{N} \;\;;\;\; ap_1 \rightarrow ap^{\text{AP}}_1(n) = \frac{(2n+1)\pi}{N} 
\end{equation}
Here, $N=d/a$ is the lattice size in the compactified spatial direction. Let the Brillouin Zone (BZ) be defined as $0 \leq ap_1 < 2\pi$, or equivalently one may also define it as $-\pi < ap_1 \leq \pi$. Note that our results for Casimir energy are independent of the choice of Brillouin zone. It is a simple exercise to check that the first Brillouin zone, which we shall use for our convention, bounds the integer $n$ in the range $0 \leq n^{\text{P, AP}} < N$. One of the momentum components is discretized in the definition of zero-point energy, as the momentum integral of the dispersion relations: (here, $\hbar = c = k_B = 1$)
\begin{align}
\label{Ninfty}
aE_0(N\rightarrow \infty) &= -Nc_{\text{deg}} \int_{\text{BZ}} \frac{d^3ap}{(2\pi)^3}aE(ap) \\
                  aE_0(N) &= -c_{\text{deg}} \int_{\text{BZ}} \frac{d^2ap_{\perp}}{(2\pi)^2} \sum_n aE(ap_{\perp},ap_1(n)) \label{Ninfty2}
\end{align}
 Note the change in dimension of the Brillouin zone in (\ref{Ninfty}). Fermions have a characteristic negative sign. The factor of ($\frac{1}{2}$) in the zero-point energy expression (\ref{Ninfty2}) is dropped due to nullifying factor of (2) from the antiparticle degrees of freedom. As seen earlier, we subtract the divergent zero-point energy in infinite volume from the one in finite volume to obtain the Casimir energy. Both the expressions here do not diverge due to the lattice cut-off.
 Therefore, the Casimir energy obtained on lattice by substracting $aE_0(N\rightarrow \infty)$ from  $aE_0(N)$ is: \cite{Ishikawa:2020ezm}
\begin{align}
\label{def3+1}
aE^{\text{3+1D}}_{\text{Cas}} &= aE_0(N) - aE_0(N \rightarrow \infty) \nonumber \\ 
              &= c_{\text{deg}}\int \frac{d^2ap_{\perp}}{(2\pi)^2}\left[-\sum_n aE(ap_{\perp},ap_1(n)) + N\int_{\text{BZ}}\frac{dap_1}{2\pi}aE(ap) \right] 
\end{align}
 Similarly, the Casimir energy for ($1 + 1$)-dimensions is:
\begin{align}
\label{def1+1}
aE^{\text{1+1D}}_{\text{Cas}} &= aE_0(N) - aE_0(N \rightarrow \infty) \nonumber \\ 
              &= c_{\text{deg}} \left[-\sum_n aE(ap_1(n)) + N\int_{\text{BZ}}\frac{dap_1}{2\pi}aE(ap) \right] 
\end{align}
 In continuum, the expressions for massless fermionic Casimir energies for the periodic and antiperiodic boundary conditions (\ref{PAP}) are obtained using the procedure mentioned in section ({\ref{photons}}) and stated from (\ref{PAP1p1}, \ref{PAP2p1}, \ref{PAP3p1}). The results for ($1+1$)-dimensions are reiterate here:
 \begin{equation}
 \label{cont}
     E^{\text{1+1D,{cont},P}}_{\text{Cas}} = \frac{\pi}{3d}\;\;;\;\;E^{\text{1+1D,{cont},AP}}_{\text{Cas}} = -\frac{\pi}{6d}
 \end{equation}
as these expressions will come in handy to compare our results on lattice to the continuum fermionic Casimir effect.
\section{Casimir effect for lattice fermion fields in slab-bag}
\label{slabbag}
The boundary conditions (\ref{allow}) we obtained on the modes allowed in the constrained direction ensures the absence of any particle current through the walls, thereby confining the fermion fields inside the slab-bag. These are realized on the lattice as follows: 
\begin{equation}
\label{allowed}
     ap_1\rightarrow ap^{\text{B}}_1(n) = \left(n+\frac{1}{2}\right)\frac{\pi a}{d} = \left(n+\frac{1}{2}\right)\frac{\pi}{N}
\end{equation}
We claim that it is necessary for these boundary conditions to be satisfied on the lattice, just as they are satisfied in the continuum for the confined massless free fermion field. Let these be called the MITBag (B) boundary conditions. The expressions for the Casimir energy of the Naive and Wilson fermion are calculated from this method analytically for ($1+1$)-dimensional spacetime, numerically for higher dimensions and plotted subsequently in Fig. \ref{allowed} \cite{Ishikawa:2020icy}. This can also be considered a novel method to calculate the Dirac fermionic Casimir energy by taking the lattice spacing limit to zero.
 \subsection{For Naive fermion}
We shall now discuss the Casimir energy $aE_{\text{Cas}}$, for Naive lattice fermion ($nf$) in $(1+1)$- dimensional spacetime calculated from the definition. In momentum space, the propagator for the Naive fermion as obtained in (\ref{naive}) is:
\begin{equation}
    aD_{nf}  = i\sum_k \gamma_k \sin(ap_k) +am_f
\end{equation}
For Naive fermion, the corresponding dispersion relation from (\ref{Drel}) is obtained as:
\begin{equation}
\label{naived}
    a^2E_{nf}(ap) = \sum_{k} \sin^2(ap_k) +(am_f)^2
\end{equation}
Numerical studies for the Naive fermion in higher dimensions are also done. Substituting (\ref{naived}) in the definition (\ref{def3+1}), the expression for Casimir energy of Naive fermion with MITBag boundary conditions in ($3+1$)-dimensional space-time is:
 \begin{align}
    \label{naivemitbagexp}
    &aE_{\text{Cas}}^{\text{3+1D,B,}nf} \equiv  aE_{\text{0}}^{\text{3+1D,B,}nf}(N) - aE_{\text{0}}^{\text{3+1D,B,}nf}(N\rightarrow\infty)\\
&= c_{\text{deg}}\int \frac{d^2ap_{\perp}}{(2\pi)^2}\Bigg[
-\sum_n \sqrt{\sin^2 \frac{(n+1/2)\pi}{N}+ \sum_{k=2,3}\sin^2 (ap_k) + (am_f)^2}\;\; +  \\&\;\;\;\;\;\;\;\;\;\;\;\;\;\;\;\;\;\;\;\;\;\;\;\;\;\;\;\;\;\;\;\;\;\;\;\;\;\;\;\;\;\;\;\;\;\;\;\;\;\;\;\;\;\;\;\;\;\;\;\;\;\;\;\;\;\;\;\;\;\;\;\;\;\;\;\;\;\;\;\;    N\int_{\text{BZ}}\frac{dap_1}{2\pi}\sqrt{ \sum_{k=1,2,3}\sin^2 (ap_k)+ (am_f)^2}\Bigg] \nonumber
\end{align}
Similarly, the expression for ($2+1$)-dimensions is used for numerical analysis. The derivations of Casimir energy in ($1+1$)-dimensional spacetime using the MITBag (B) momenta boundary conditions for the massless Naive fermion are presented in Appendix B (\ref{Naivemitbag}) using the Abel- Plana formulae for finite range. Series expansions from (\ref{csc}) are substituted, and the results are calculated to be:
\begin{align}
     aE_{\text{Cas}}^{\text{1+1D,B,}nf} & =\frac{2N}{\pi} -\csc(\frac{\pi}{2N})\label{mitbagnaive}\\
   & =-4\left[-\frac{N}{2\pi} + \frac{1}{4}\left\{\frac{2N}{\pi} + \frac{1}{6}\frac{\pi}{2N} +\frac{7}{360}\left(\frac{\pi}{2N}\right)^3 + ...\right\} \right]\nonumber\\
  \Rightarrow E_{\text{Cas}}^{\text{1+1D,B,}nf} & = -\frac{\pi}{12d} - \frac{7\pi^3a^2}{2880d^3} + \mathcal{O}(a^4) 
 \end{align} 
 Therefore, the Casimir energy obtained per unit area in the continuum limit $a\rightarrow 0$, is:
\begin{equation}
\lim_{a \to 0}E_{\text{Cas}}^{\text{1+1D,B},nf} = -\frac{\pi}{12 d}
\end{equation}
As expected, the above result in continuum limit is twice the expression obtained for Dirac fermion. This is because of the Naive fermion doubling in one spatial dimension. 
 \subsection{For Wilson fermion}
\begin{figure}[t!]
\centering
    \subfloat\footnotesize{(a)}{{\includegraphics[width=7.4cm]{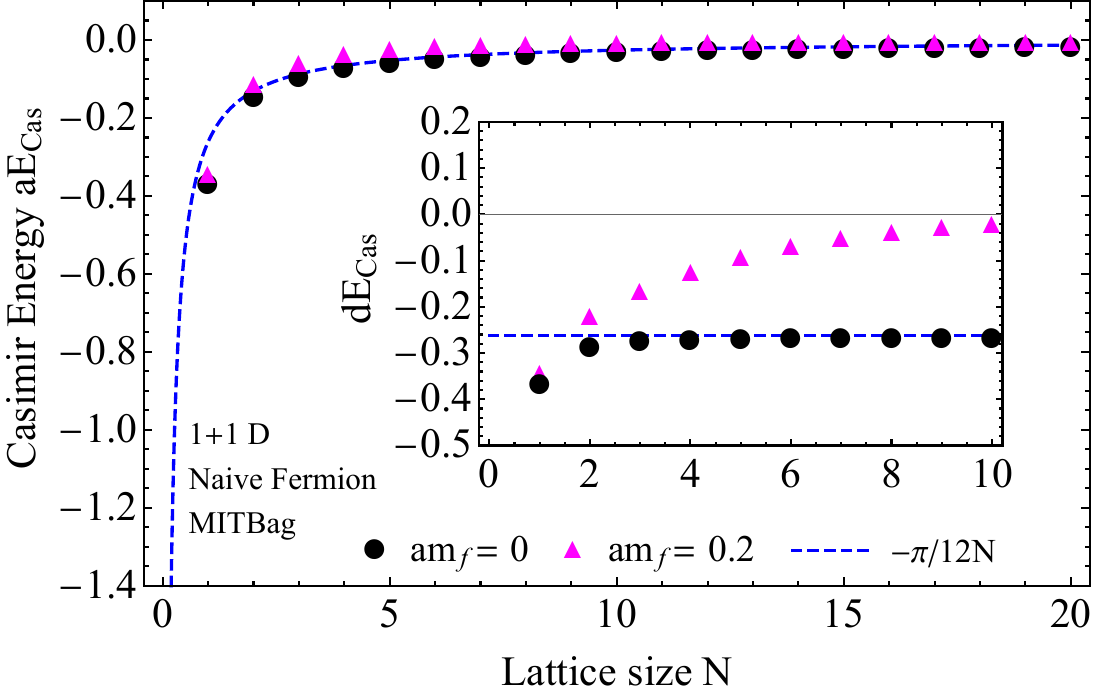} }} %
    \subfloat\footnotesize{(b)}{{\includegraphics[width=7.4cm ]{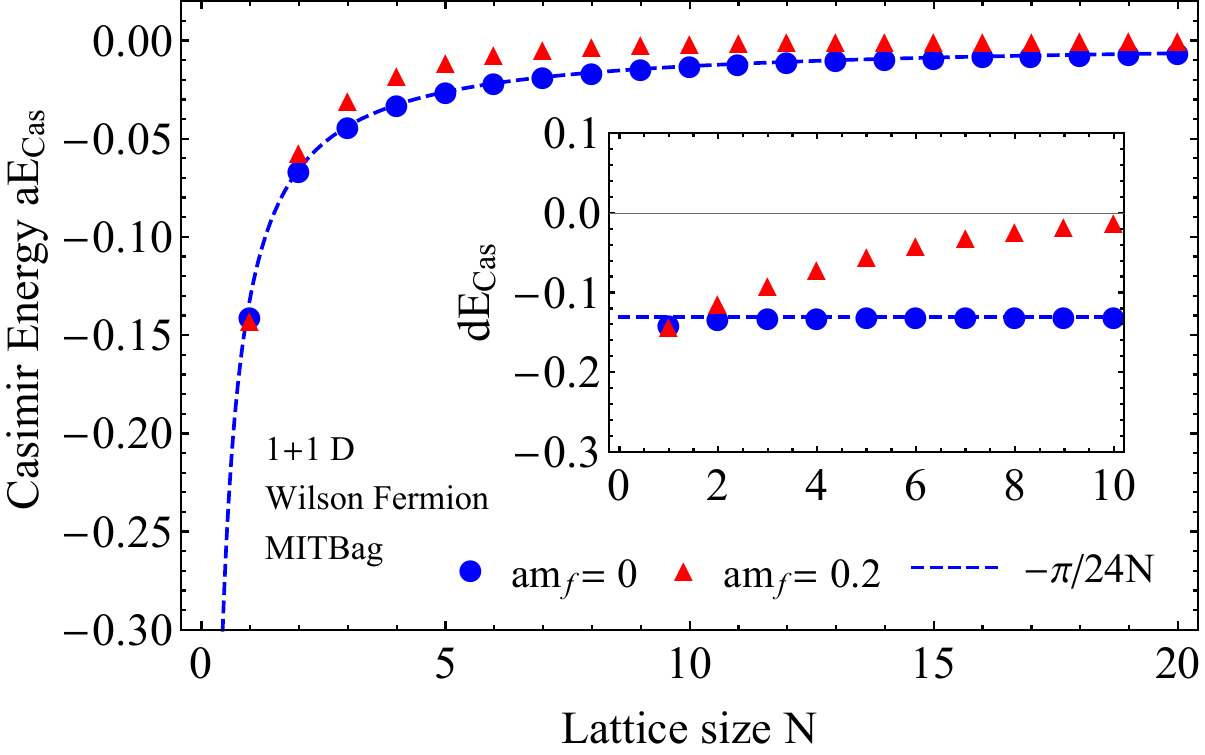} }}%
     \subfloat\footnotesize{(c)}{{\includegraphics[width=7.4cm]{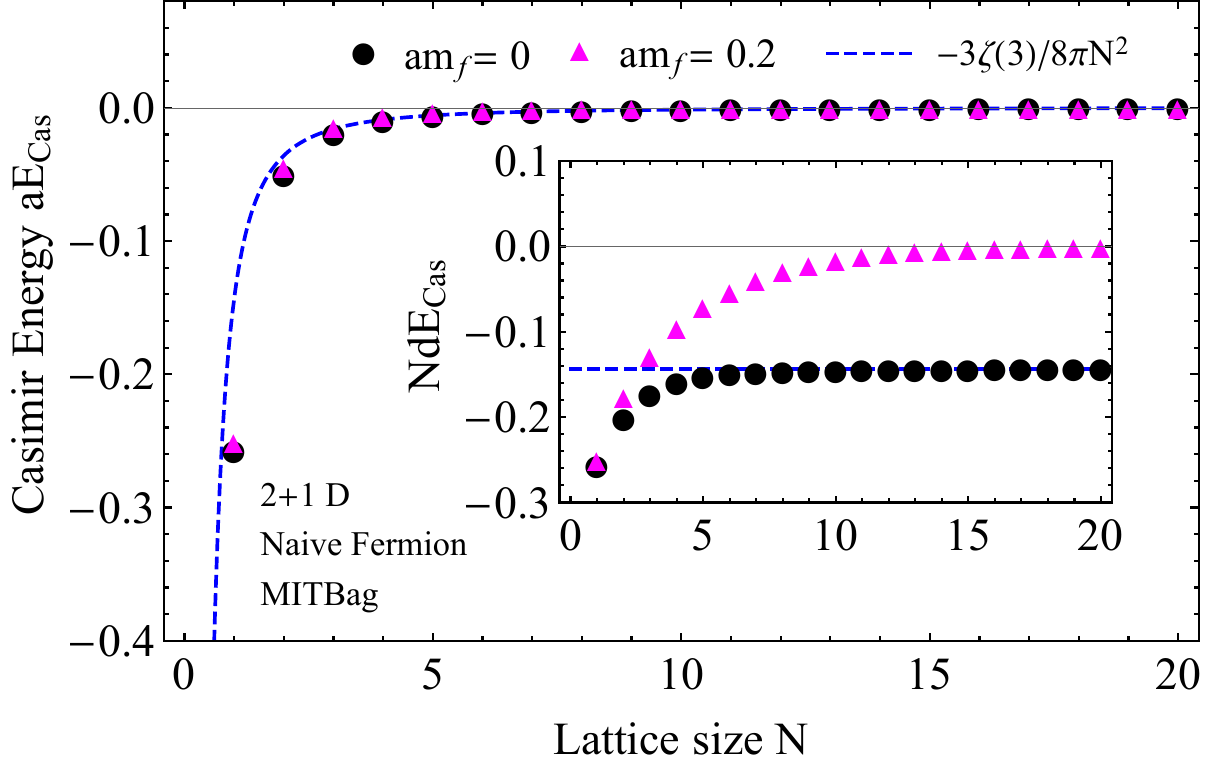}}} 
    \subfloat\footnotesize{(d)}{{\includegraphics[width=7.4cm]{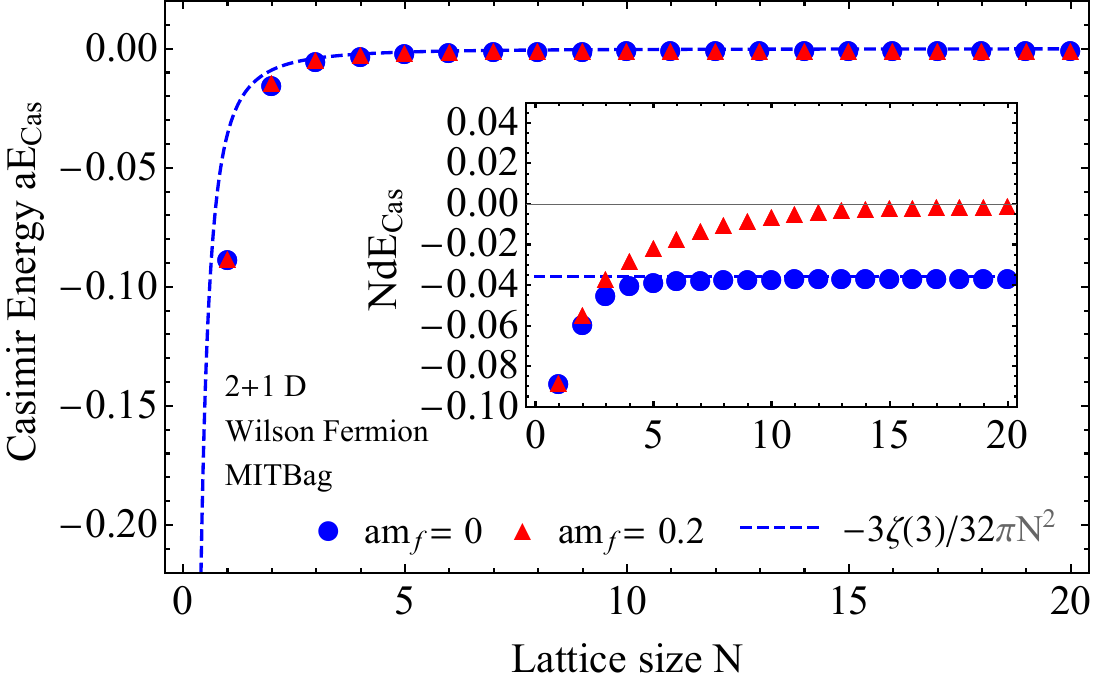}}}
    \subfloat\footnotesize{(e)}{{\includegraphics[width=7.4cm]{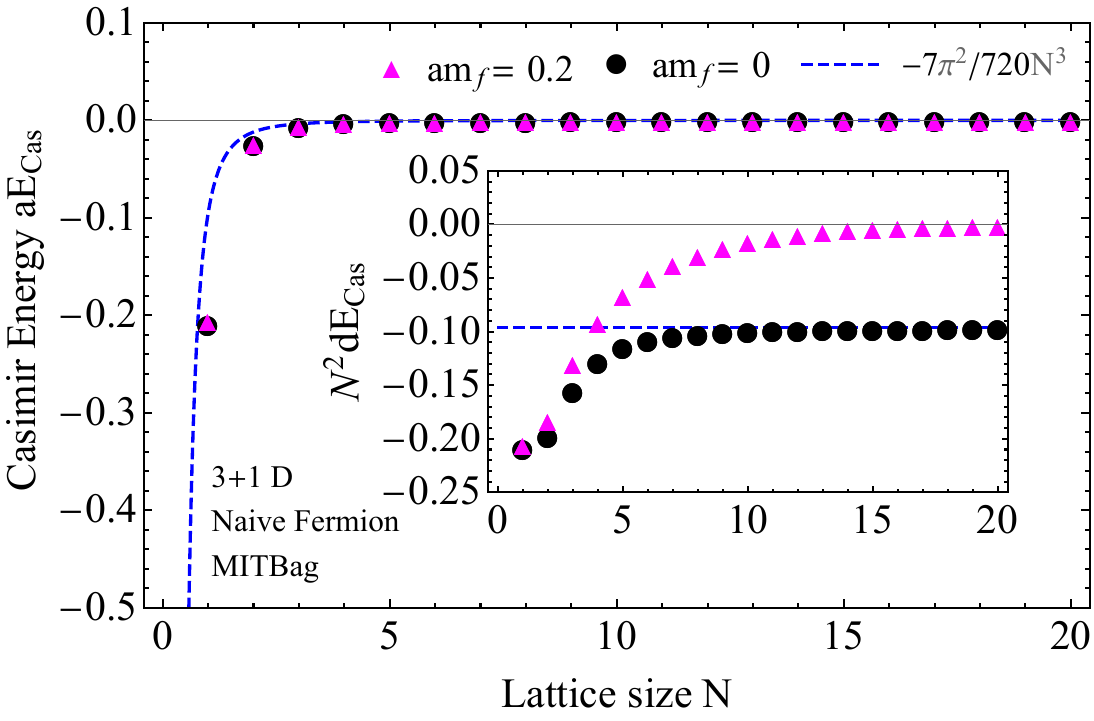}}} %
    \subfloat\footnotesize{(f)}{{\includegraphics[width=7.4cm]{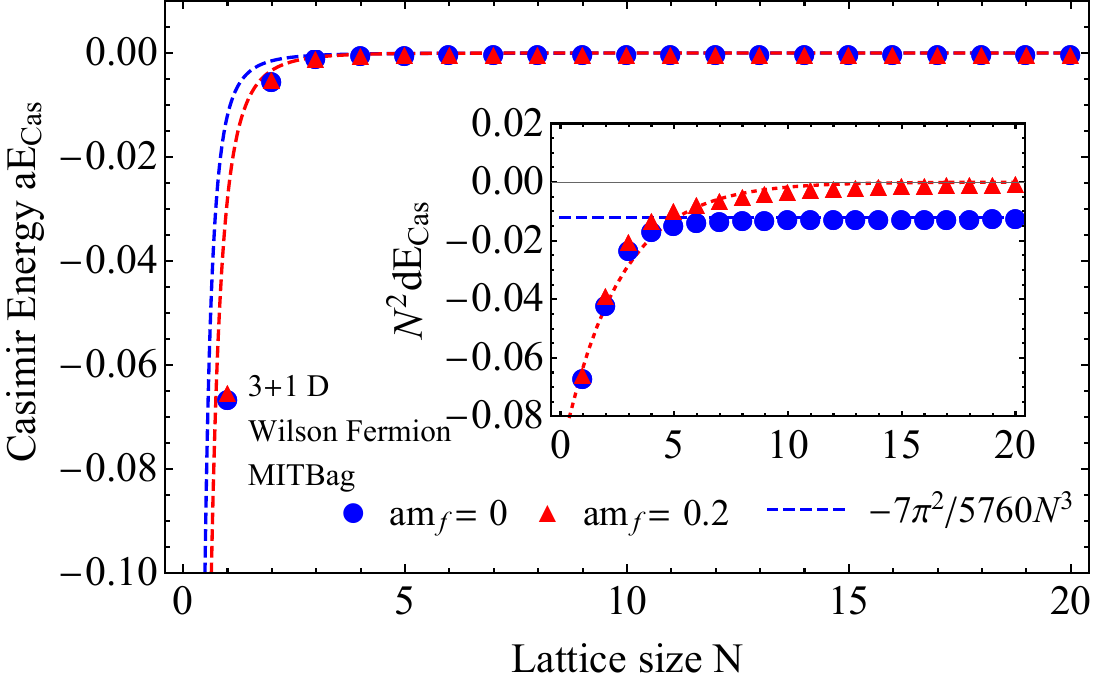}}}%
    
    \caption{\footnotesize{The Casimir energy $aE_{\text{Cas}}$ for massless and positive massive Wilson and Naive fermion with MITBag boundary conditions. [(a),(b)], [(c),(d)] and [(e),(f)] represent the MITBag results for Naive and Wilson fermion in ($1+1$)-, ($2+1$)-, and ($3+1$)-dimension respectively. Subfigure (f) also confirms the exponential decay of $aE_{\text{Cas}}$ for massive fermions with the lattice size $N$.}}%
\label{mitbagplot1}
\end{figure}

 We shall now discuss the Casimir energy for Wilson lattice fermion in $(1+1)$-dimensional spacetime. In (\ref{Dp}), the Dirac operator obtained for Wilson fermion with fermion mass $m_f$ and the Wilson parameter $r$ in momentum space is:
\begin{equation}
\label{Wilsonoperator}
    aD_{\text{W}}  =  i\sum_k \gamma_k \sin(ap_k) + r\sum_k (1- \cos(ap_k) + am_f
\end{equation}
The Wilson term proportional to $r$ in the Dirac operator breaks the chiral symmmetry and acts as a momentum dependent mass term introduced to solve fermion doubling as seen in section (\ref{Doublingwilson}). The corresponding dispersion relation for Wilson femrion is:
\begin{equation}
    \label{Wilsonrelation}
    a^2E^2_{\text{W}}(ap) = \sum_k \sin^2(ap_k) +\left[r\sum_k(1-\cos(ap_k)+am_f\right]^2
\end{equation}
The expression for Casimir energy of Wilson fermion in ($3+1$) is similar to the one of Naive fermion, where the Wilson Dispersion relation (\ref{Wilsonrelation}) is substituted in (\ref{def3+1}). The detailed derivations of Casimir energy in $(1+1)$-dimensional spacetime for MITBag (B) boundary conditions using the Abel-Plana formulae for finite range are presented in Appendix B (\ref{Wilsonmitbag}). Series expansions from (\ref{csc}) are substituted, and the following results are obtained:
\begin{align}
    aE_{\text{Cas}}^{\text{1+1D,B,W}}& =\frac{4N}{\pi} -\csc(\frac{\pi}{4N})\label{mitbagwilson} \\
   & =-4\left[-\frac{N}{\pi} + \frac{1}{4}\left\{\frac{4N}{\pi} + \frac{1}{6}\frac{\pi}{4N} +\frac{7}{360}\left(\frac{\pi}{4N}\right)^3 + ...\right\} \right]\nonumber\\
 \Rightarrow E_{\text{Cas}}^{\text{1+1D,B,W}} & = -\frac{\pi}{24d} - \frac{7\pi^3a^2}{23040d^3} + \mathcal{O}(a^4)
 \end{align}
Therefore, In the limit $a\rightarrow 0$, The Casimir energy obtained per unit area is:
\begin{equation}
\lim_{a \to 0}E_{\text{Cas}}^{\text{1+1D,B,W}} = -\frac{\pi}{24 d}
\end{equation}
The exact expressions obtained for the massless Naive and Wilson fermion using MITBag (B) boundary conditions in $(1+1)$-dimension using Abel-Plana formulae are (\ref{mitbagnaive}) and (\ref{mitbagwilson}) respectively.
   The leading terms of the order $1/N$ in the large $N$-limit for the analytic results obtained for Naive fermion in ($1+1$)-dimension using periodic (\ref{naivep}) and antiperiodic (\ref{naivea}) boundary conditions are plotted in Fig. \ref{mitbagplot1}. 
   Additionally, the numerical results for Naive lattice fermion in ($2+1$)- and ($3+1$)- dimensions, for both, massless and massive cases are also plotted for comparison. The smaller windows in each plot represent the coefficient of Casimir energy, $N^{D-1} dE_{\text{Cas}}$, which is constant in continuum and thus, is used to compare the lattice and continuum theories. Note that the exponential suppression of Casimir energy for massive fermionic fields using the boundary conditions in (\ref{allowed}), which was calculated for the fermion mass limit $md\ll1$ in (\ref{massivefer}), was also verified on a lattice and plotted along with the numerical results in Fig. \ref{mitbagplot1}(f). 
 \begin{figure}[t!]
\centering
    \subfloat\footnotesize{(a)}{{\includegraphics[width=7.4cm]{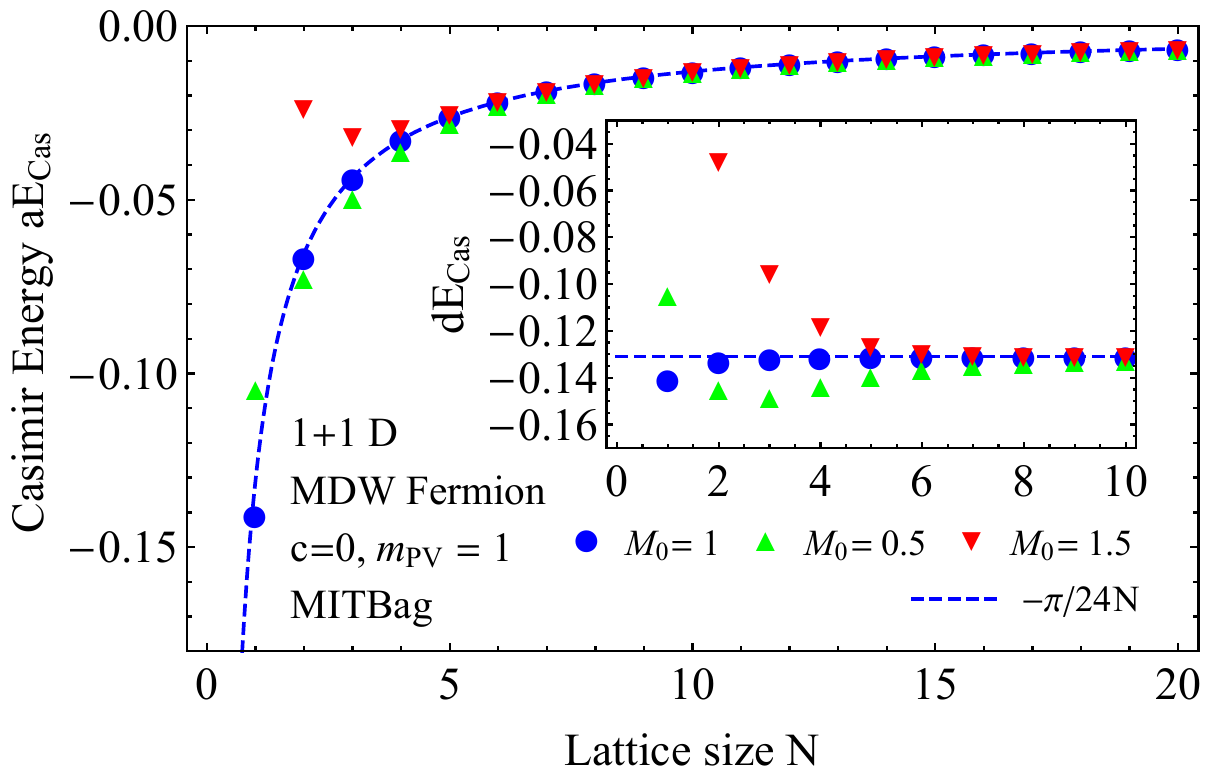} }} 
    \subfloat\footnotesize{(b)}{{\includegraphics[width=7.4cm ]{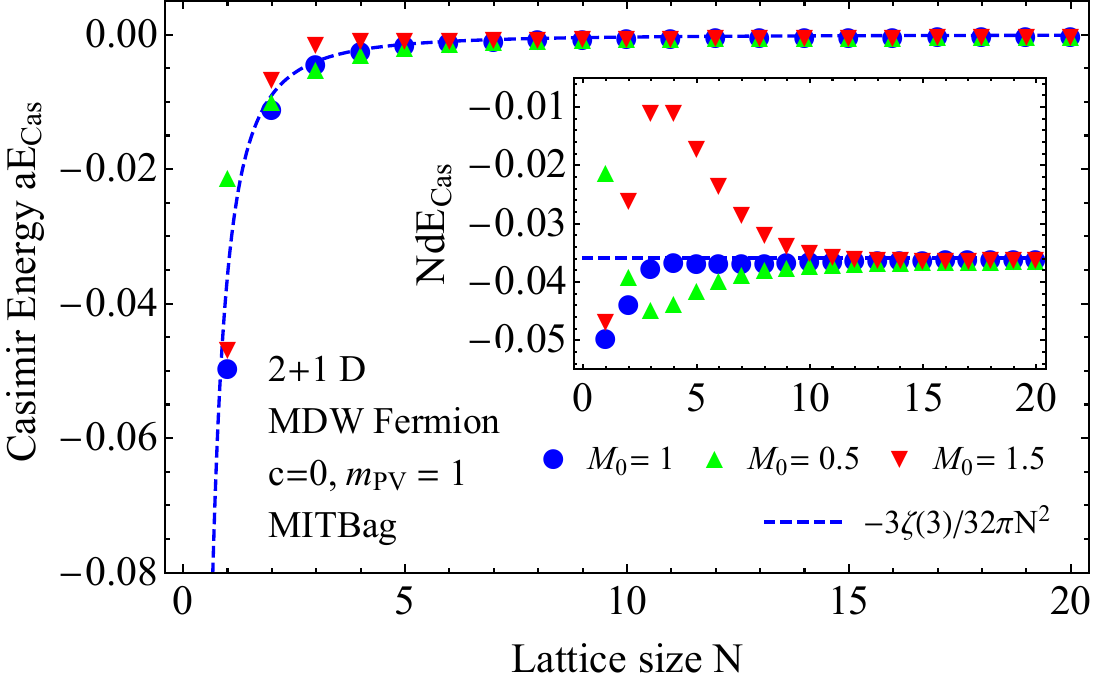} }}%
     \subfloat\footnotesize{(c)}{{\includegraphics[width=7.4cm]{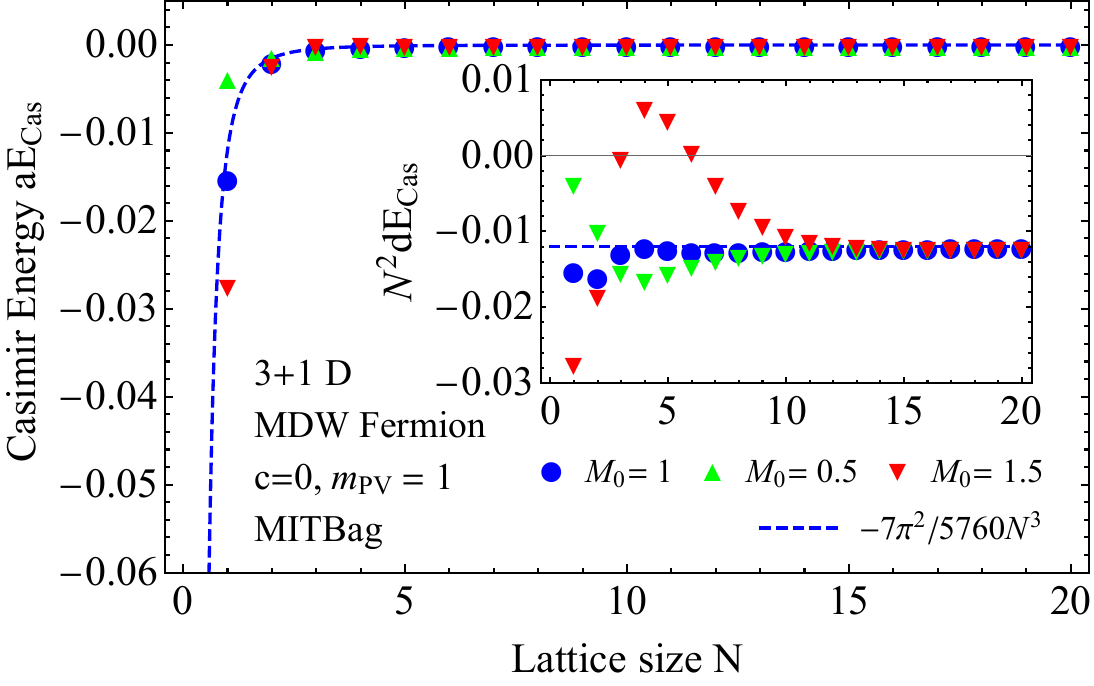}}}
    
    \caption{\footnotesize{The Casimir energy calculated numerically for Overlap fermion with MDW kernel with MITBag boundary conditions. (a),(b) and (c) represent the Casimir energy in ($1+1$)-, ($2+1$)-, and ($3+1$)-dimension respectively. }}%
\label{mdwmitbag}
\end{figure}The suppression of Casimir energy for massive naive fermion, as compared to the massless case can also be verified from Fig. \ref{dispersionrelations}. Again, in the ($1+1$)- and ($2+1$)-dimensional case, the expression for Naive fermion is twice and four times the continuum result, respectively, due to the fermion doubling in spatial directions. Also remember that the degeneracy factor $c_{\text{deg}} = 1$ in these cases, and only a factor of two ($N_{{D}} = 2$ for both cases) accounting for the particle-antiparticle degeneracy is considered. 
The coefficient $\frac{1}{2}$ in the definition (\ref{def1+1}, \ref{def3+1}) involving the zero-point energy is cancelled by this factor. 

The result obtained for the Wilson fermion in ($1+1$)- and ($2+1$)-dimension match with the continuum result (\ref{fercont}) exactly. Although, in the ($3+1$)-dimension, we have to take into account that the degeneracy factor $c_{\text{deg}} = 2$ arising from the spin degrees of freedom. Thus, in this case, the overall factor is four (as $N_D$ = 4). Once this additional factor of two is taken care of in the numerical results obtained on the lattice, the result we obtained for the Wilson fermion for ($3+1$)-dimension matches exactly with the continuum result in (\ref{cont}). In this case, the result for the Naive fermion is eight times the expression for the Wilson fermion. Thus, the Casimir energy for the Naive fermion is $2^D$ times the result for the Wilson fermion. We conclude that the reason for this discrepancy is definitely the phenomenon of doubling multiplicity for the Naive fermion.
\subsection{For Overlap fermion with MDW kernel}
\label{overlapmitbag}
The Casimir effect is numerically calculated for the Overlap fermion using the MITBag (B) boundary conditions  by substituting (\ref{EOV}) into the definitions (\ref{def1+1}) and (\ref{def3+1}) and varying domain wall height $M_0 = 0.5$, $1.0$ and $1.5$. Note that $c=0$, $m_{\text{PV}}=1.0$ are kept as fixed parameters in the dispersion relation throughout our calculation. The results obtained for $(1+1)$-, $ (2+1)$-, and $(3+1)$-dimensions are plotted in Fig \ref{mdwmitbag}.

It is easy to notice in ($1+1$)- dimensions that the Overlap fermion case $M_0 = 1.0$ is equivalent to the massless ($am_f = 0$) Wilson fermion case. This can be verified from the equal dispersion relations for both these cases. The expressions obtained numerically for Overlap fermions with MDW kernel match precisely with the Wilson fermions in the zero lattice spacing limit $a\rightarrow0$ and, subsequently, also with the continuum result.
\section{For the Naive lattice fermion}
We now discuss the Casimir energy expressions for Naive lattice fermion ($nf$) in $(1+1)$- dimensional spacetime using periodic and antiperiodic boundary conditions.
 Further, the numerically evaluated Casimir energy results for the massive lattice fermions and for lattice fermions in higher dimensions are also discussed.
\subsection{Periodic Boundary}
The expression for the Casimir energy of Naive fermion with periodic (P) boundary conditions, used to numerically study the Casimir energy in higher dimensions is:
   \begin{align}
   \label{naiveexp}
       &aE_{\text{Cas}}^{\text{3+1D,P,}nf} \equiv  aE_{\text{0}}^{\text{3+1D,P,}nf}(N) - aE_{\text{0}}^{\text{3+1D,P,}nf}(N\rightarrow\infty)\\
       &=  c_{\text{deg}}\int \frac{d^2ap_{\perp}}{(2\pi)^2}\Bigg[-\sum_{n=0}^{N-1} \sqrt{\sin^2 \frac{2n\pi}{N}+\sum_{k=2,3}\sin^2(ap_k) + (am_f)^2} \\& \;\;\;\;\;\;\;\;\;\;\;\;\;\;\;\;\;\;\;\;\;\;\;\;\;\;\;\;\;\;\;\;\;\;\;\;\;\;\;\;\;\;\;\;\;\;\;\;\;\;\;\;\;\;\;\;\;\;\;\;+ N\int_{\text{BZ}}\frac{dap_1}{2\pi}\sqrt{ \sum_{k=1,2,3}\sin^2(ap_k) + (am_f)^2}  \Bigg]\nonumber
  \end{align}
 Similar to the previous section, exact expressions for the Casimir energy $aE_{\text{Cas}}$ of massless Naive fermion on lattice in ($1+1$)-dimensional spacetime are obtained. For detailed derivation of these expressions, refer to Ref. \cite{Ishikawa:2020icy}. They are as follows:
\begin{equation} 
aE^{\text{1+1D,P},nf}_{\text{Cas}} = \begin{dcases}
                \frac{2N}{\pi}- 2\cot(\frac{\pi}{N}) & \text{if $N=\text{even}$}\\ 
                \frac{2N}{\pi}- \cot(\frac{\pi}{N})- \csc(\frac{\pi}{N})  = \frac{2N}{\pi}- \cot(\frac{\pi}{2N})& \text{if $N=\text{odd}$}\\
               \end{dcases}  
               \label{naivep}
\end{equation}
We have encountered different expressions for Casimir energy on odd and even lattice sizes. Although, if one calculates the difference between these two quantities at a single lattice point, i.e $aE^{\text{1+1D,P},nf}_{\text{Cas}}(N=\text{even})-aE^{\text{1+1D,P},nf}_{\text{Cas}}(N=\text{odd})$, the difference is $\tan(\frac{\pi}{2N})$ which is a rapidly converging function. Series expansions from (\ref{cot}) are substituted and the following results are obtained:
 \begin{align}
     \label{1/Nexp}
    E^{\text{1+1D,P},nf}_{\text{Cas}} &= 
    \begin{dcases}
    \frac{\pi}{6d} + \frac{\pi^3a^2}{360d^3} +\mathcal{O}(a^4)\;\;\;\;\;& \text{if }N=\text{odd}\\
     \frac{2\pi}{3d} + \frac{2\pi^3a^2}{45d^3} +\mathcal{O}(a^4)\;\;\;\;\;& \text{if }N=\text{even}
     \end{dcases}
\end{align} 
Thus, expressions obtained for the Casimir effect of Naive fermion in the continuum limit for odd and even lattice sizes, individually are:
\begin{equation}
\label{naivecont}
\lim_{a \to 0}E_{\text{Cas}}^{\text{1+1D,P,}nf} = \frac{\pi}{6d}\;\;\;\;\;(\text{if }N=\text{odd})\;\;;\;\;\lim_{a \to 0}E_{\text{Cas}}^{\text{1+1D,P,}nf} = \frac{2\pi}{3d}\;\;\;\;\;(\text{if }N=\text{even})
\end{equation}
\subsection{Antiperiodic Boundary}
The expression for the Casimir energy of Naive fermion with antiperiodic (AP) boundary conditions in ($3+1$)-dimensional spacetime is:
   \begin{align}
   \label{naiveexp2}
       &aE_{\text{Cas}}^{\text{3+1D,AP,}nf} \equiv  aE_{\text{0}}^{\text{3+1D,AP,}nf}(N) - aE_{\text{0}}^{\text{3+1D,AP,}nf}(N\rightarrow\infty)\\
       &=  c_{\text{deg}}\int \frac{d^2ap_{\perp}}{(2\pi)^2}\Bigg[-\sum_{n=0}^{N-1} \sqrt{\sin^2 \frac{(2n+1)\pi}{N}+\sum_{k=2,3}\sin^2(ap_k) + (am_f)^2} \;\;+\\&\;\;\;\;\;\;\;\;\;\;\;\;\;\;\;\;\;\;\;\;\;\;\;\;\;\;\;\;\;\;\;\;\;\;\;\;\;\;\;\;\;\;\;\;\;\;\;\;\;\;\;\;\;\;\;\;\;\;\;\;\;\;\;\;\;\;\;\;\;\;\;\;\;\;\;\;\;\;\;\;  N\int_{\text{BZ}}\frac{dap_1}{2\pi}\sqrt{ \sum_{k=1,2,3}\sin^2(ap_k) + (am_f)^2}  \Bigg]\nonumber
  \end{align}
 \begin{figure}[t!]
\centering
    \subfloat\footnotesize{(a)}{{\includegraphics[width=7.4cm]{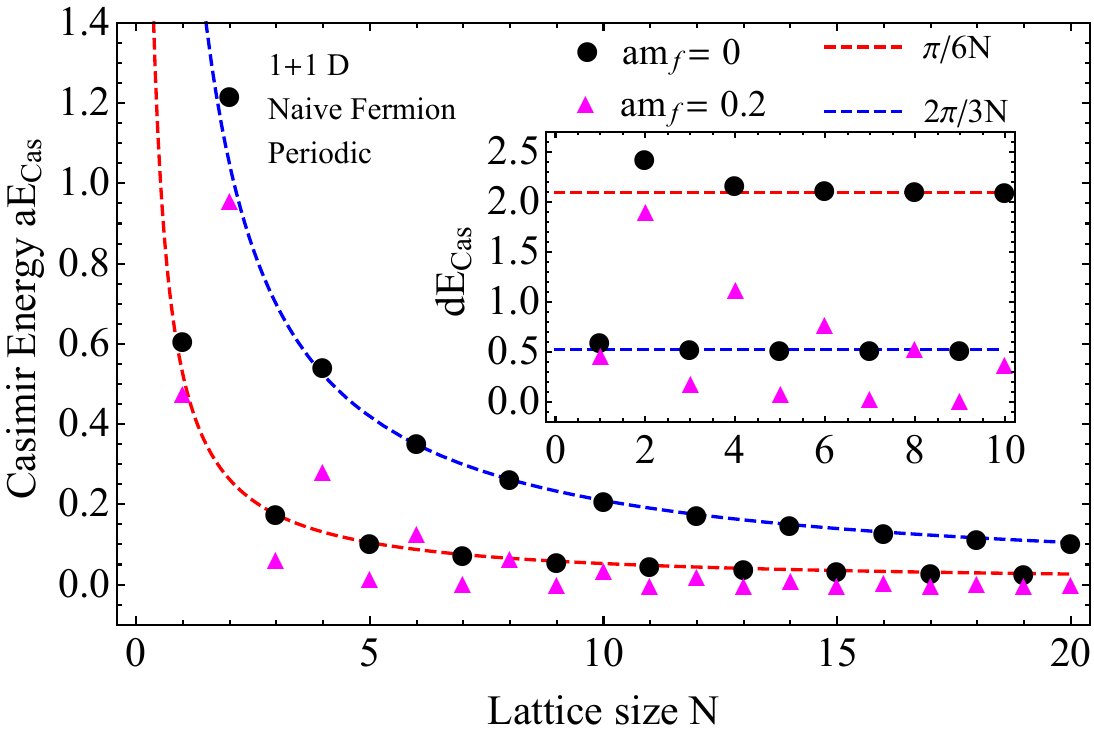} }} 
    \subfloat\footnotesize{(b)}{{\includegraphics[width=7.4cm ]{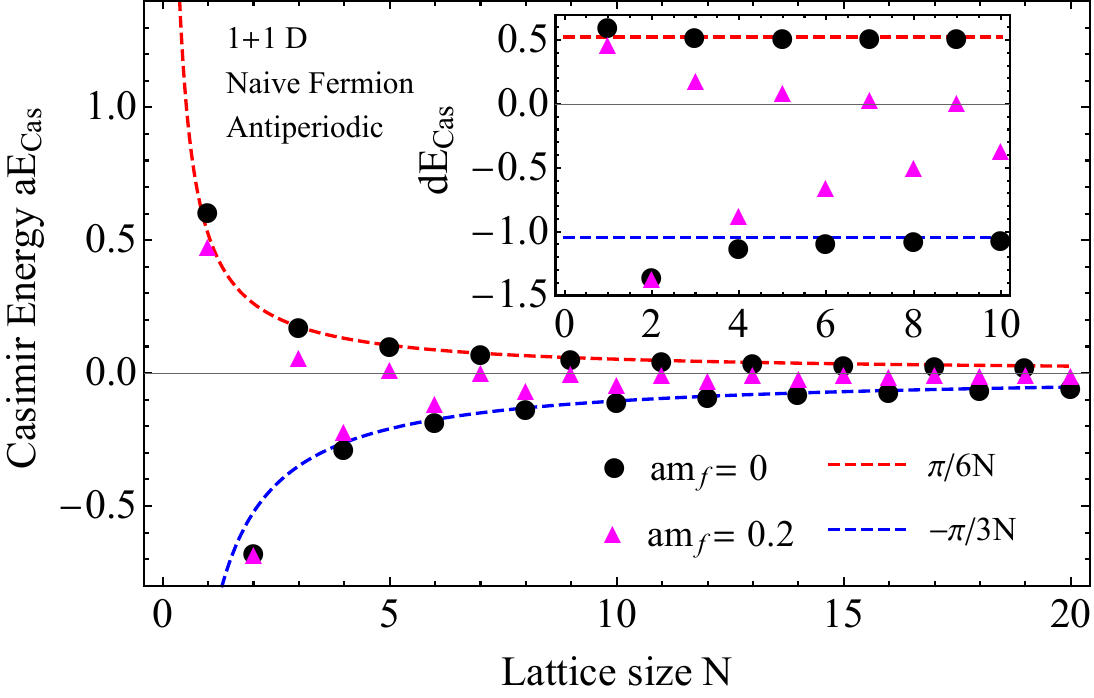} }}%
     \subfloat\footnotesize{(c)}{{\includegraphics[width=7.4cm]{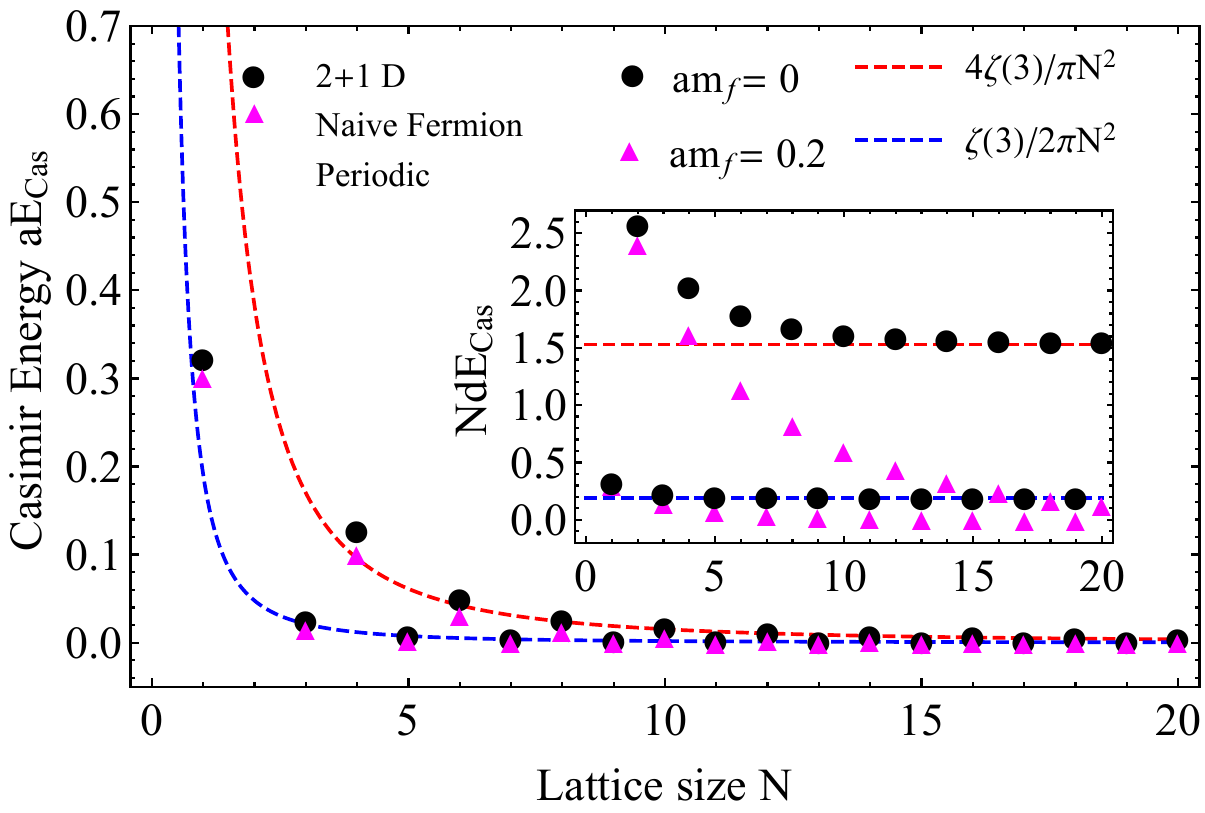}}} %
    \subfloat\footnotesize{(d)}{{\includegraphics[width=7.4cm]{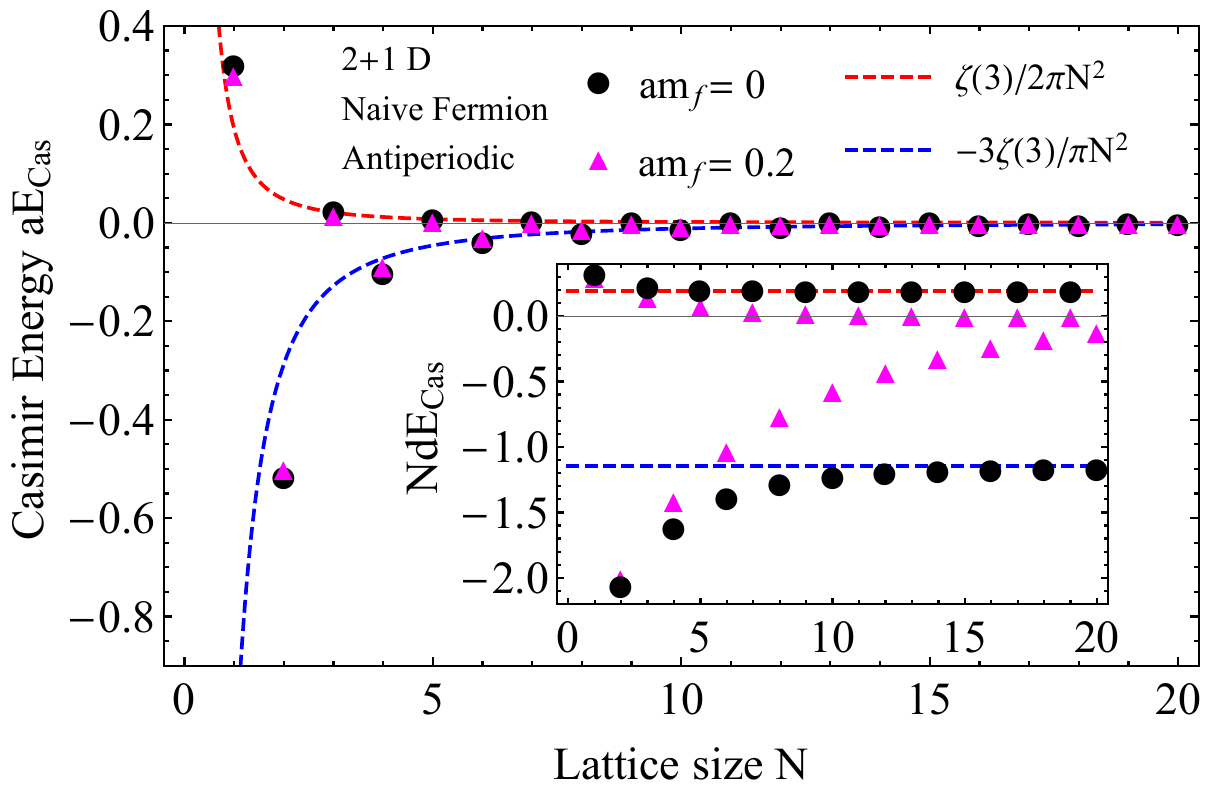}}}
    \subfloat\footnotesize{(e)}{{\includegraphics[width=7.4cm]{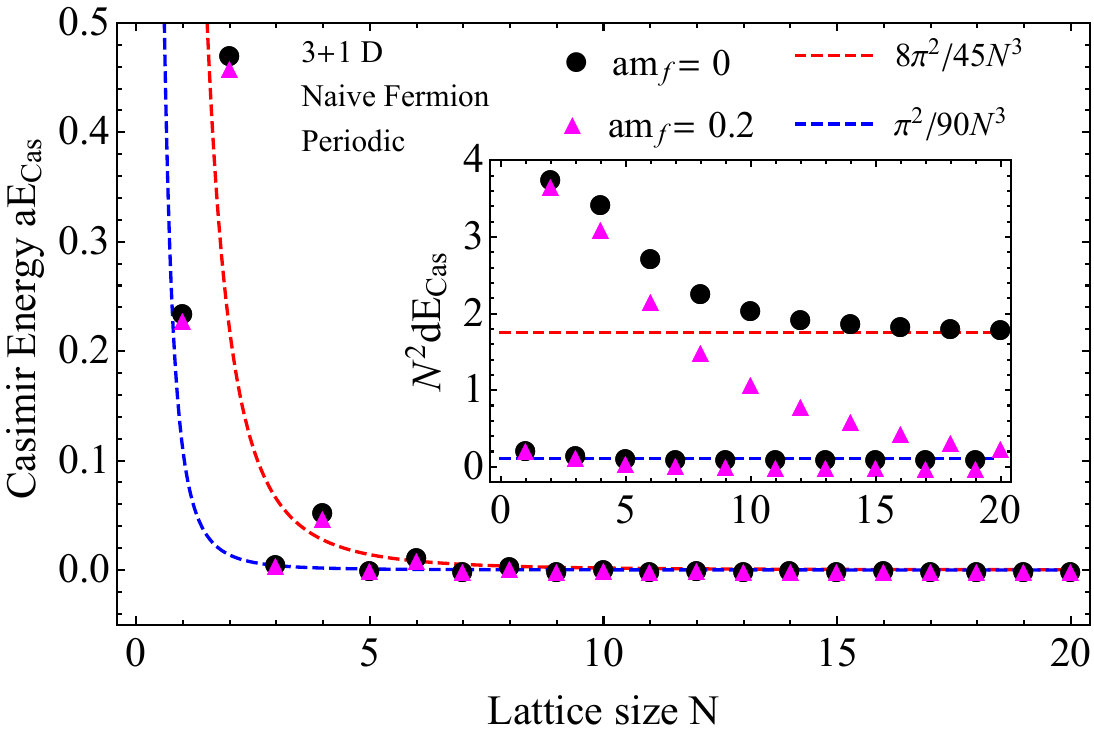}}} %
    \subfloat\footnotesize{(f)}{{\includegraphics[width=7.4cm]{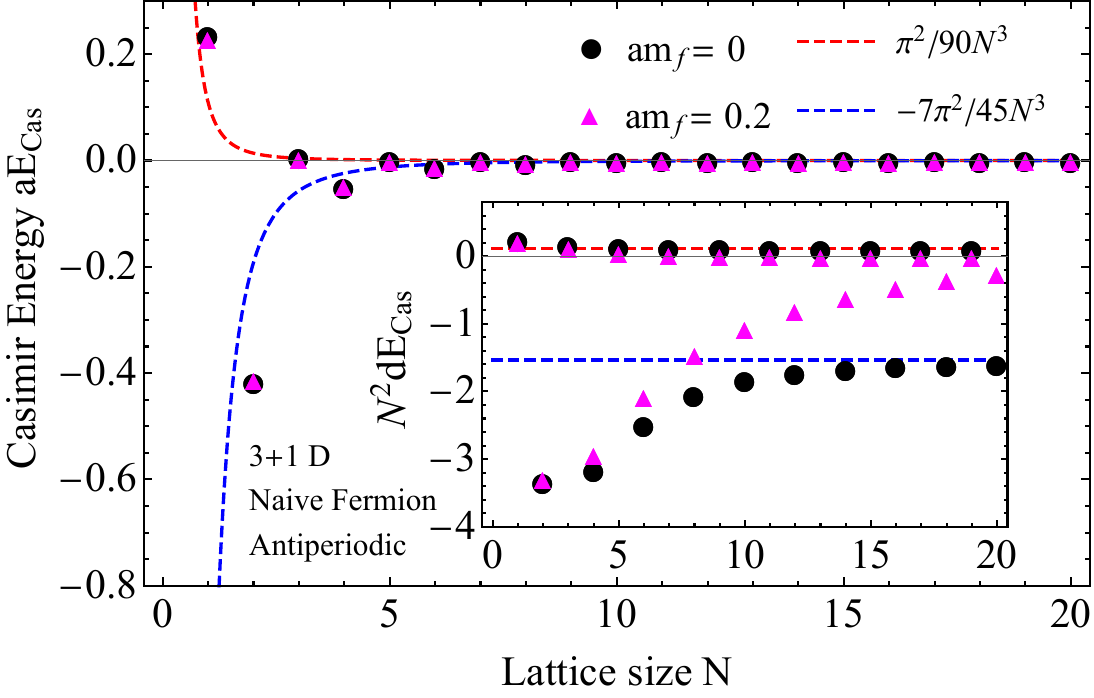}}}%
\caption{\footnotesize{Casimir energy for massless and positively massive Naive fermion.[(a),(b)], [(c),(d)] and [(e),(f)] represent the periodic and antiperiodic Casimir energy for Naive fermion in ($1+1$)-, ($2+1$)-, and ($3+1$)-dimensions respectively. The lines represent the leading order terms of $1/N$ expansion (\ref{1/Nexp}) for the massless fermion in the large $N$-limit.}}%
\label{naivefig}
\end{figure}
This is used for the numerical analysis of Casimir energy in higher dimensions. Detailed derivation for exact expressions in ($1+1$)-dimensional space-time can be found in Ref. \cite{Ishikawa:2020icy}. The results are as follows:
\begin{equation}
 aE_{\text{Cas}}^{\text{1+1D,AP},nf} =   \begin{dcases}
                \frac{2N}{\pi}- 2\csc(\frac{\pi}{N}) & \text{ if } N=\text{even}\\ 
                \frac{2N}{\pi}- \cot(\frac{\pi}{2N}) & \text{ if } N=\text{odd}
               \end{dcases} 
               \label{naivea}
\end{equation}
Similar to the periodic case, the difference between the above two expressions on a single lattice point $aE^{\text{1+1D,P},nf}_{\text{Cas}}(N=\text{even})-aE^{\text{1+1D,P},nf}_{\text{Cas}}(N=\text{odd})$ is $-\tan(\frac{\pi}{2N})$ and converges rapidly. The small $a$ expansions of formulas (\ref{naivep}, \ref{naivea})  by $1/N$ are obtained in \cite{Ishikawa:2020ezm} for odd and even lattice sizes ($N$) separately by substitution the trigonometric series (\ref{cot}, \ref{csc}). 
\begin{align}
      E^{\text{1+1D,AP},nf}_{\text{Cas}} &= 
     \begin{dcases}
   \frac{\pi}{6d} + \frac{\pi^3a^2}{360d^3} +\mathcal{O}(a^4)\;\;\;\;\;& \text{if }N=\text{odd}\\
    -\frac{\pi}{3d} - \frac{7\pi^3a^2}{180d^3} +\mathcal{O}(a^4)\;\;\;\;\;&\text{if }N=\text{even}
    \end{dcases}
\end{align}
The continuum limit ($a\rightarrow0)$ expressions for the original Dirac fermion are also obtained subsequently as:
\begin{equation}
\lim_{a \to 0}E_{\text{Cas}}^{\text{1+1D,AP,}nf} = \frac{\pi}{6d}\;\;\;\;\;(\text{if }N=\text{odd})\;\;;\;\;\lim_{a \to 0}E_{\text{Cas}}^{\text{1+1D,AP,}nf} = -\frac{\pi}{3d}\;\;\;\;\;(\text{if }N=\text{even})
\end{equation}
 which are surprisingly different from the continuum expressions obtained in (\ref{cont}).
 These results are plotted in Fig. \ref{naivefig}. They suggested an apparent violation of universality for the lattice fermions and needed review. This issue for the Naive fermion has been addressed in section (\ref{extrapol}).
\subsection{Splitting the Brillouin zone for Naive fermion}
\label{BZsplit}
 We earlier speculated the origins of the oscillation in the Casimir energy for Naive fermion to be in the fermion doubling, as both of these phenomena are unique to the Naive discretization. We sought to investigate the behaviour of our calculations by splitting the Brillouin zone into two regions $R_1$: $0\leq ap_k <\pi$ and $R_2$: $\pi \leq ap_k <2\pi$. The corresponding expressions for Casimir energy of Naive fermion (\ref{def3+1}) in ($1+1$)- and higher dimensions were split and were simplified in terms of the integral limit from $0\leq ap_k <\pi$ ($R_1$). Let us briefly discuss the analytic results obtained in the ($1+1$)-dimension for periodic boundary conditions.

Considering the case of $N$ being even, $R_1$ limits $n \in[0,\frac{N}{2}-1]$. Solving this case, we observe that the Casimir energy expression for momenta in the complete Brillouin zone in (\ref{naiveexp}) is exactly twice the expression obtained for momenta restricted to $R_1$. Although, performing a similar exercise and solving further for $N=$odd lands us into a problem. If $N$ is odd, we get $n \in[0,\frac{N-1}{2}]$. In this case, one is unable to write the expression (\ref{naiveexp}) corresponding to the complete Brillouin zone in terms of the partial integral form $R_1$. An antiperiodic boundary contribution emerges out of the summation of $n$ in (\ref{naiveexp}) for the periodic boundary condition. Likewise, on performing a similar exercise for the Naive fermion Casimir energy expression using antiperiodic boundary condition (\ref{naiveexp2}), a periodic contribution emerges from the expression for the $N=$ odd case. We realized that if one expresses odd $N$ as $2k+1$, where $k\in\mathbb{Z}$, the summation in the expression over the complete Brillouin zone is over an odd number of $n$ values and thus cannot be equally split between the two regions $R_1$ and $R_2$, having $k+1$ and $k$ digits in the sum respectively.

Note that the intermediate calculations 
done by limiting the range of momenta from $ap_k\in[0,\pi)$ 
depends on the lattice size $N$ being odd or even. Thus, when one considers the complete Brillouin zone, the final result is also dependent on $N$ being odd or even.
\section{For the Wilson lattice fermion}
\label{WilsonPAP}
 We shall now discuss the analytic and numerical Casimir energy results for the Wilson lattice fermion for periodic and anti-periodic boundary conditions. 
  \begin{figure}[t!]
\centering
    \subfloat\footnotesize{(a)}{{\includegraphics[width=7.4cm]{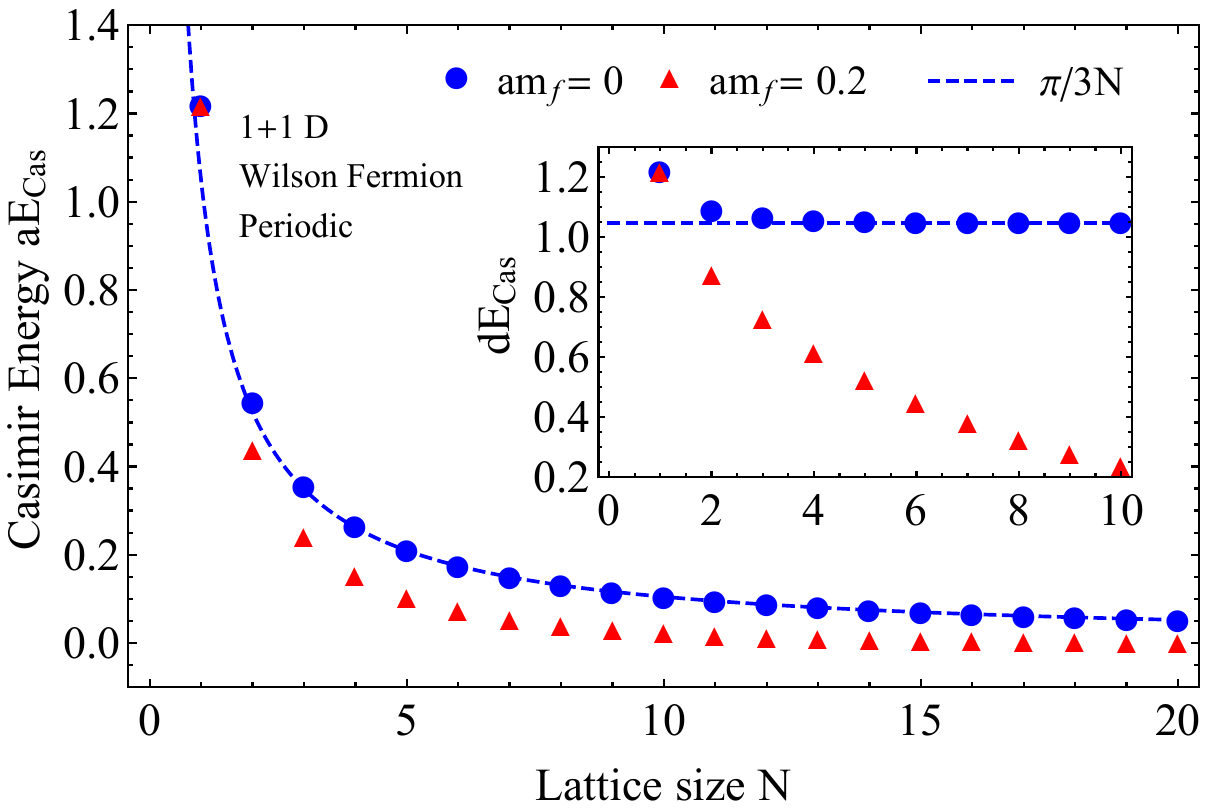} }} %
    \subfloat\footnotesize{(b)}{{\includegraphics[width=7.4cm ]{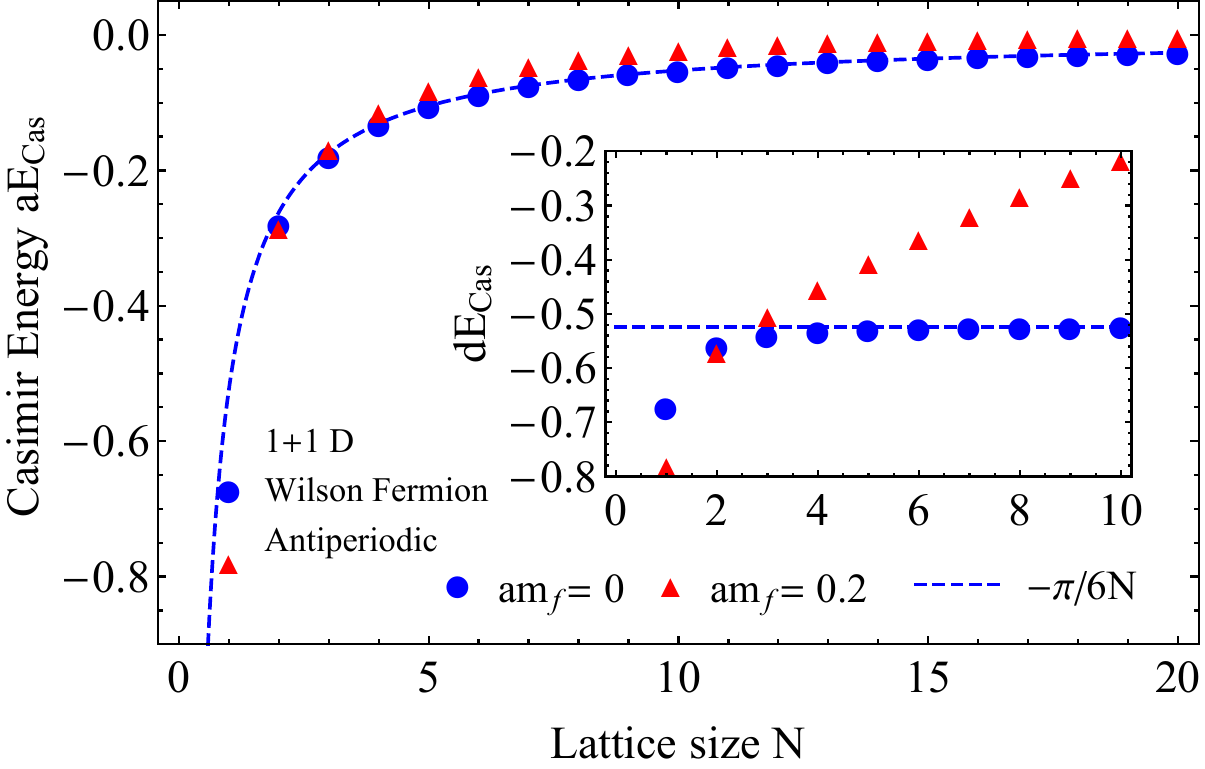} }}%
     \subfloat\footnotesize{(c)}{{\includegraphics[width=7.4cm]{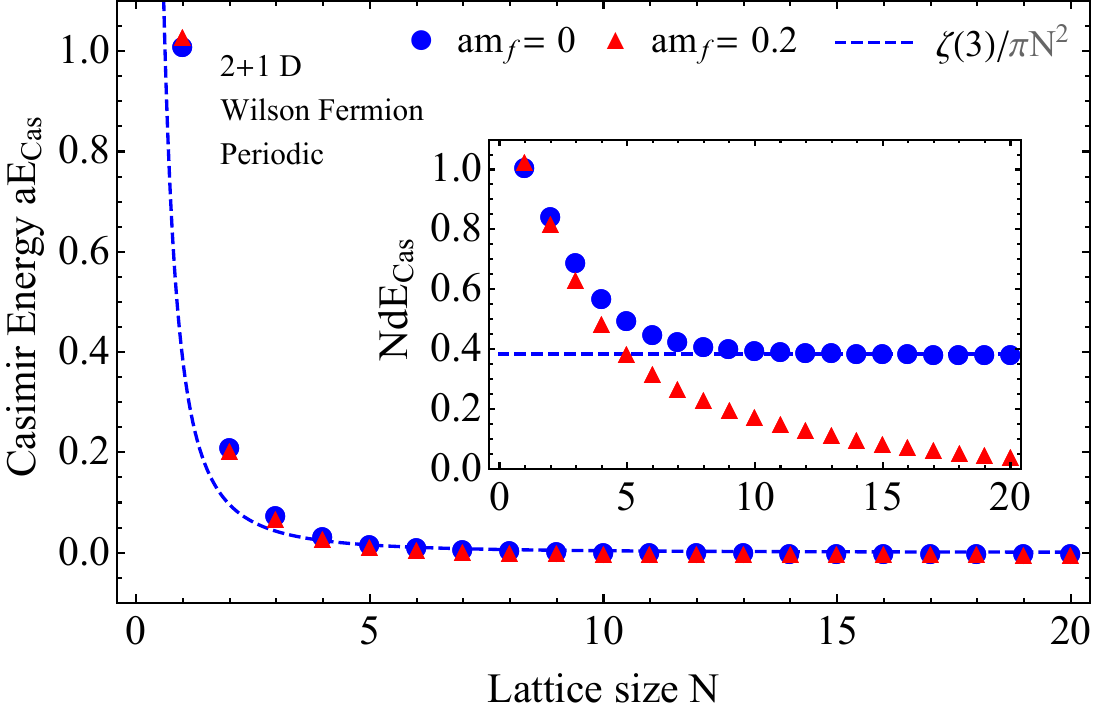}}} 
    \subfloat\footnotesize{(d)}{{\includegraphics[width=7.4cm]{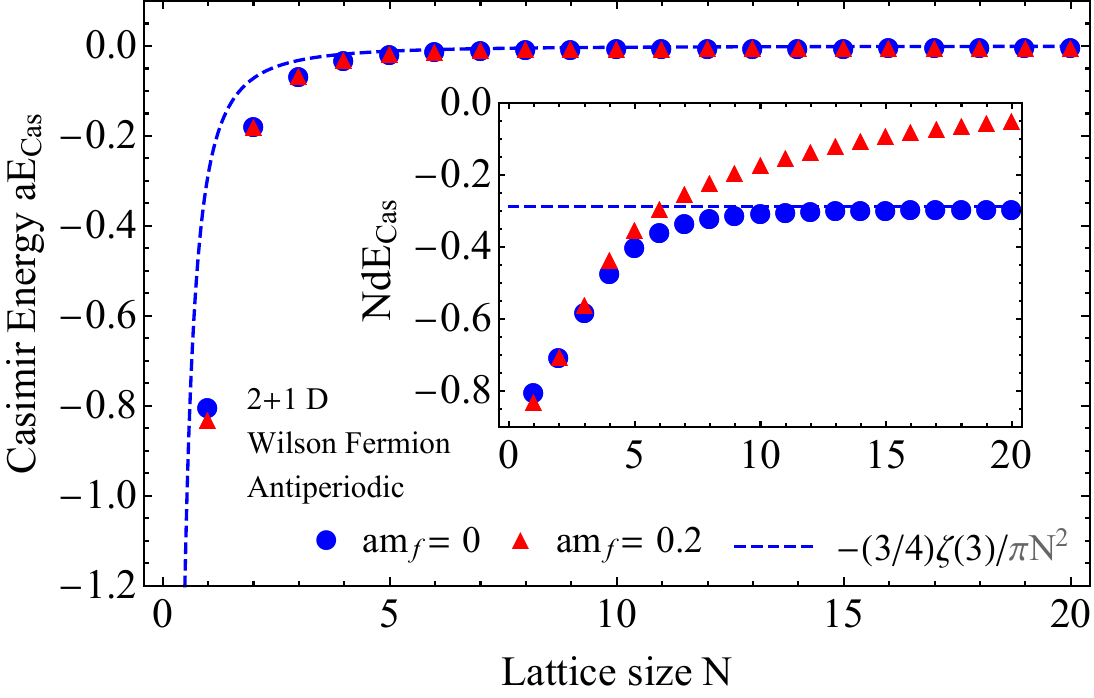}}}
    \subfloat\footnotesize{(e)}{{\includegraphics[width=7.4cm]{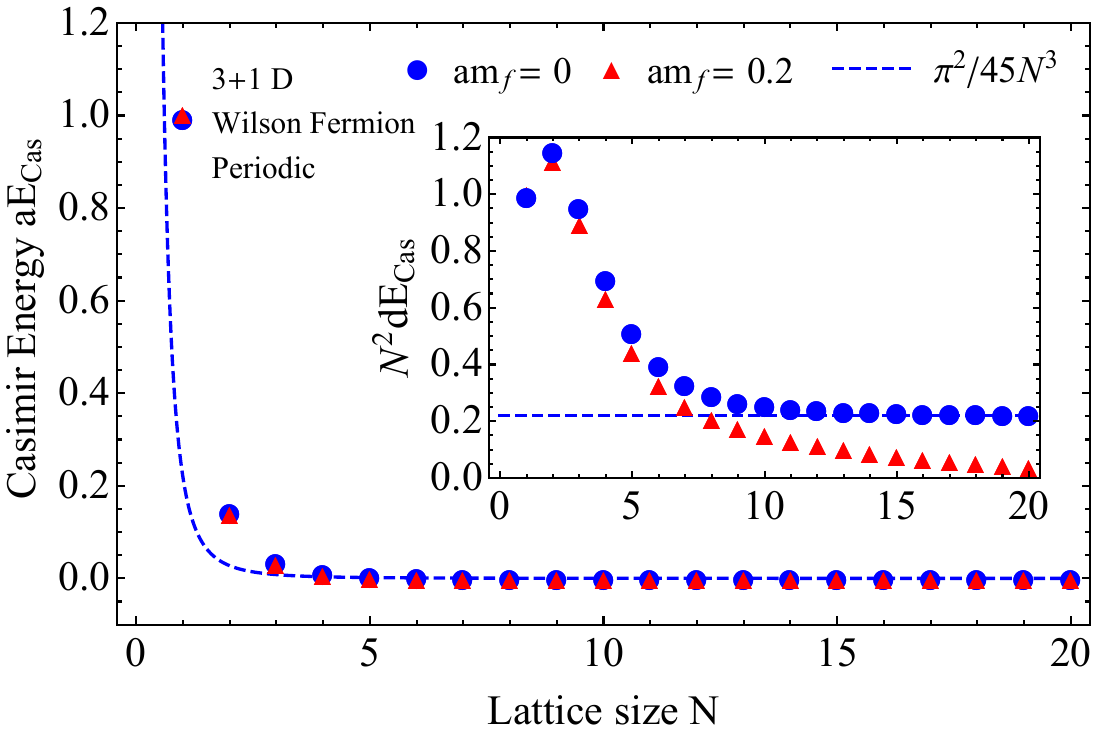}}} %
    \subfloat\footnotesize{(f)}{{\includegraphics[width=7.4cm]{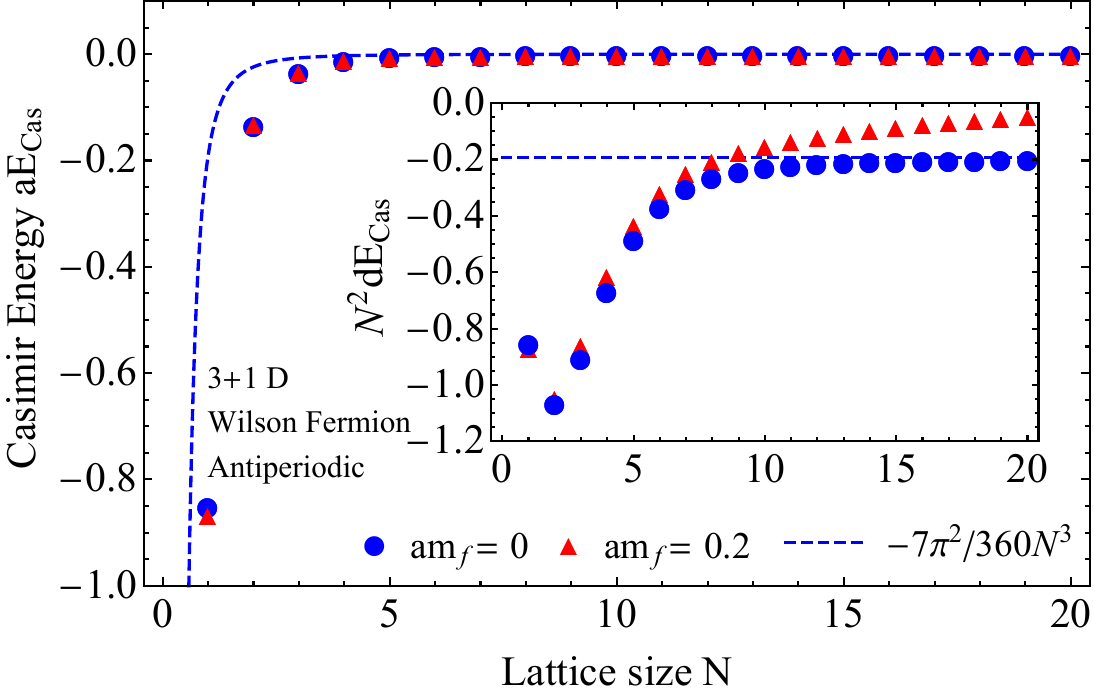}}}%
    
    \caption{\footnotesize{The Casimir energy for Wilson fermion, both massless and positive massive cases are plotted. [(a),(b)], [(c),(d)] and [(e),(f)] represent the periodic and anti-periodic boundary conditions for Wilson fermion in ($1+1$)-, ($2+1$)-, and ($3+1$)-dimension respectively.}}%
\label{wilsonfig}
\end{figure}
\subsection{Periodic Boundary}
We use the integral Abel-Plana formulae in finite range for periodic boundary (\ref{int}). Following the detailed derivation for these expressions given in Ref. \cite{Ishikawa:2020icy}, the Casimir energy is:
 \begin{align} 
 \label{wilsonperi}
aE_{\text{Cas}}^{\text{1+1D,P,W}} &=  \frac{4N}{\pi}- 2\cot(\frac{\pi}{2N})\\\Rightarrow E_{\text{Cas}}^{\text{1+1D,P,W}} &= \frac{\pi}{3d}+\frac{\pi^3a^2}{180d^3}+\mathcal{O}(a^4) 
\end{align}
Therefore, in the zero lattice spacing limit $a\rightarrow 0$, the expression obtained per unit area is:
\begin{equation}
\lim_{a \to 0}E_{\text{Cas}}^{\text{1+1D,P,W}} = \frac{\pi}{3 d}
\end{equation}
This is equivalent to the expression for Casimir energy calculated for Dirac fermions in continuum for periodic boundary conditions mentioned in (\ref{PAP1p1}). Thus, this method can effectively be regarded as an alternative derivation for the fermionic Casimir effect. The Casimir energy for Wilson fermion using periodic and antiperiodic boundary conditions is plotted in Fig. \ref{wilsonfig}. 
\subsection{Antiperiodic Boundary}
 
We use the half-integral Abel-Plana formulae in finite range for antiperiodic boundary (\ref{nonint}). Following the detailed derivation for these expressions given in Ref. \cite{Ishikawa:2020icy}, we get Casimir energy as:
\begin{align}
\label{wilsonantiperi}
aE_{\text{Cas}}^{\text{1+1D,AP,W}} &=  \frac{4N}{\pi}- 2\csc(\frac{\pi}{2N})\\\Rightarrow E_{\text{Cas}}^{\text{1+1D,AP,W}} &= -\frac{\pi}{6d}-\frac{7\pi^3a^2}{1440d^3}+\mathcal{O}(a^4)      
\end{align}
Similary, the Casimir energy obtained on lattice in the limit $a\rightarrow 0$ per unit area is:
\begin{equation}
\lim_{a \to 0}E_{\text{Cas}}^{\text{1+1D,AP,W}} = -\frac{\pi}{6d}
\end{equation}
This, again, is equivalent to the Casimir energy expressions calculated for Dirac fermions in continuum for antiperiodic boundary conditions mentioned in (\ref{PAP1p1}). The Casimir energy expressions of Naive and Wilson fermion for periodic and antiperiodic boundaries disagree with the zero lattice spacing limit result, thus, challenging the notion of universality for lattice fermions. 
\section{Series Extrapolation in large $N$-Limit}
\label{extrapol}
We saw in section (\ref{slabbag}) that the Casimir energy for all lattice fermions with MIT Bag boundary conditions matched the results for the Dirac fermion in continuum when lattice spacing was brought to zero. There was no oscillation observed over lattice size $N$ in this case. After studying the Casimir energy expression for Naive fermion using periodic and antiperiodic conditions by splitting the Brillouin zone in section (\ref{BZsplit}), we sought to understand the behaviour of Casimir energy for large lattice sizes $N$. On extrapolating the Casimir energy expression using suitable numerical techniques, we found that the results for Naive fermion match the continuum expression for fermionic Casimir energy precisely in the limit lattice spacing $a\rightarrow0$. The techniques of series acceleration are often applied in numerical analysis to improve the speed of numerical convergence. The Euler-Maclaurin summation formulae in (\ref{EMFormula}), Wynn's epsilon ($\epsilon$) method and Richardson Extrapolation are some examples of series acceleration/ convergence techniques. The Euler-Maclaurin method is used in cases where the last term tends to 0 as $n\rightarrow\infty$, and an infinite series of higher-order derivative differences can then be obtained. In such cases, sums may be expressed in terms of integrals by inverting the formula to obtain the Euler-Maclaurin sum\footnote{The routine attribute of numerical summation in Wolfram Mathematica, $\texttt{"NSum"}$ uses the Euler-Maclaurin summation formulae using $\texttt{"Method}\rightarrow\texttt{EulerMaclaurin"}$ to extrapolate the series after a specified number of terms in the summation using the $\texttt{"NSumTerms"}$ attribute. The expression used for this is mentioned in Appendix (\ref{EMMathematica}).}.
 \begin{figure}[H]
\centering
    \subfloat\footnotesize{(a)}{{\includegraphics[width=7.4cm]{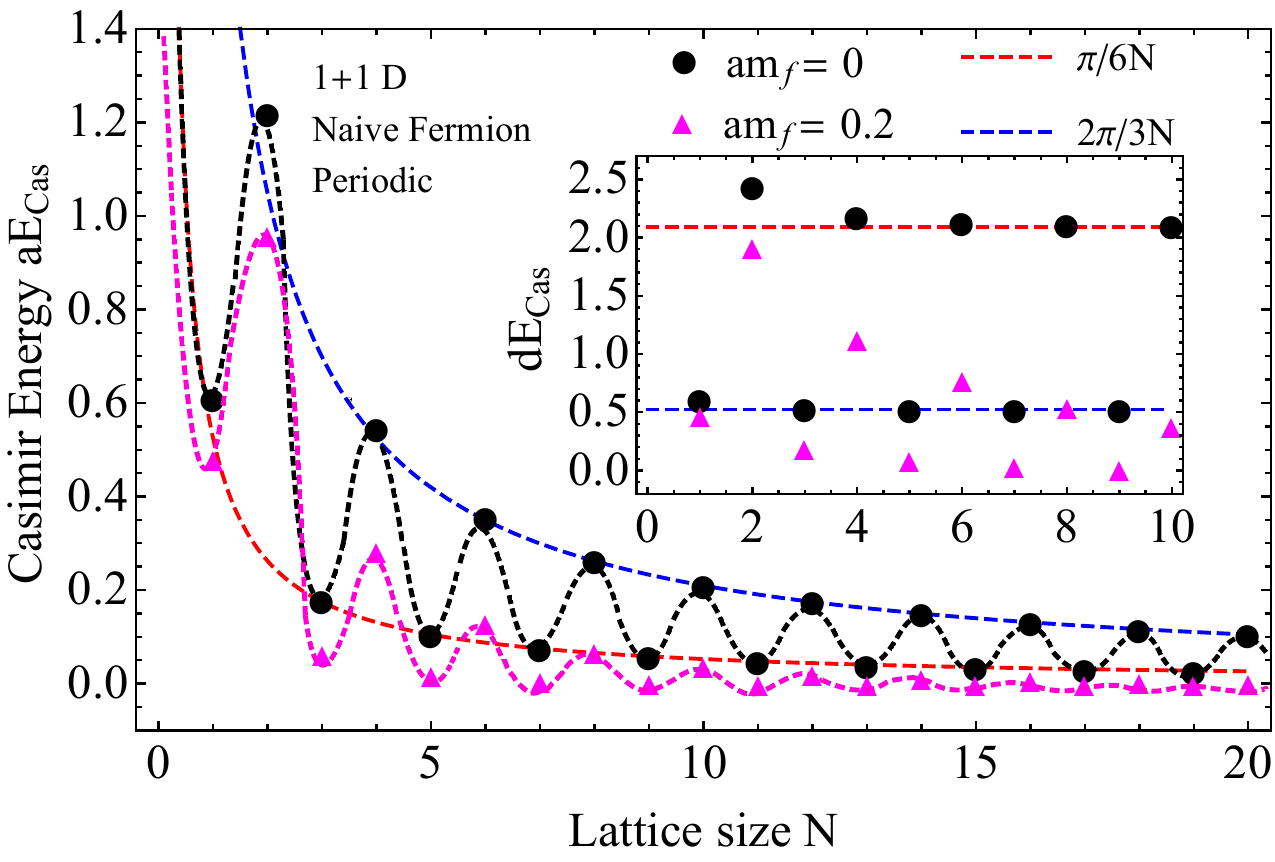} }} 
    \subfloat\footnotesize{(b)}{{\includegraphics[width=7.4cm ]{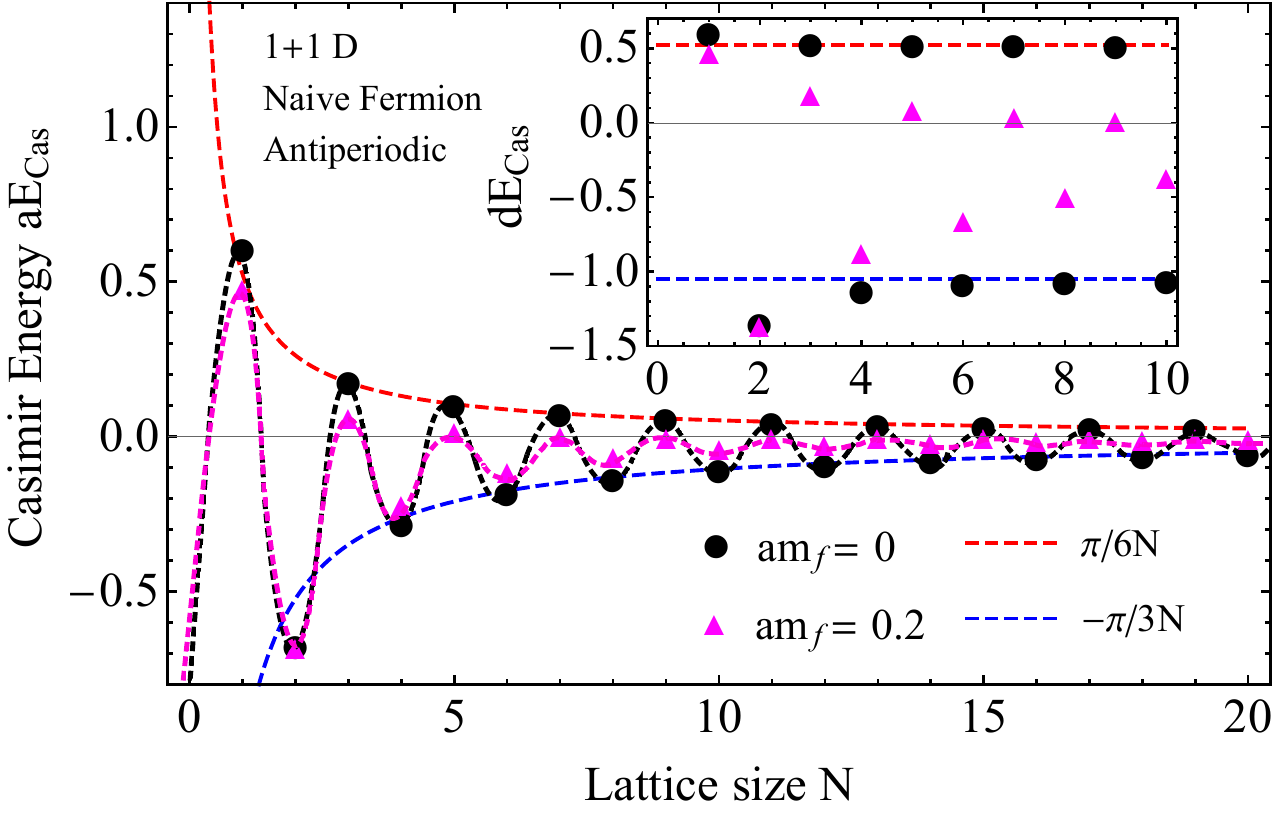} }}%
\caption{\footnotesize{Oscillation of Casimir energy for positively massive and massless Naive fermion for odd and even $N$. [(a),(b)] represent the periodic and antiperiodic boundary conditions  in ($1+1$)-dimension respectively.}}%
\label{osci}
\end{figure}

\begin{figure}[t!]
\centering
    \subfloat\footnotesize{(a)}{{\includegraphics[width=7.4cm]{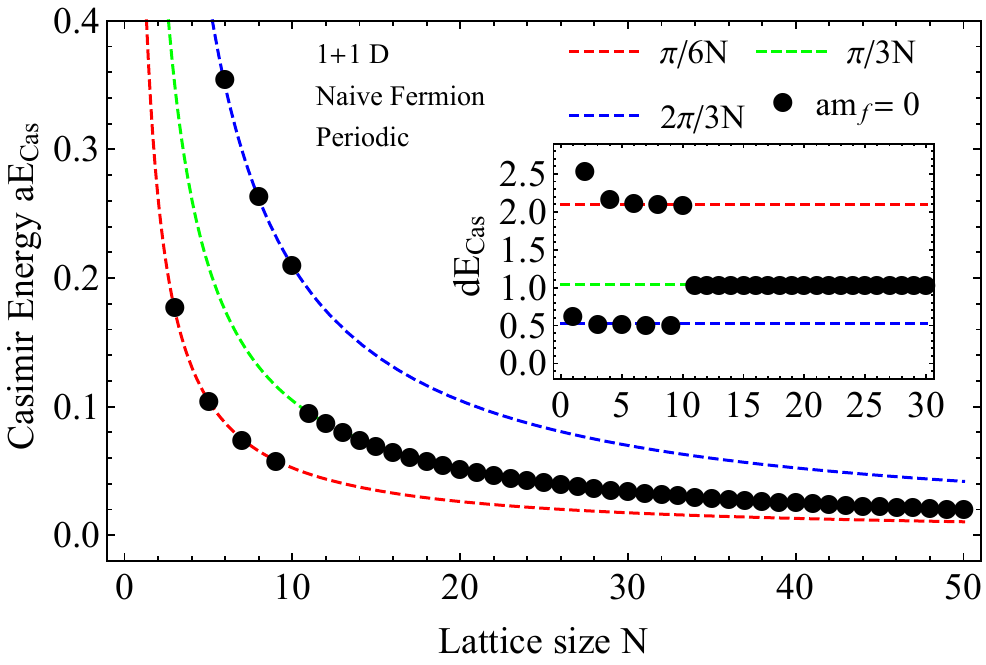} }} %
  \subfloat\footnotesize{(b)}{{\includegraphics[width=7.4cm ]{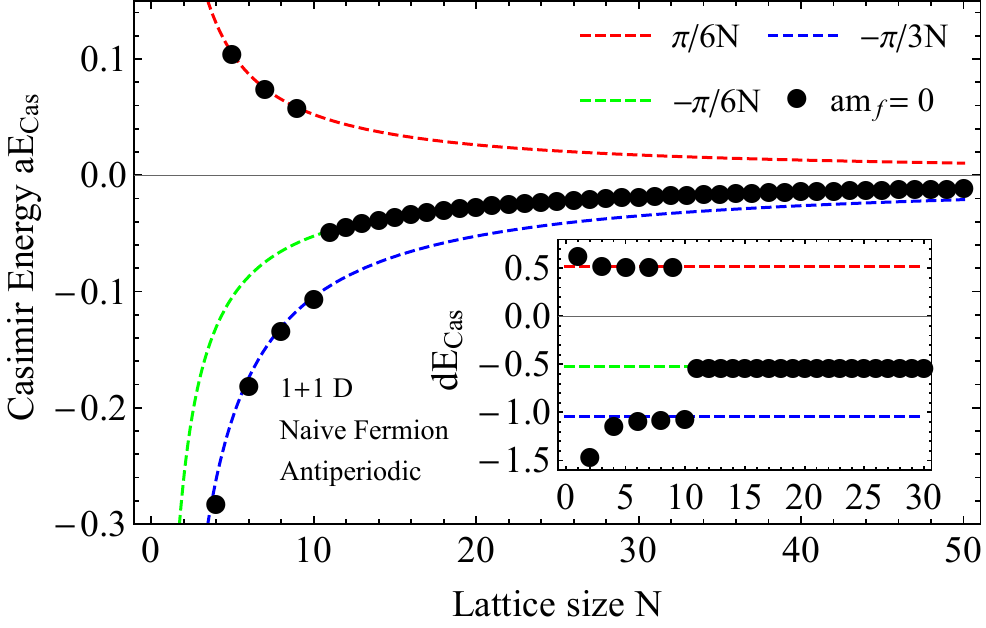} }}%
     \subfloat\footnotesize{(c)}{{\includegraphics[width=7.4cm]{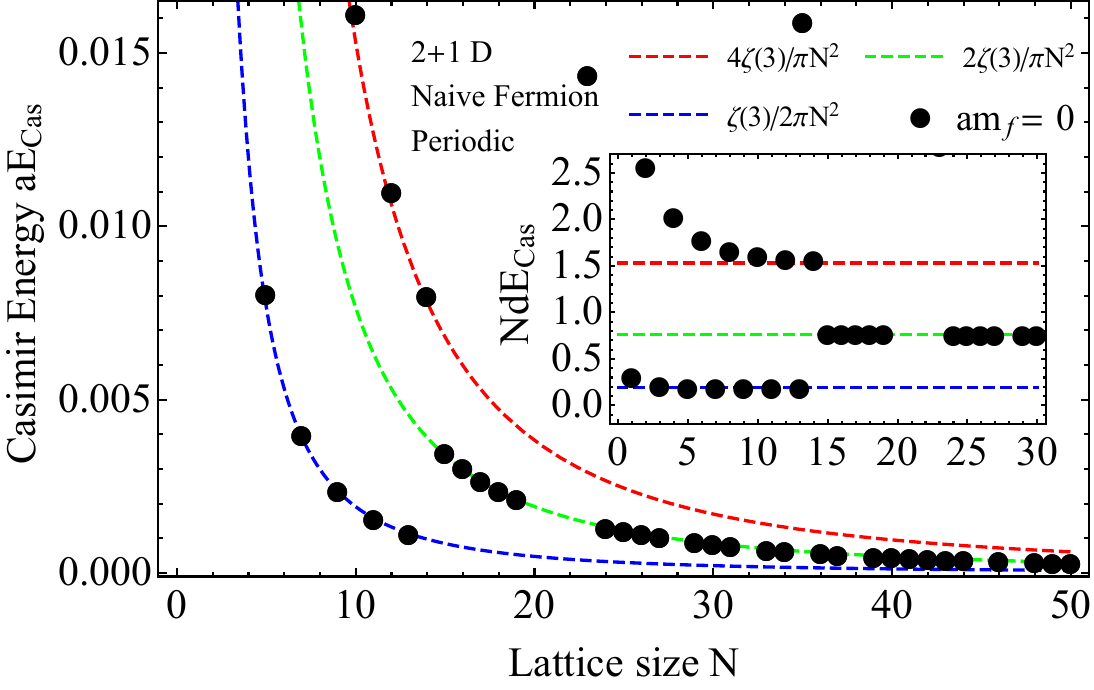}}} %
    \subfloat\footnotesize{(d)}{{\includegraphics[width=7.4cm]{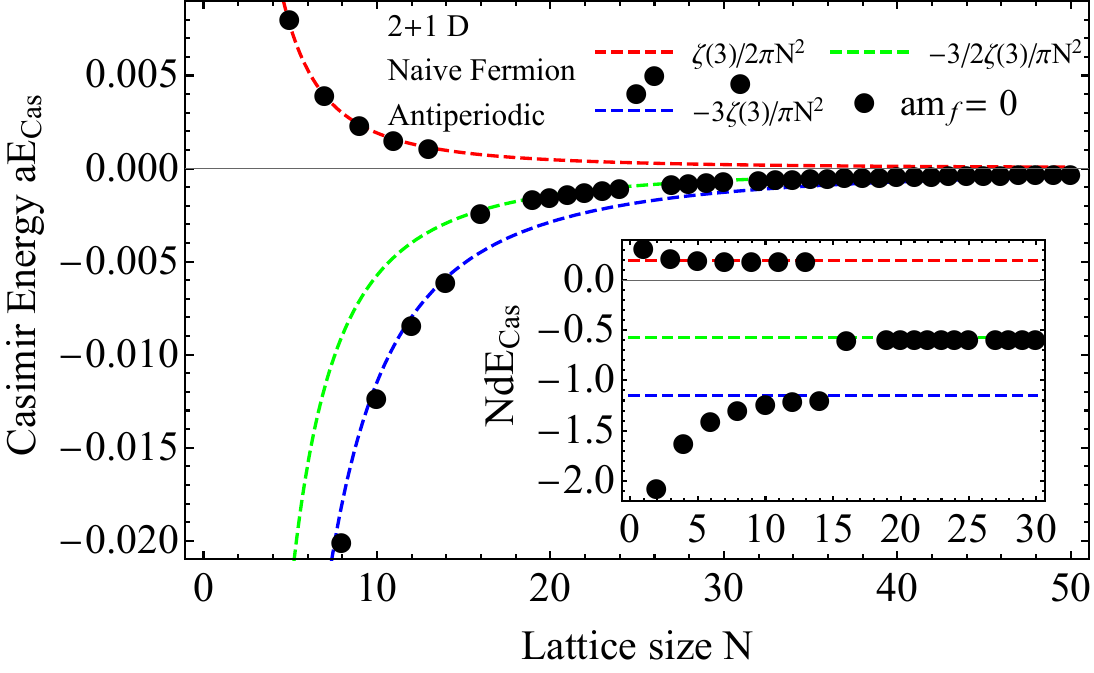}}}
 \caption{\footnotesize{Convergence of the Casimir energy expressions for massless Naive fermion to the continuum expression after extrapolation is demonstrated by series acceleration methods. [(a),(b)] and [(c),(d)] represent the periodic and anti-periodic boundary conditions for Naive fermion in ($1+1$)- and ($2+1$)-dimension respectively.}}%
\label{extrapolation}
\end{figure}

Authors in Ref. \cite{Ishikawa:2020ezm} claim that one cannot determine the original Dirac fermionic Casimir effect expressions from the zero lattice spacing $a\rightarrow0$ limit of the Naive fermion results. The Casimir energy calculations for the Naive fermion using periodic and anti-periodic boundary conditions are done with the Abel-Plana formulae in finite range (\ref{int}, \ref{nonint}) in ($1+1$)-dimensions, and numerically in higher dimensions gave us different continuum expressions for odd and even lattice sizes ($N$) in (\ref{naivep}, \ref{naivea}). As a result, the continuum $a\rightarrow0$ limit of the two expressions, for odd and even lattice sizes ($N$), give different limiting values for the Casimir energy as calculated in (\ref{naivecont}). We realized that from these expressions, one must not treat the Casimir energy for odd and even $N$ as different cases and express it by a single expression, rapidly oscillating between the values for odd and even $N$ as shown in Fig. \ref{osci}. This oscillatory expression on lattice after extrapolation, in turn, converges to a single continuum expression at large $N$, as shown in Fig. \ref{extrapolation}. Moreover, this expression obtained here is precisely the same as the one obtained in the continuum (\ref{cont}), from the Wilson fermion in section (\ref{WilsonPAP}) and Overlap fermions in section (\ref{mdw}) for both the periodic and antiperiodic boundary conditions.

Using the above-mentioned series acceleration techniques, we have shown that the oscillating Naive Casimir energy converges to a single expression in the continuum limit for sufficiently large $N$ and does not violate the universality of lattice fermions in the continuum limit. Thus, the claim in Ref. \cite{Ishikawa:2020ezm} has been shown to be incorrect, as it is possible to derive the Dirac fermionic Casimir energy using the Naive fermion results, irrespective of the boundary condition, just as we did for Wilson and Overlap fermion in this thesis. In Fig. \ref{extrapolation}, one can see that the expression for Casimir energy of Naive fermion in (\ref{naivep}, \ref{naivea}) converges to a single expression after extrapolation. This ensures that the universality of lattice fermions is not violated. The default number of terms it uses before extrapolation depends on the spacetime dimensions and is attributed by $\texttt{"NSumTerms"}$. In the extrapolation for $(2+1)$-dimensions, we observe from Fig. \ref{extrapolation}(c),(d) (that some lattice points are off the mark as its numerical computation involves calculating the integral of a  sum approximated by an expression of derivatives \ref{EMFormula} of the function $|\sin(2\pi n/N)|$, over a continuous variable. The function $|\sin(2\pi n/N)|$ is not differentiable at the lattice points and thus leads to this discrepancy. Nonetheless, it is sufficiently apparent that the oscillating series indeed converges to the same continuum expression as $N$ increases.
  
In section (\ref{slabbag}), we realized the MIT Bag boundary conditions for lattice fermions and calculated the corresponding Casimir energy. We observe that the Naive Casimir energy oscillation, as seen for the periodic and antiperiodic boundary conditions in the Naive fermion case, disappears. The expression obtained in the continuum limit for the MIT Bag boundary conditions with a doubling parameter of two in ($1+1$)-dimensions matches the continuum expression exactly. This leads us to conclude that the oscillation for odd and even lattice sizes is dependent on the boundary conditions and is not a consequence of Naive fermion doubling. 
\chapter{Applications to Topological Insulators}
 
Topological insulators have attracted immense interest as a new class of materials over the past decade. Their surface contains conducting states, with electrons only moving along the surface of the material while they act as an insulator in their interior. They exhibit time-reversal symmetry and topologically protected surface states. 

A Chern insulator is a new two-dimensional topological state of matter, a category of topological insulators with broken time-reversal symmetry. They realize the Integer Quantum Hall Effect (also called the Quantum Anomalous Hall Effect) on a two-dimensional lattice without an external applied magnetic field. The topological invariant of such a system is called the Chern number $C \in\mathbb{Z}$ and this gives the number of chiral edge states in the Brillouin zone. Chern insulators are characterized by their non-zero Chern number. So, when one has a non-trivial Chern insulator, this means it has edge states. One can transit from the trivial phase to the topological phase by changing parameters in the corresponding lattice model, such as the on-site or hopping energy.

In two dimensions, several models for topological insulators like the Kane-Mel\'e model for Quantum Spin Hall (QSH) insulators have been formulated in terms of massive lattice fermions. This model consists of the spin-orbit coupling term $H_{SO}$, which leads to the QSH effect. The topological insulators keep the time-reversal symmetry while breaking the spin symmetry. The spin-orbit coupling term splits the degeneracy between two valleys and acts as momentum dependent mass term in addition to the normal mass term $H_M$, called the staggered Zeeman term. This is similar to the degeneracy of valleys in Fig. \ref{Negmass2+1}, which indicates fermion doubling by the Neilsen-Ninomiya theorem. The combined contribution of these terms determines the topological phase structure of the insulator. If the magnitude of the contribution of the spin-orbit coupling is greater than the staggered Zeeman (mass) term, the system is dominated by the QSH effect. Whereas, if the magnitude of the staggered Zeeman term is greater than the spin-orbit coupling, the system behaves like a normal insulator \cite{Araki:2013qva}. Similar to the terms determining the topological phase structure of an insulator, the Wilson term acts as a momentum dependent mass, where the Wilson parameter `$r$' corresponds to the magnitude of the spin-orbit coupling. In the subsequent analysis, it is observed that the Wilson fermions with masses in the range $am_f>0$ or $am_f<-2$ behave normally and can model the normal insulators without any non-trivial topology. Whereas the Wilson fermions with masses in the range $-2<am_f<0$ model topological insulators with winding number $\nu=1$. \cite{tongqhe}

The study of the Casimir effect for these systems is also important for bulk modes in topological insulators, as it will also cause thermodynamic effects in the dynamics of the system. It is found that seemingly tiny Casimir interactions play a significant role in the mechanical operation of nano- and micro-electromechanical devices, unlike the predominantly classical world we live in. For example, very strong attractive forces offer resistance and cease the gear wheels to rotate freely, which affects the efficiency of such systems. Thus, an active area of research is looking for suitable materials and geometrical setups where Casimir effects and interactions will have a negligible impact or be repulsive. The Casimir interaction between usual dielectrics is attractive. Although, it has been observed that the Chern insulators demonstrate a repulsive Casimir effect. \cite{Rodriguez}  A system of oppositely signed parallel Chern Insulator plates ($C_1C_2<0$) realizes a repulsive Casimir effect at long distances in vacuum, which can be tuned to attraction by simply turning over plates. The recent discoveries of materials acting as Chern insulators like the Cr-doped $\text{(Bi, Sb)}_2 \text{Te}_3$ \cite{chang} with quantized Hall conductivity adds to the exciting possibility of overcoming theoretical bounds on Casimir repulsion using Chern insulators.
\section{For Negative mass Wilson fermion }
It can be seen that the dispersion relations such as $E=\pm\sqrt{p^2+m^2}$ for a Dirac fermion are not affected by the sign of mass term in Lagrangian. Although, a negative fermion mass $am_f$ in the Wilson fermion dispersion relation (\ref{Wilsonrelation}) on lattice shifts the Wilson term negatively and affects the Lagrangian of the system. In topological insulators,  parameters like the fermion mass $am_f$ and the Wilson parameter $r$ are related to intrinsic properties of the material like the spin-orbit interaction strength and the band-structure of the material in absence of spin-orbit interaction. \cite{Araki:2013qva}
In this section, we shall study the Casimir energy for negative mass Wilson fermions from the dispersion relation varying with the fermion mass plotted in Fig. \ref{NegmassDispersionRelations}. We study its properties by  distributing the negative fermion masses in ranges:
\begin{enumerate}[(I)]
    \item $-1<am_f\leq0$ : As the fermion mass $am_f$ decreases below zero for Wilson fermion, we observe that the Casimir energy $aE_{\text{Cas}}$ behaves similar to Wilson fermions with a positive mass and is exponentially suppressed as lattice size $N$ increases. This similarity in properties can also be seen by comparing the dispersion relations for the $am_f=0.5$ and $am_f=-0.5$.
\begin{figure}[t!]
\centering
    \subfloat\footnotesize{(a)}{{\includegraphics[width=7.4cm]{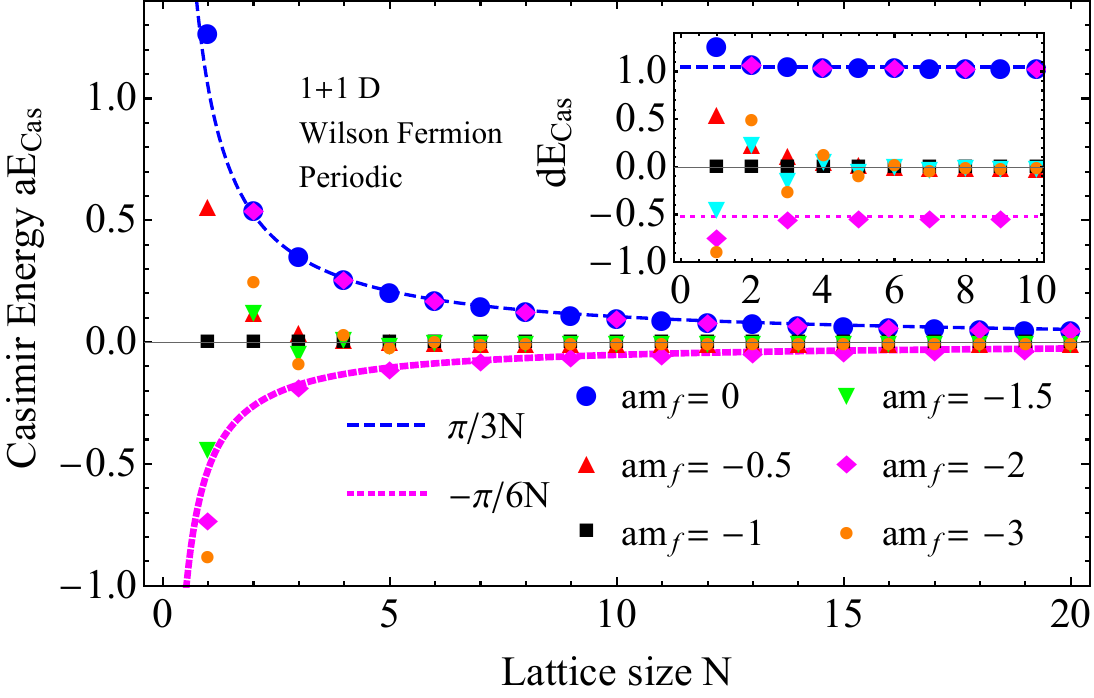} }} %
  \subfloat\footnotesize{(b)}{{\includegraphics[width=7.4cm ]{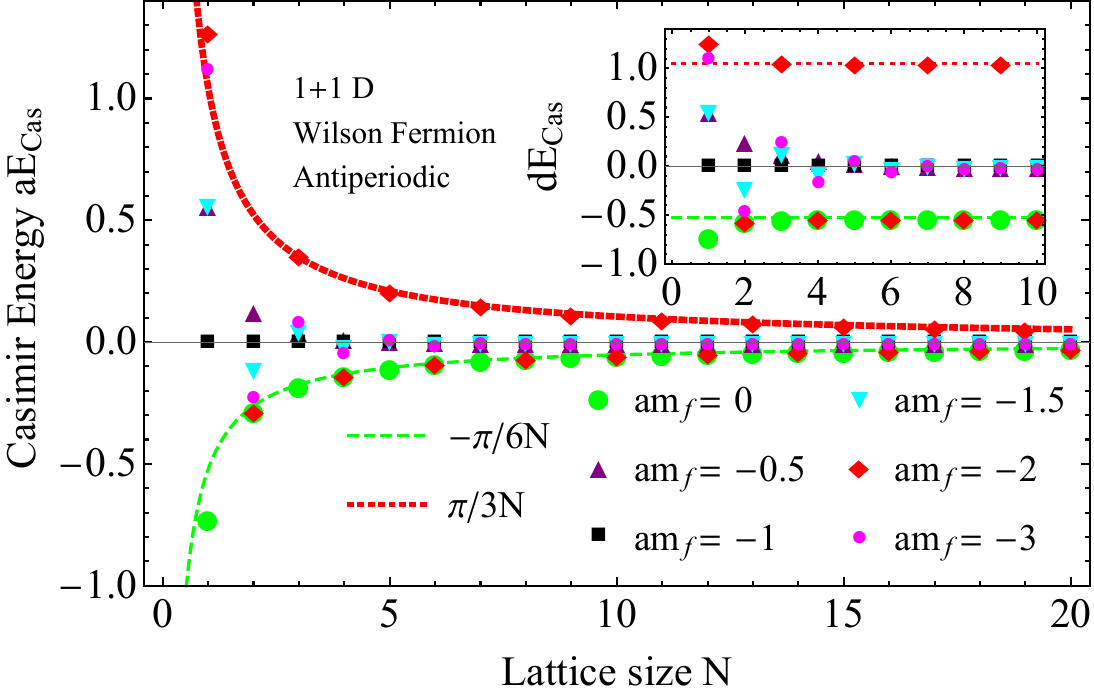} }}%
     \subfloat\footnotesize{(c)}{{\includegraphics[width=7.4cm]{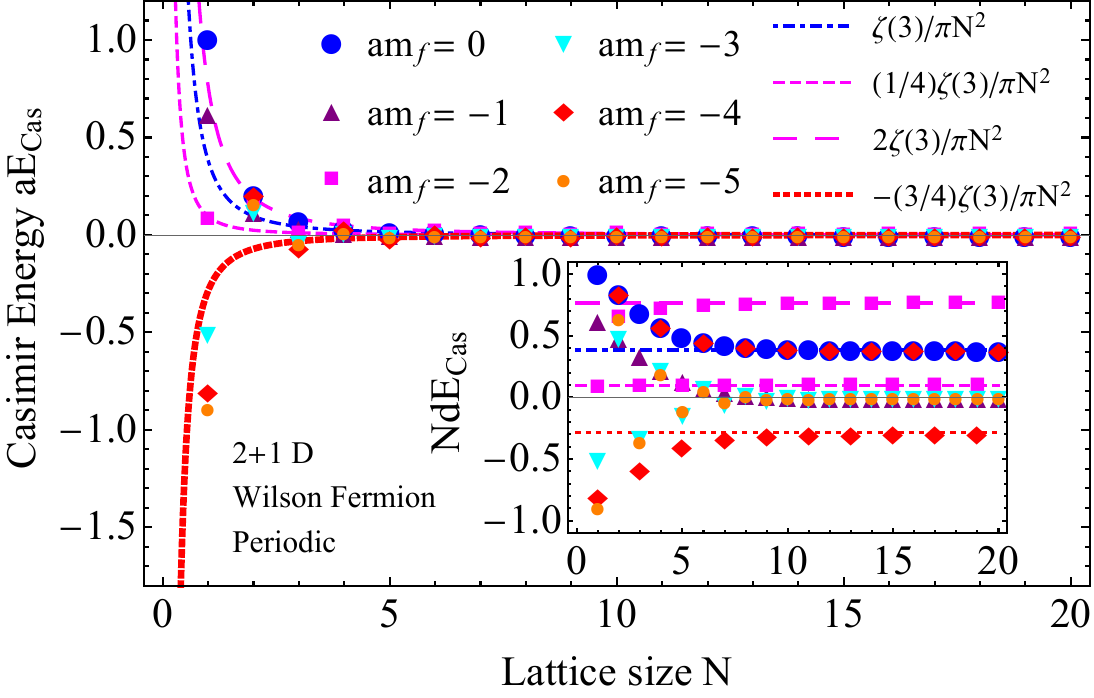}}} %
    \subfloat\footnotesize{(d)}{{\includegraphics[width=7.4cm]{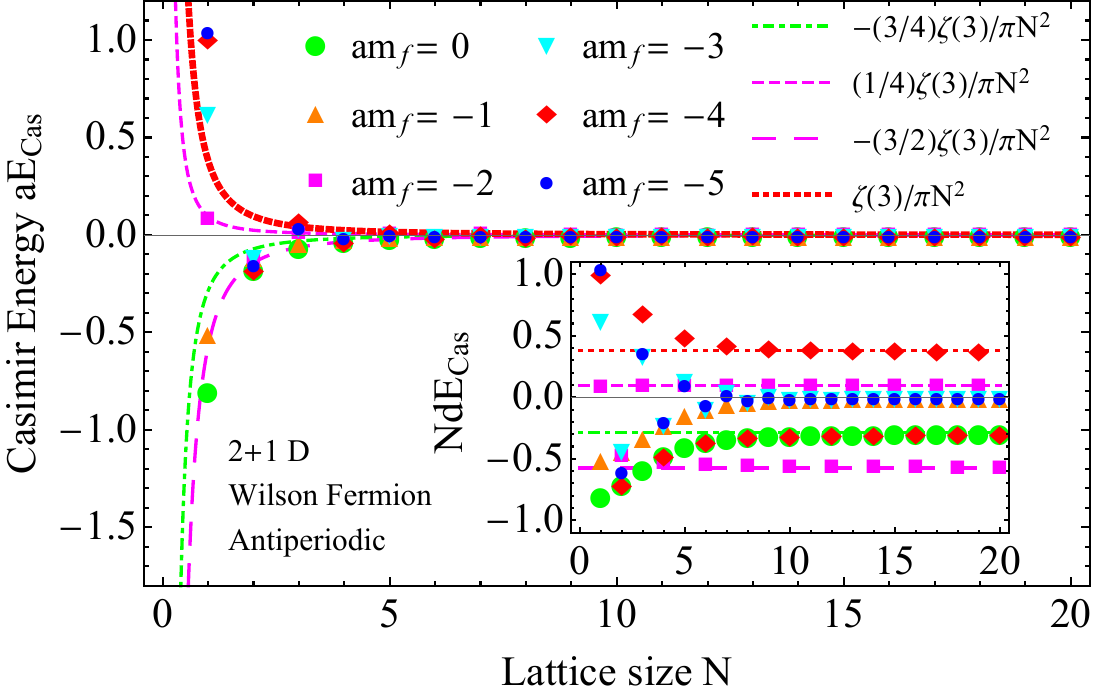}}}
 \caption{\footnotesize{Casimir energy for negative masses Wilson fermions with different mass ranges. [(a),(b)] and [(c),(d)] represent the periodic and anti-periodic boundary conditions for Wilson fermion in ($1+1$)- and ($2+1$)-dimension respectively.}}%
\label{Negmassplots}
\end{figure}
   \item $am_f=-1$ : This is an interesting parameter, where no Casimir effect takes place in $(1+1)$-dimensions. From the corresponding Wilson fermion Dirac operator (\ref{Wilsonoperator}) the dispersion relation is found to be constant, and independent of the momentum component. 
    \begin{equation}
    \label{nocasimir}
        a^2E^2_{\text{1+1D}}(ap) = 1
    \end{equation}
  \begin{figure}
     \centering
     \includegraphics[width=10cm]{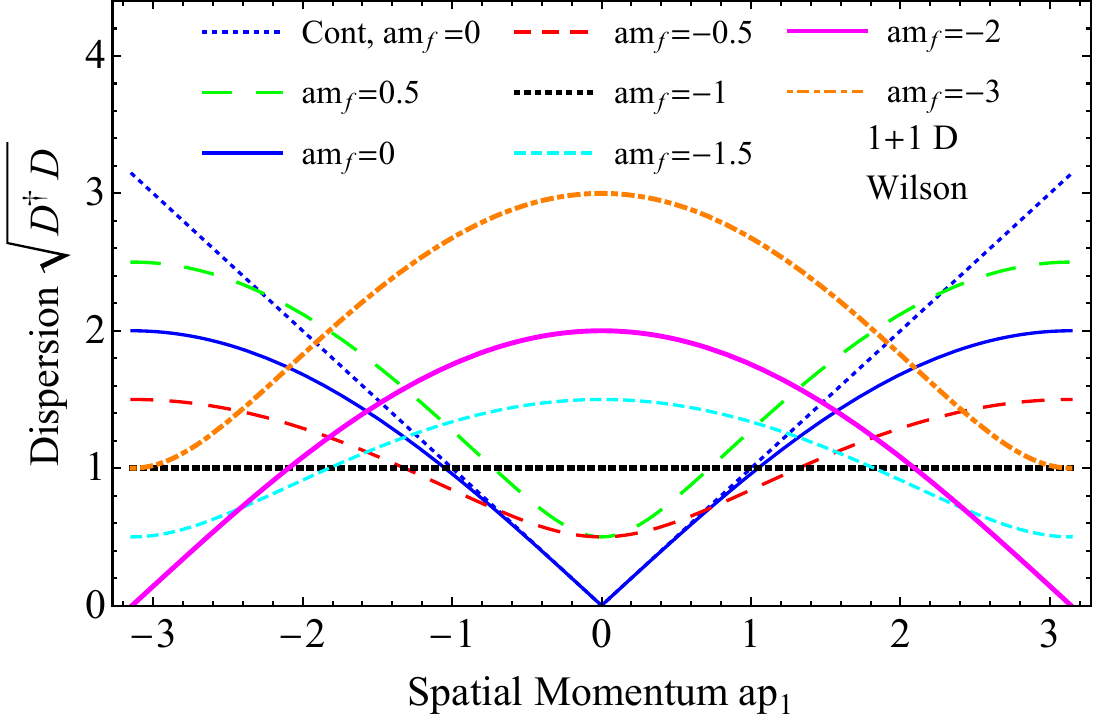}
     \caption{\footnotesize{Dispersion relations for different values of negative masses for Wilson fermions in ($1+1$)- dimension.}}
     \label{NegmassDispersionRelations}
 \end{figure}
    Thus, in Fig. \ref{NegmassDispersionRelations} the $am_f=-1$ dispersion relation is represented by a ``flat band". As discussed in section (\ref{photons}), the difference between zero-point energies in finite and infinite lattice sizes leads to the Casimir effect. As the dispersion relation is constant, the Casimir energy for a flat band is zero ($aE^{\text{1+1D}}_{\text{Cas}}=0$). Thus, no Casimir effect occurs in this case irrespective of the boundary conditions.
    \item $-2<am_f<-1$ : It can be seen from Fig. \ref{Negmassplots} that the Casimir energy exponentially decays and oscillates between the odd and even $N$. For $am_f$, it is seen that the non-zero modes around $ap_1=\pi$ dominate this oscillation, relatively suppressing the contribution from modes around $ap_1=0$. 
    \item $am_f=-2$ : One finds that the Casimir energy oscillations for even and odd lattice sizes are retained, but the parameter $dE_{\text{Cas}}$ is not suppressed in the large lattice size limit and approaches a constant value. This is because, at $ ap_1=\pi$, as seen in Fig. \ref{NegmassDispersionRelations}, the dispersion relation has an ultraviolet zero mode, and massless degrees of freedom dominate the Casimir energy. The Wilson Dirac operator for $am_f=-2$ is:
    \begin{equation}
    \label{wilson-2}
        aD_{\text{W}}^{\text{1+1D}} = i\gamma_1 \sin{ap_1} + r(-1-\cos{ap_1})
    \end{equation}
    The method used in Appendix B is employed to obtain exact expressions for Casimir energies from the Dirac operator (\ref{wilson-2}) in Ref. \cite{Ishikawa:2020icy}. An interesting pattern is observed in the expressions:
\begin{align}
 aE^{\text{1+1D,P,W}}_{\text{Cas}} &=
  \begin{dcases}
  \frac{4N}{\pi}-2\csc(\frac{\pi}{2N})\;\;\;\;&\text{if }N=\text{odd} \\  \frac{4N}{\pi}-2\cot(\frac{\pi}{2N})\;\;\;\;  &\text{if }N=\text{even}\label{2}
  \end{dcases}
  \end{align}
 \begin{align}
 aE^{\text{1+1D,AP,W}}_{\text{Cas}} &=
\begin{dcases}
\frac{4N}{\pi}-2\cot(\frac{\pi}{2N})\;\;\;\;&\text{if }N=\text{odd} \\ \frac{4N}{\pi}-2\csc(\frac{\pi}{2N})\;\;\;\;&\text{if }N=\text{even}\label{4}
\end{dcases}
 \end{align}
  Note that the expressions $(\ref{2}\;;\;N= \text{even})$ and $(\ref{4}\;;\;N= \text{odd})$ are the same as expression for massless Wilson fermion ($am_f=0$) with periodic boundary conditions in (\ref{wilsonperi}) and expressions $(\ref{2}\;;\;N= \text{odd})$ and $(\ref{4} \;;\;N= \text{even})$ are same as (\ref{wilsonantiperi}) with antiperiodic boundary conditions. Thus, for even lattice sizes, the Casimir energies are equivalent to each other for $am_f =0$ and $am_f =-2$. On the other hand, for odd lattice sizes, the Casimir energies with periodic (anti-periodic) boundary conditions at $am_f=-2$ are equivalent with the anti-periodic (periodic) boundary conditions at $am_f=0$. When calculated  for negative Wilson fermions, using the MIT Bag boundary conditions, the Casimir energy for cases $am_f=0$ and $am_f=-2$ are completely equal. 
   \begin{figure}[t!]
\centering
    \subfloat(a){{\includegraphics[width=5.5cm]{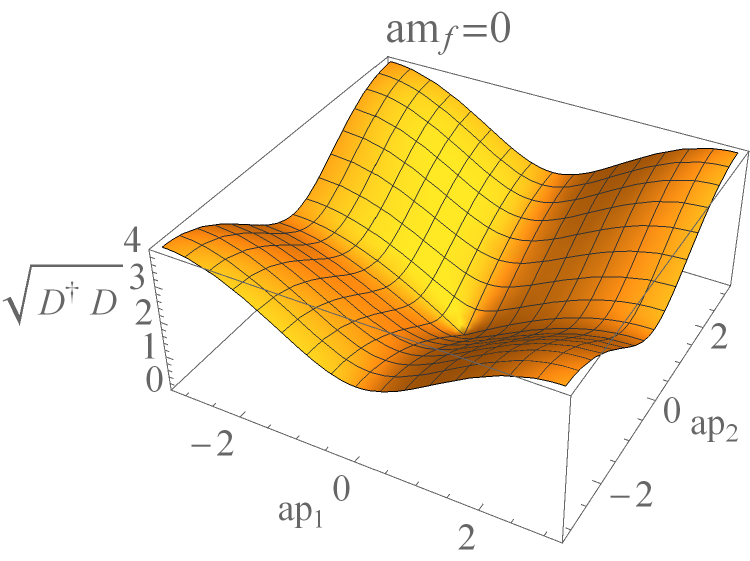} }} %
    \qquad
    \subfloat(b){{\includegraphics[width=5.5cm ]{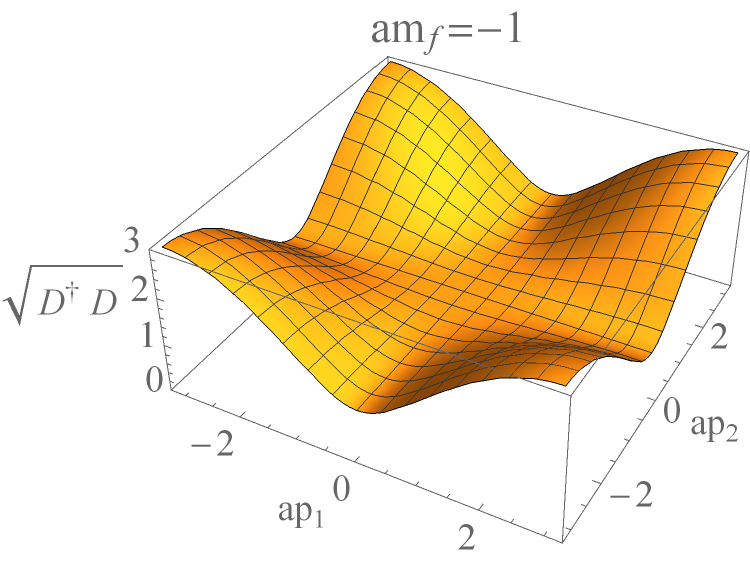} }}%
     \subfloat(c){{\includegraphics[width=5.5cm]{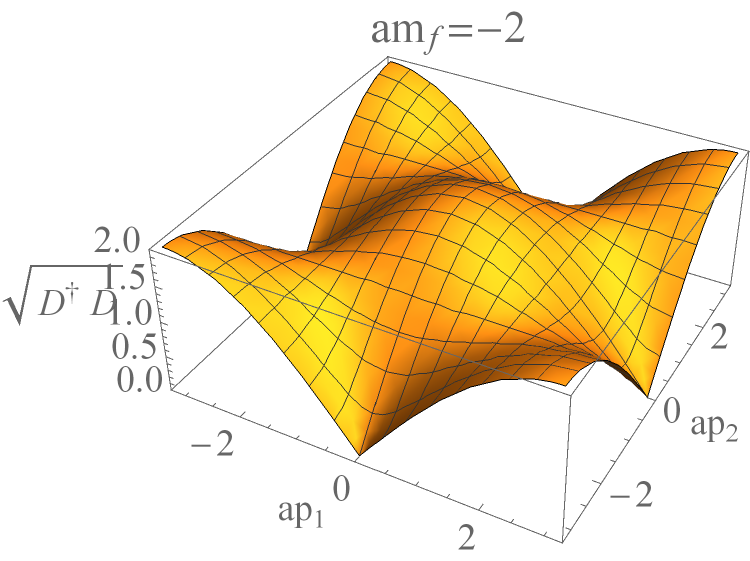}}} %
    \qquad
    \subfloat(d){{\includegraphics[width=5.5cm]{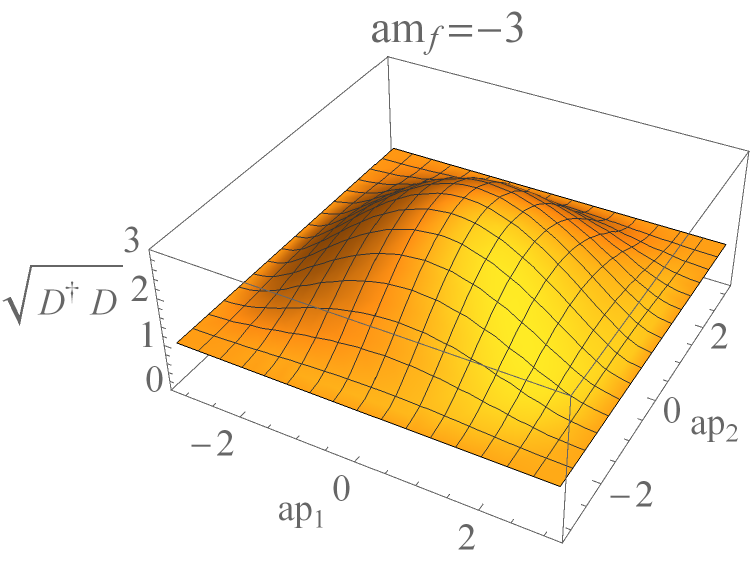}}}
    \subfloat(e){{\includegraphics[width=5.5cm]{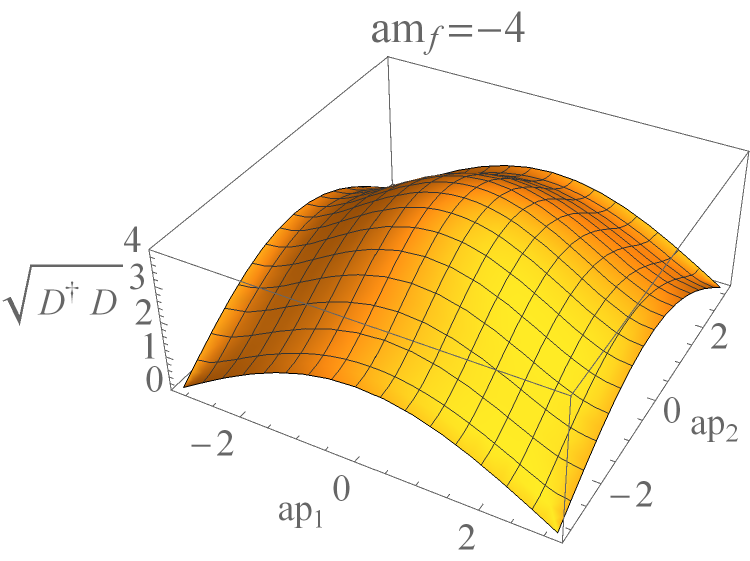}}} %
    \qquad
    \subfloat(f){{\includegraphics[width=5.5cm]{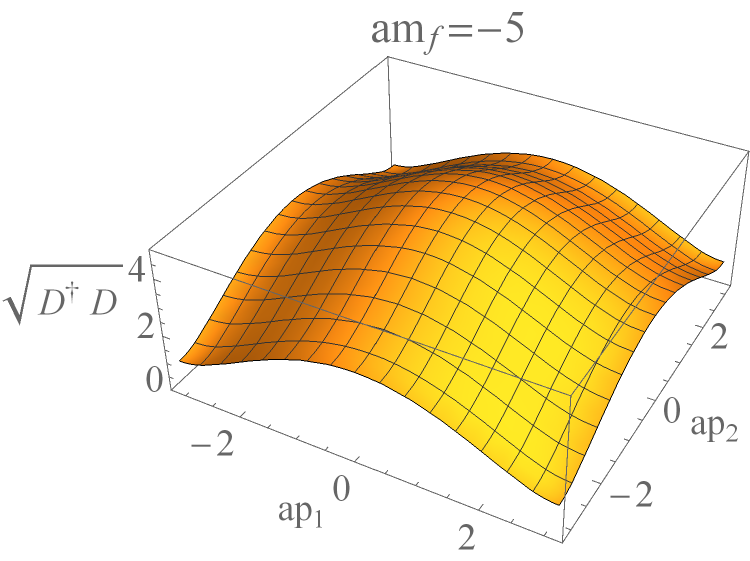}}}%
    \caption{\footnotesize{Dispersion relations for different values of negative masses for Wilson fermions in ($2+1$)- dimension. The temporal direction is not latticized.} }%
    \label{Negmass2+1}
\end{figure}
  \item $am_f<-2$ : The Casimir energy behaves similar to the case $-2<am_f<-1$ where it oscillates between odd and even lattice sizes and $dE_{\text{Cas}}$ is exponentially suppressed. 
\end{enumerate}
 The negative mass dependence of Casimir energy is now studied in $(2+1)$-dimensions where one spatial direction has either periodic, or antiperiodic boundary conditions and the other spatial direction has an open boundary. We study the behaviour of negative mass Wilson fermions in the following range of masses:
\begin{enumerate}[(I)]
    \item $-2<am_f\leq0$ : It is observed that at $am_f=0$, the massless Casimir energy is not suppressed. In Fig. \ref{Negmass2+1}(a), an infrared zero-mode present at $(ap_1,ap_2)\equiv(0,0)$ is responsible for $1/N^2$, inverse squared dependence of the Casimir energy. For lower masses, the massive degrees of freedom are dominant and thus, $NdE_{\text{Cas}}$ is exponentially suppressed. From the dispersion relation plotted in Fig. \ref{Negmass2+1}(b), no zero mode with energy greater than 1 is present for all modes in this range.
    \item $am_f = -2$ : One observes an oscillatory behaviour of Casimir energy for odd and even $N$. The massless degrees of freedom dominate the Casimir energy and thus, $NdE_{\text{Cas}}$ is not suppressed. In Fig. \ref{Negmass2+1}(c), two zero modes present at $(ap_1,ap_2)\equiv(0,\pi)$ and $(\pi,0)$ in our Brillouin zone contribute to the Casimir effect.
\item $-4<am_f<-2$ : Here, the Casimir energy oscillates between odd and even $N$ and the coefficient $NdE_{\text{Cas}}$ is suppressed exponentially. Although we do not observe any zero modes in Fig. \ref{Negmass2+1}(d), the lowest modes in the ultraviolet momentum region contribute to this Casimir energy oscillation with odd and even $N$.
\item $am_f=-4$ : This case is similar to the $am_f=-2$ case where we observed an oscillatory behaviour with odd or even $N$ and $NdE_{\text{Cas}}$ not suppressed in large $N$. Also, from Fig. \ref{Negmass2+1}(e) we have only one zero-mode at $(ap_1,ap_2)\equiv(\pi,\pi)$ in the Brillouin zore, contributing to the Casimir energy.
\item $am_f<-4$ : A similar behaviour to the $-4<am_f<-2$ case is observed, where Casimir energy is oscillatory with odd or even $N$ and $NdE_{\text{Cas}}$ is suppressed exponentially. Although there is no zero mode present, from Fig. \ref{Negmass2+1}(f), we see that the dispersion relation corresponding to $am_f=-5$ is similar to the $am_f=-4$ case with all modes being greater than 1.
\end{enumerate}

\section{For Overlap fermion with M\"{o}bius domain wall kernel }
\label{mdw}
\begin{figure}
    \centering
    \includegraphics[width=10cm]{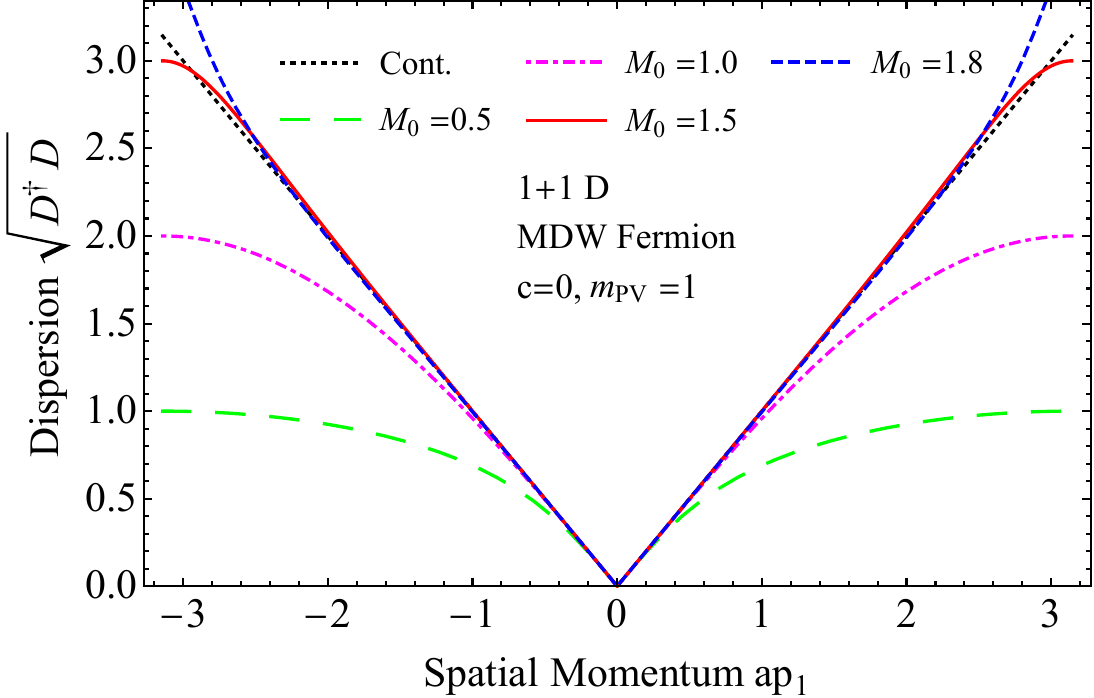}
    \caption{\footnotesize{Dispersion relations for different heights of the domain wall for Overlap fermions with a MDW kernel in ($1+1$)-dimension.}}
    \label{mdwdispersion}
\end{figure}
In the conventional domain wall formulation, the \textit{bulk} fermion is defined in $(D+1)$-dimensional Euclidean space, which also becomes the kernel operator, projected into the chiral \textit{surface} fermion in $D$-dimensional space. 
The extra dimension is considered infinite length for simplicity, making the domain wall fermion equivalent to the overlap fermion. \cite{Kaplan}

\begin{figure}[t!]
\centering
    \subfloat\footnotesize{(a)}{{\includegraphics[width=7.4cm]{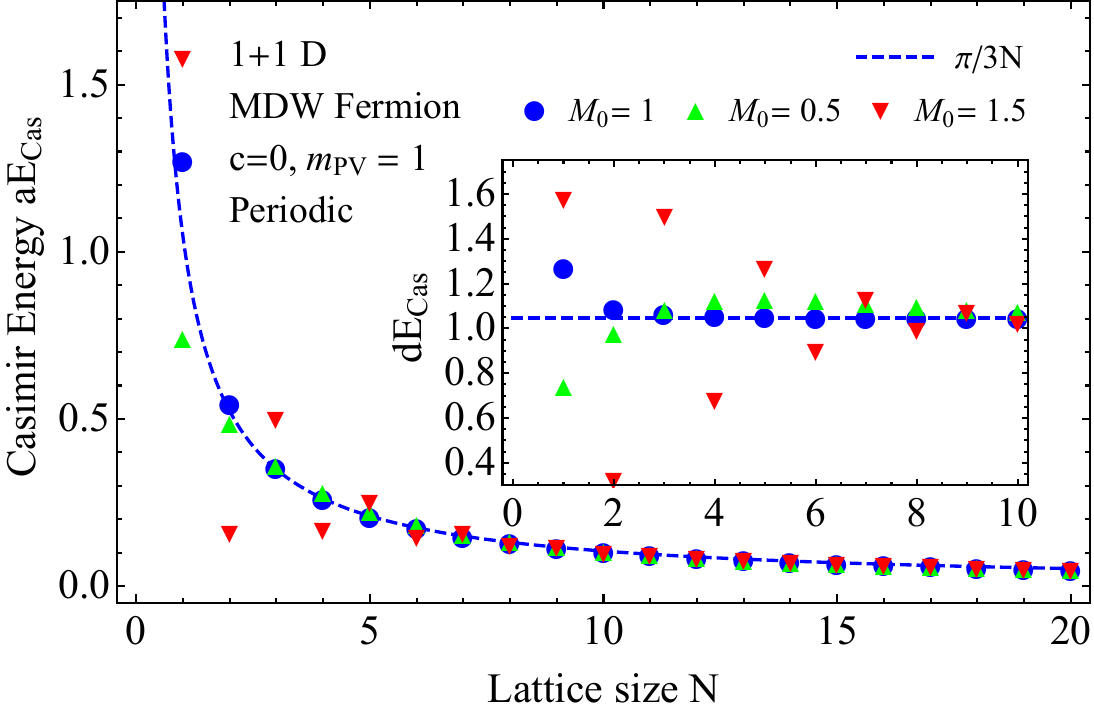} }} %
    \subfloat\footnotesize{(b)}{{\includegraphics[width=7.4cm ]{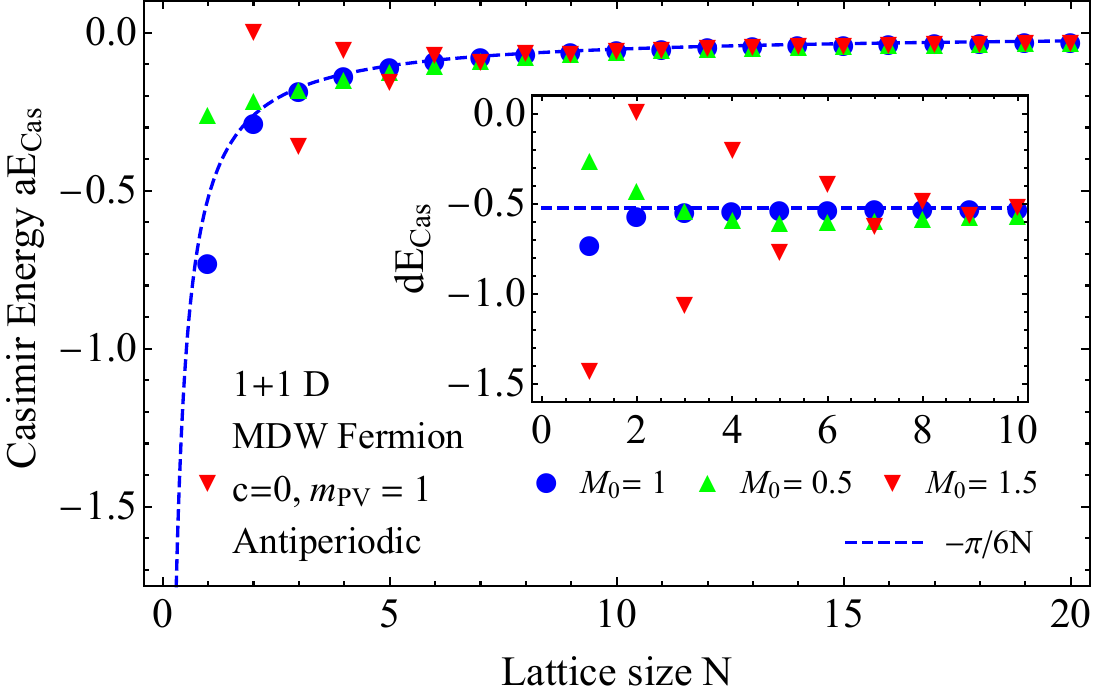} }}%
     \subfloat\footnotesize{(c)}{{\includegraphics[width=7.4cm]{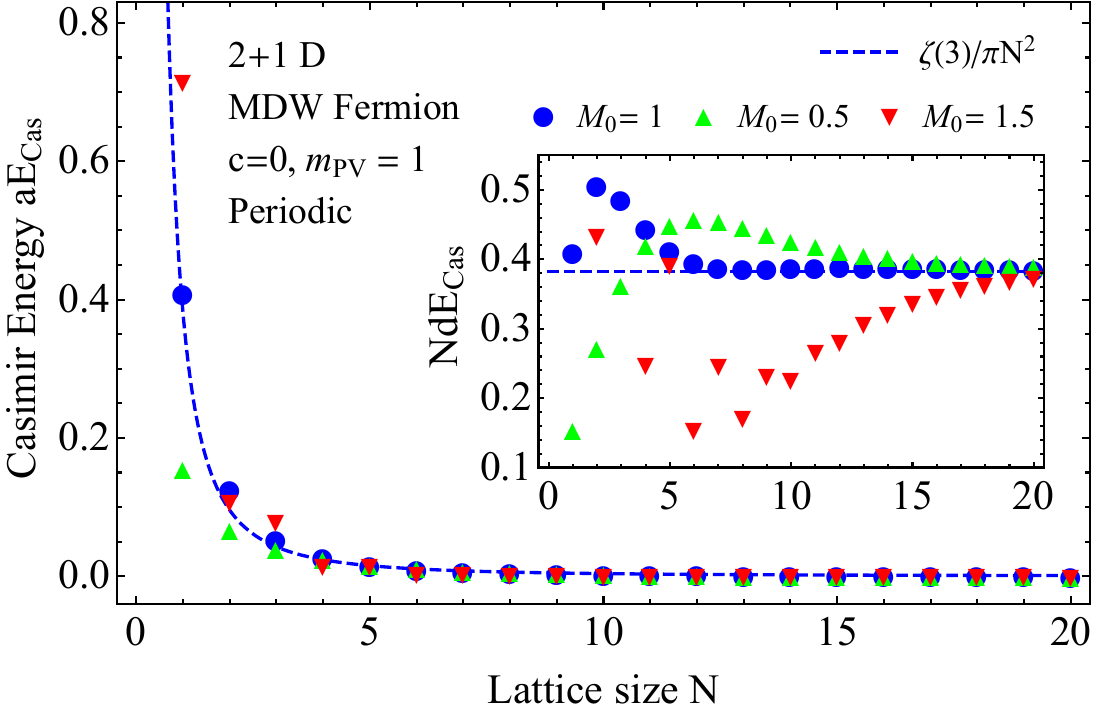}}} %
    \subfloat\footnotesize{(d)}{{\includegraphics[width=7.4cm]{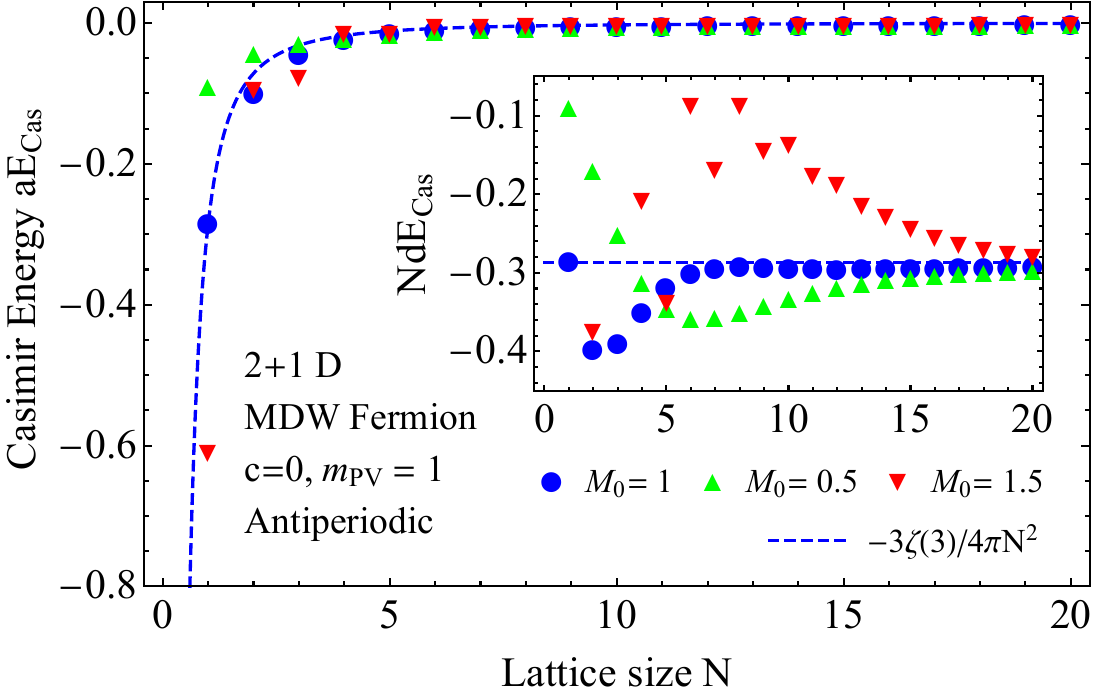}}}
    \subfloat\footnotesize{(e)}{{\includegraphics[width=7.4cm]{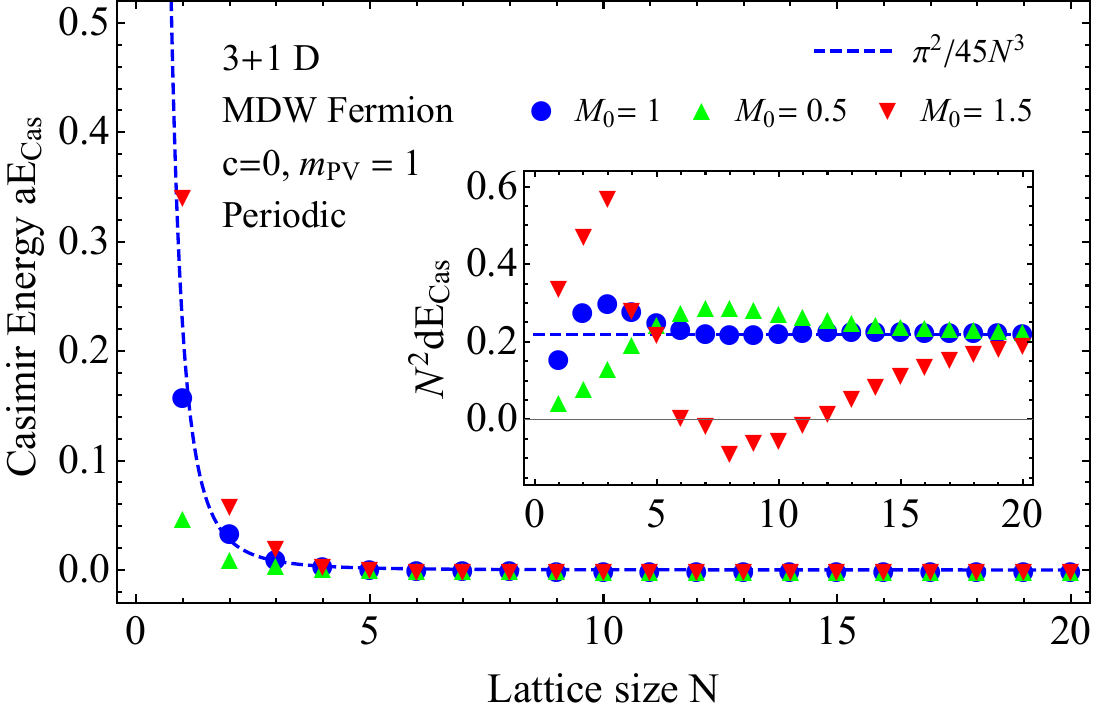}}} %
    \subfloat\footnotesize{(f)}{{\includegraphics[width=7.4cm]{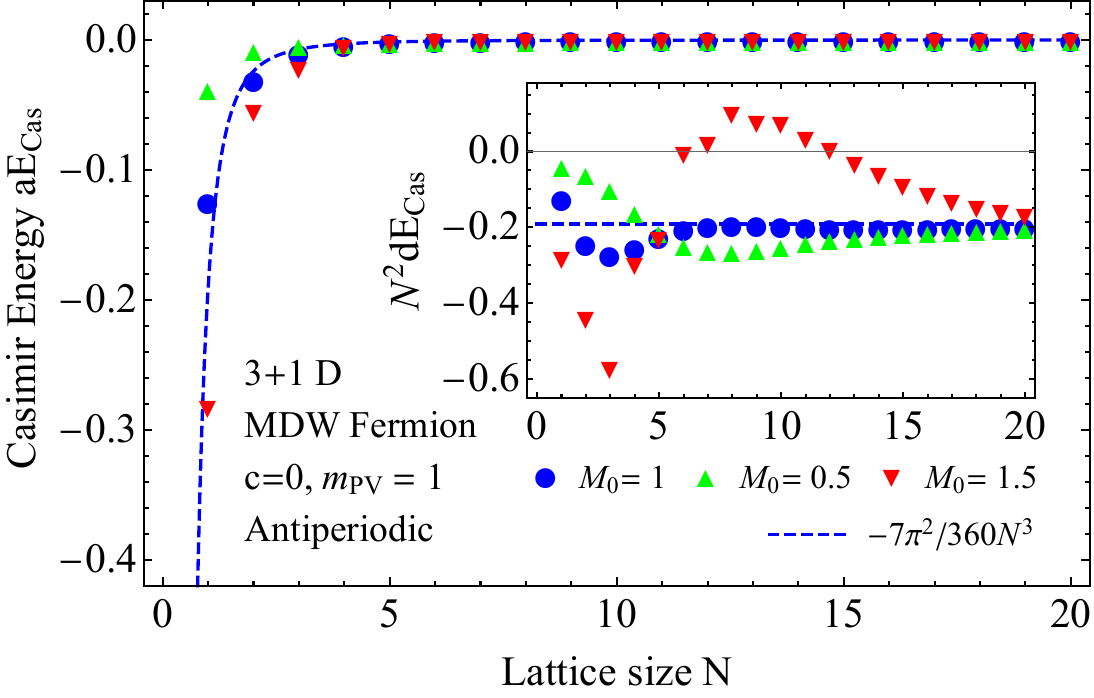}}}%
    \caption{\footnotesize{The Casimir energy for Overlap fermion with MDW kernel for different domain wall heights using perodic and anti-periodic boundary conditions. [(a),(b)], [(c),(d)] and [(e),(f)] represent the Casimir energy in ($1+1$)-, ($2+1$)-, and ($3+1$)-dimension respectively.}}%
\label{mdwperiantiperi}
\end{figure}

 We have already numerically calculated the Casimir energy for the Overlap fermion using bag (B) boundary conditions in section (\ref{overlapmitbag}). Likewise, we perform suitable substitutions for periodic and antiperiodic boundary conditions. The domain wall height $M_0$, introduced in section (\ref{Overlap}), can be interpreted as the negative mass of the bulk fermions. The dependence of $M_0$ on the Casimir energy for the Overlap fermion with MDW kernel operator is studied by varying it and keeping the parameters $c=0$, $m_{\text{PV}}=1.0$ fixed in the dispersion relation. \cite{Ishikawa:2020icy} The results obtained for $(1+1)$-, $ (2+1)$-, and $(3+1)$-dimensions are plotted in Fig. \ref{mdwperiantiperi}. Also, the corresponding dispersion relations in ($1+1$)-dimensional spacetime are plotted in Fig. \ref{mdwdispersion} for comparison.
 These plots will not only help us study the dependence of $M_0$ on the Casimir energy but also discuss the behaviour of numerical values obtained for different ranges of lattice points. 
 
 We saw in section (\ref{overlapmitbag}) that the overlap fermion case $M_0 = 1.0$, only in the ($1+1$)- dimensions is equivalent to the massless Wilson fermion ($am_f = 0$) case, as verified from the equal dispersion relations for both these cases. However, despite this equivalence not being present in higher dimensions, the Casimir energy for overlap fermions with $M_0=1.0$ on lattice agrees very well with the continuum expression in higher dimensions, especially for larger $N$. The infrared part of the dispersion relation contributes to the Casimir energy for large $N$. It is interesting to note that the dispersion relation for $M_0>1.0$ in Fig. \ref{mdwdispersion} agrees well with the continuum linear dispersion relation. For these cases ($M_0>1.0$), the Casimir energy oscillates between odd and even lattice sizes. The domain wall fermions are such that the doubling ultraviolet momentum modes with a heavy mass still exist, called the massive doublers. 
Ref. \cite{Ishikawa:2020icy} suggests that such massive doublers lead to the oscillation of Casimir energy for odd and even lattice sizes, unlike the continuum behaviour. Similar oscillation is also seen for overlap fermions with domain-wall kernel in free fermions at finite temperature on the Euclidean lattice, where the temporal direction is subjected to antiperiodic boundary conditions \cite{Gavai1, Gavai2}.

No such oscillation takes place for overlap fermions with domain wall height $0<M_0\leq1$ due to the absence of massive doublers. For small enough $M_0\gtrsim0$, we clearly observe that the Casimir energy is suppressed for small lattice size $N$ and enhanced for larger $N$.

Thus, the negative mass Wilson fermions behave like the bulk, whereas the projected domain wall fermions correlate to the surface fermions of the topological insulator. Thus, the Casimir effect for topological insulators can be studied using these lattice fermionic formulations. These observations can also be experimentally verified, subject to the availability of thin films for small lattice sizes. As mentioned in the introduction, observing a repulsive Casimir effect for Chern insulators has immense possibility to be applied in nano and micro-mechanical devices.
 
\chapter{Conclusion and future prospects}
\section{Conclusion}

In this thesis, we first studied the causes and effects of the Casimir effect for photons and fermionic fields in the continuum. The Fermionic Casimir effect includes modelling a perfectly conducting parallel plate slab with the MIT Bag Model and corresponding boundary conditions to calculate the Casimir energy. These MIT Bag boundary conditions were realized on a lattice in this thesis for the first time and used to calculate the Casimir effect for lattice fermions as per the formalism developed in Ref. \cite{Ishikawa:2020ezm}. Analytic expressions obtained for the Casimir energy of Naive (with doubling correction), Wilson fermions and numerical results for Overlap fermions with MDW kernel in ($1+1$)-dimensions and higher dimensional spacetime matched the continuum results exactly. Results in higher dimensions are also almost similar to the ($1+1$)-dimensions, but in the case of negative mass Wilson fermions, we studied certain special properties like (\ref{nocasimir}) specific to the ($1+1$)- dimensions. 

In accordance with the studies done in Ref. \cite{Ishikawa:2020ezm}, exact expressions for Casimir energy in continuum, for Naive and Wilson lattice fermions using periodic and antiperiodic boundary conditions were also studied, keeping in mind its applications to condensed matter systems. We have shown the author's claim that `the Casimir energy for Naive fermion in continuum limit cannot be used to obtain Casimir energy for the Dirac fermion' is incorrect. Using suitable series acceleration and extrapolation methods, we showed the convergence of the oscillating analytic expression between odd and even lattice sizes to the same expression as continuum. No such oscillation in $N$ for Naive fermion was observed for the MIT Bag Boundary conditions. Additionally, the expressions for Naive fermion agreed with the Casimir energy of Dirac fermions in continuum after a doubling correction. Thus, the Casimir energy oscillation is dependent on the boundary conditions and not related to the doubling. 
On the other hand, in the case of the Overlap fermions, it is claimed in literature Ref. \cite{Ishikawa:2020icy} that the oscillation in $N$ is due to the massive doublers for the periodic and antiperiodic boundary conditions.
\section{Future Prospects}
The aim is to study further and generalize several aspects of calculating and exploring the role of Casimir energy for lattice fermions and use this formalism to measure its consequences in various systems.
\subsection{Alternative interpretations of Casimir effect}
\begin{itemize}
    \item Robert Jaffe in Ref. \cite{Jaffe:2005vp} asserts that the Casimir effects can be formulated as inter-atomic interactions and computed without any reference to zero-point energies. He shows them to be relativistic, quantum forces between charges and currents. Thereby, it is argued that the existence of the Casimir effect cannot be considered proof of the existence of quantum vacuum fluctuations. Like any other observable effects in quantum electrodynamics, the Casimir force (per unit area $A$) between parallel plates vanishes as $\alpha$, the fine structure constant goes to zero. He states that the standard Casimir force (per unit area $A$) expression is independent of $\alpha$ because it actually corresponds to the $\alpha\rightarrow\infty$ limit. According to this alternate derivation, the Casimir force is simply the relativistic, retarded van der Waals force between the two parallel perfect conductor plates.
    From this, the condition obtained on $\alpha$ for the Casimir effect to take place is roughly 
    \begin{equation}
        \alpha \gg \frac{mc}{4\pi \hbar n d^2}
    \end{equation} 
    where, $n$ is the \textit{number density} and $m$ is the \textit{effective mass} of electrons \cite{Jaffe:2005vp}.  Using this approach will help us understand the dependence of the Casimir effect on the coupling constant, which is essential to generalize this phenomenon for different gauge fields. Comparing these results to the existing ones may also help us establish an equivalence between the two approaches and conclude something substantial.
\end{itemize}
\subsection{Thermodynamics of Casimir effect on lattice}
\begin{itemize}
    \item Several such approaches using the Lagrangian approach to define the Casimir effect in U(1) \cite{Chernodub:2016owp}, Yang-Mills gauge theories \cite{ChernodubYangMills}, and other non-perturbative aspects of Casimir effect \cite{ChernodubNonPert} make it easy to generalize these studies in the finite temperature regime. However, the approach presented in this thesis is based on the Hamiltonian approach. One might be able to do the same by discretizing the temporal dimension using antiperiodic boundary conditions for the fermions using the Matsubara formalism. Applications of the contribution of the Casimir effect for lattice fermions to Lattice QCD simulations would also require the temporal component of momentum to be discretized.  
    \item  
    In the finite-temperature regime of the Casimir effect, we can estimate the change of the specific heat caused by the Casimir effect by taking the temperature derivative of the potential. It would be interesting to study the influence of the Casimir effect on various experimental observables like the specific heat and magnetic susceptibility of materials. 
\item Studying the effects of results at finite temperature on the condensed matter systems using the free lattice fermions will also be interesting.
\end{itemize}
\subsection{Further applications to condensed matter systems}
\begin{itemize}
\item  Casimir effects for fermions living on 2D honeycomb lattices such as graphene nanoribbons, different structures of carbon nanotubes with the periodic boundary condition, 3D diamond lattices and topological insulators can be theoretically predicted, experimentally measured in micro-sized materials and applied to solving related problems in the industry. 
\item The next obvious step to the study for free fermions would be to study the Casimir effect by introducing an interaction between fermions. We know from literature that interacting negative mass Wilson fermions exhibit the spontaneous parity-broken `Aoki' phase. A similar `tilted antiferromagnetic (TAF)' phase is observed in the case of the transition from Normal insulators to the Quantum Spin Hall phase in Ref.\cite{Araki:2013qva}. It would be interesting to study the negative-mass Casimir effect and phase structure after switching on interaction and see if an analogy can be drawn.
\item  The topological insulator thin films with higher Chern numbers offer a promising route to control the Casimir repulsion. Exploring this effect by effectively modelling band structures of Chern insulators using Wilson fermions on the lattice would be interesting.
\end{itemize}

\appendix
 
\chapter{Mathematical tools \& Calculations}
\renewcommand{\thesection}{\Roman{section}} 
\section{Euler-Maclaurin Summation formulae}
\label{AEuler}The Euler–Maclaurin summation formula holds when the function $f(z)$ is analytic in the integration region
\begin{equation}
\label{EMFormula}
   \sum_{k=1}^{n-1}=\int_0^n f(k) dk-\frac{1}{2}[f(0)+f(n)]+\sum_{k=1}^\infty\frac{B_{2k}}{(2k)!}[f^{(2k-1)}(n)-f^{(2k-1)}(0)]
\end{equation}
It relates sums to associated integrals, proving to be a highly important result that is used in numerical practice to this day, and is implemented as the method \texttt{EulerMaclaurin} in Mathematica’s \texttt{NSum} routine. The formula reads:
\begin{equation}
\label{EMMathematica}
    \sum_{n=a+1}^{b}f(n)=\int_{a+\delta}^{b+\delta}f(n)dn-\sum_{k=0}^{l}\frac{(-1)^k}{(k+1)!}B_{k+1}(y)f^{(k)}(y)\Big|_{y=a+\delta}^{y=b+\delta}+\int_{a+\delta}^{b+\delta}\frac{(-1)^l}{(l+1)!}B_{l+1}(y)f^{(l+1)}(n)dn
\end{equation}
The sum is approximated by an integral with an arbitrary offset $\delta \in (0, 1]$. Higher-order corrections to the integral approximation are described by derivatives of the summand function at the boundaries of integration. The coefficients of these derivatives are formed by the periodic Bernoulli functions $B_k$.
\section{Abel Plana Formulae in Finite Range}
\label{AAbel}
Under the mentioned structure of cut-off function $f(k)$, one can show that the Euler-Maclaurin and the Abel-Plana formulae are equivalent through the approximation theorem by Weierstrass, by connecting the analytic constraints of Abel-Plana and the $f \in C^{(2)}[0,\infty]$ constraint of the Euler-Maclaurin formula \cite{math_casimir}. The Abel Plana formula is as follows:\\
Let $f : \mathbb{C} \rightarrow \mathbb{C} $ satisfy the following conditions \cite{math_casimir}:
\begin{enumerate}
    \item $F(z)$ is analytic for $\mathbb{R}(z) \geq 0$ (not necesssarily at infinity).
    \item $\lim_{|y|\to \infty}|F(x+iy)|e^{-2\pi|y|}=0$ uniformly in $x$ in every finite interval.
    \item $\int_{-\infty}^{\infty}|F(x+iy)-F(x-iy)|e^{-2\pi|y|}dy$ exists for every $x \geq 0$ and tends to zero for $x \rightarrow \infty$.
    \item $\int_{0}^{\infty}F(t)dt$ is convergent, and $\lim_{n \to \infty} F(n) = 0$. Then,
\end{enumerate}
\begin{equation}
    \sum_{n=1}^\infty F(n) -\int_{0}^{\infty}F(\kappa)d\kappa +\frac{1}{2}F(0) = i\int_0^\infty \frac{\left(F(it)-F(-it)\right)dt}{e^{2\pi t}-1} 
\end{equation}
For half-integers, the formula modifies as :
\begin{equation}
    \sum_{n=0}^\infty F\left(n+\frac{1}{2}\right) = \int_0^\infty F(\kappa)d\kappa - i\int_0^\infty \frac{\left(F(it)-F(-it)\right)dt}{e^{2\pi t}+1} 
\end{equation}
where $F(n)$ is as defined in (\ref{subs})
 \subsection{Abel-Plana Formulae in finite range}
As seen earlier, the momentum space of the lattice fermion is restricted to a Brillouin Zone. Thus, the Abel-Plana Formula must also be modified and obtained for a finite range, indefinite limits. The derivation of these formulas is based on the generalized Abel Plana formula. The results obtained for integer values is \cite{Saharian:2007ph} :
\begin{dmath}
\label{int}
    \sum_{n=\lceil a \rceil}^{\lfloor b \rfloor} F(n) -\int_a^b dxF(x) -\left(\frac{1}{2}F(a) + \frac{1}{2}F(b) \text{ if  } a,b\in\mathbb{Z}\right) = i\int_0^\infty dy\frac{F(a+iy)}{e^{2\pi(y-ia)}-1} - i\int_0^\infty dy\frac{F(a-iy)}{e^{2\pi(y+ia)}-1} - i\int_0^\infty dy\frac{F(b+iy)}{e^{2\pi(y-ib)}-1} + i\int_0^\infty dy\frac{F(b-iy)}{e^{2\pi(y+ib)}-1}
\end{dmath}
The result for the half-integer values is :
\begin{dmath}
\label{nonint}
    \sum_{n=\lceil a-\frac{1}{2} \rceil}^{\lfloor b-\frac{1}{2} \rfloor} F\left(n+\frac{1}{2}\right) -\int_a^b dxF(x) -\left(\frac{1}{2}F(a) + \frac{1}{2}F(b) \text{ if  } a-\frac{1}{2},b-\frac{1}{2}\in\mathbb{Z}\right) = -i\int_0^\infty dy\frac{F(a+iy)}{e^{2\pi(y-ia)}+1} + i\int_0^\infty dy\frac{F(a-iy)}{e^{2\pi(y+ia)}+1} + i\int_0^\infty dy\frac{F(b+iy)}{e^{2\pi(y-ib)}+1} - i\int_0^\infty dy\frac{F(b-iy)}{e^{2\pi(y+ib)}+1}
\end{dmath}
where $\lceil a \rceil$ and $\lfloor b \rfloor$ are the ceiling and the floor functions respectively.
\subsection{Trigonometric series}
The Taylor series for relevant trigonometric functions is:
\begin{align}
    \cot(x) &= \frac{1}{x}-\frac{x}{3}-\frac{x^3}{45}-\frac{2x^5}{945} -\dots-\frac{2^{2n}B_nx^{2n-1}}{(2n)!}-\dots\;\;\;0<|x|<\pi\label{cot}\\
    \csc(x) &= \frac{1}{x}+\frac{x}{6}+\frac{7x^3}{360}+\frac{31x^5}{15120} +\dots+\frac{2\left(2^{2n-1}-1\right)B_nx^{2n-1}}{(2n)!}+\dots\;\;\;0<|x|<\pi \label{csc}
\end{align}
where $B_n$ is the $n^{\text{th}}$ Bernoulli number.
 \section{Intermediate Derivations for Fermionic Casimir effect}
 \label{derx}
We first apply the massless boundary conditions (\ref{bound}) on the plate located at $x^1 = 0$. Note that the outward unit normal for this plate would be in the negative $x^1$ direction. Thus, we have:
 \begin{align}
 &\Rightarrow (1 - i \gamma^1)\psi^{(+)} = 0\\
 &\Rightarrow \Bigg[ \left( \begin{array}{cc}
\mathbbm{1} & 0 \\
0 & \mathbbm{1}
\end{array} \right) -
i\left( \begin{array}{cc}
0 & \sigma_1\\
-\sigma_1 & 0
\end{array} \right) \Bigg]
e^{-i\omega t}\left( \begin{array}{cc}
\varphi^{(+)} \\
\dfrac{-i\bm{\sigma\cdot\nabla}\varphi^{(+)}}{\omega +m}
\end{array} \right)\\
     &\Rightarrow \left( \begin{array}{cc}
\mathbbm{1} & -i\sigma_1\\
i\sigma_1 & \mathbbm{1}
\end{array} \right) \nonumber
e^{-i\omega t}\left( \begin{array}{cc}
\varphi^{(+)} \\
\dfrac{-i\bm{\sigma\cdot\nabla}\varphi^{(+)}}{\omega +m}
\end{array} \right)
    \end{align}
    We obtain the following equation from both the elements after calculating the matrix, since $\sigma_1^2 = \mathbbm{1}$. Then, substituting (\ref{varphi2}) at $x^1=0$ in (\ref{phiexp1}) leads us to the final expression.
\begin{align}
    &\Rightarrow \varphi^{(+)} - \dfrac{\sigma_{1}\bm{\sigma\cdot\nabla}\varphi^{(+)}}{\omega +m} =0 \label{phiexp1}
   \\
    &\Rightarrow \left[(\varphi_+^{(+)} e^{ik_1x^1} + \varphi_-^{(+)} e^{-ik_1x^1}) -  \dfrac{\sigma_{1} \bm{\sigma\cdot\nabla}}{\omega +m} ( \varphi_+^{(+)} e^{ik_1x^1} + \varphi_-^{(+)} e^{-ik_1x^1}) \right]\exp{ i\sum_{j=2}^D k_jx^j} = 0
    \\
    &\Rightarrow \varphi_+^{(+)}\left[1-i\frac{k_1 +\sigma_1  \bm{\sigma}_\parallel \cdot \bm{k}_\parallel}{m+\omega} \right] +
    \varphi_-^{(+)}\left[1+i\frac{k_1 -\sigma_1   \bm{\sigma}_\parallel \cdot \bm{k}_\parallel}{m+\omega} \right] = 0
    \\
    &\text{Note that here, $\bm{\sigma_{\parallel}} = (\sigma_2,...,\sigma_D)$ and $\bm{k_{\parallel}} = (k_2,...,k_D)$}.\\
     &\therefore \;\;\varphi_+^{(+)} =\left[ \frac{1+i\frac{k_1 -\sigma_1   \bm{\sigma}_\parallel \cdot \bm{k}_\parallel}{m+\omega}}{1-i\frac{k_1 +\sigma_1  \sigma_\parallel \cdot k_\parallel}{m+\omega}}\right]\varphi_{-}^{(+)} 
    \\
    &= -\frac{(m+\omega) + i(k_1 - \sigma_1  \bm{\sigma}_\parallel \cdot \bm{k}_\parallel)}{(m+\omega) - i(k_1 + \sigma_1  \bm{\sigma}_\parallel \cdot \bm{k}_\parallel)}\varphi_{-}^{(+)}
    \\
    &=  \frac{[(m+\omega) + i(k_1 - \sigma_1 \bm{\sigma}_\parallel \cdot \bm{k}_\parallel)][(m+\omega) - i(k_1 - \sigma_1 \bm{\sigma}_\parallel \cdot \bm{k}_\parallel)]}{[(m+\omega) - i(k_1 + \bm{\sigma}_\parallel \cdot \bm{k}_\parallel)][(m+\omega) + i(k_1 - \sigma_1 \bm{\sigma}_\parallel \cdot \bm{k}_\parallel)]} \varphi_{-}^{(+)}
    \\
    &= -\frac{(m+\omega)^2 + (k_1 - \sigma_1  \bm{\sigma}_\parallel \cdot \bm{k}_\parallel)^2}{(m+\omega)^2 -2i(m+\omega)k_1 -  ({k_1}^2 - {\sigma_1}^2 ( \bm{\sigma}_\parallel \cdot \bm{k}_\parallel)^2)}\varphi_{-}^{(+)}
    \\
    &= -\frac{m^2 + 2m\omega + \omega^2 + {k_1}^2 - 2\sigma_1 k_1  \bm{\sigma}_\parallel \cdot \bm{k}_\parallel + (\bm{\sigma}_\parallel \cdot \bm{k}_\parallel)^2 }{m^2 + 2m\omega + \omega^2 -2i(m+\omega)k_1 - {k_1}^2 + (\bm{\sigma}_\parallel \cdot \bm{k}_\parallel)^2)}\varphi_{-}^{(+)} \label{nirna}
    \\
    &= -\frac{2(m^2 + m\omega + {k_1}^2 - \sigma_1 k_1  \bm{\sigma}_\parallel \cdot \bm{k}_\parallel)}{2(m^2 + m\omega- ik_1(m+\omega))}\varphi_{-}^{(+)}
    \\
    &= -\frac{m(m+\omega) + {k_1}^2 - \sigma_1 k_1  \bm{\sigma}_\parallel \cdot \bm{k}_\parallel}{(m-ik_1)(m+\omega)}\varphi_{-}^{(+)}
\end{align}
 In (\ref{nirna}), we use of the energy-momentum dispersion relation given by $\omega^2 = \sum_{j=1}^{D}k_j^2 + m^2$ and the property of Pauli matrices $\{\sigma_j,\sigma_l\}=\sigma_j\sigma_l + \sigma_l\sigma_j = 2\delta_{lj}$.
 Similarly, we can as well solve for the relation between $\varphi_{+}^{(-)}$ and $\varphi_{-}^{(-)}$. Combining both the results we get:
 \begin{equation}
      \varphi_{+}^{(\pm)}= -\frac{m(m+\omega) + {k_1}^2 - \sigma_1 k_1  \bm{\sigma}_\parallel \cdot \bm{k}_\parallel}{(m\mp ik_1)(m+\omega)}\varphi_{-}^{(\pm)}
 \end{equation}
 Now, we similarly compute a relation between $ \varphi_{+}^{(\pm)}$ and $\varphi_{-}^{(\pm)}$ using the boundary condition at $x^1 = d$ with the corresponding outward unit normal along the positive $x^1$ direction, we get:
 \begin{align}
 &\Rightarrow (1 + i \gamma^1)\psi^{(+)} = 0\\
 &\Rightarrow \Bigg[ \left( \begin{array}{cc}
\mathbbm{1} & 0 \\
0 & \mathbbm{1}
\end{array} \right) +
i\left( \begin{array}{cc}
0 & \sigma_1\\
-\sigma_1 & 0
\end{array} \right) \Bigg] \nonumber
e^{-i\omega t}\left( \begin{array}{cc}
\varphi^{(+)} \\
\dfrac{-i\bm{\sigma\cdot\nabla}\varphi^{(+)}}{\omega +m}
\end{array} \right)\\
     &\Rightarrow \left( \begin{array}{cc}
\mathbbm{1} & i\sigma_1\\
-i\sigma_1 & \mathbbm{1} \nonumber
\end{array} \right) 
e^{-i\omega t}\left( \begin{array}{cc}
\varphi^{(+)} \\
\dfrac{-i\bm{\sigma\cdot\nabla}\varphi^{(+)}}{\omega +m}  \nonumber
\end{array} \right) \nonumber
\end{align}
    We obtain the following equation from both the elements after calculating the matrix, since $\sigma_1^2 = \mathbbm{1}$. Then, substituting (\ref{varphi2}) at $x^1=d$ in (\ref{phiexp2}) leads us to the final expression.
\begin{align}
    &\Rightarrow \varphi^{(+)} + \frac{\sigma_{1}\bm{\sigma\cdot\nabla}\varphi^{(+)}}{\omega +m} =0 \label{phiexp2}
   \\
    &\Rightarrow \left[(\varphi_+^{(+)} e^{ik_1x^1} + \varphi_-^{(+)} e^{-ik_1x^1}) + \frac{\sigma_{1} \bm{\sigma\cdot\nabla}}{\omega +m} ( \varphi_+^{(+)} e^{ik_1x^1} + \varphi_-^{(+)} e^{-ik_1x^1}) \right]\exp{ i\sum_{j=2}^D k_jx^j} = 0
    \\
    &\Rightarrow \left[1+i\left(\frac{k_1 +\sigma_1 \bm{\sigma}_\parallel \cdot \bm{k}_\parallel}{m+\omega}\right)  \right] \varphi_+^{(+)} e^{ik_1 d}+
    \left[1-i\left(\frac{k_1 -\sigma_1  \bm{\sigma}_\parallel \cdot \bm{k}_\parallel}{m+\omega}\right)  \right]\varphi_-^{(+)}e^{-ik_1 d} = 0
    \\
     &\therefore \;\;\varphi_+^{(+)} = \left[\frac{1-i(\frac{k_1 -\sigma_1 \bm{\sigma}_\parallel \cdot \bm{k}_\parallel}{m+\omega})}{1+i(\frac{k_1 +\sigma_1  \bm{\sigma}_\parallel \cdot \bm{k}_\parallel}{m+\omega}) } \right]\varphi_{-}^{(+)} e^{-2ik_1 d}
    \\
    &= -\frac{(m+\omega) - i(k_1 - \sigma_1  \bm{\sigma}_\parallel \cdot \bm{k}_\parallel)}{(m+\omega) + i(k_1 + \sigma_1  \bm{\sigma}_\parallel \cdot \bm{k}_\parallel)}\varphi_{-}^{(+)}e^{-2ik_1 d}
    \\
    &= -\frac{[(m+\omega) - i(k_1 - \sigma_1  \bm{\sigma}_\parallel \cdot \bm{k}_\parallel)][(m+\omega) + i(k_1 - \sigma_1  \bm{\sigma}_\parallel \cdot \bm{k}_\parallel)]}{[(m+\omega) + i(k_1 + \sigma_1 \bm{\sigma}_\parallel \cdot \bm{k}_\parallel)][(m+\omega) + i(k_1 - \sigma_1  \bm{\sigma}_\parallel \cdot \bm{k}_\parallel)]} \varphi_{-}^{(+)}e^{-2ik_1 d}
    \\
    &= -\frac{(m+\omega)^2 + (k_1 - \sigma_1 \bm{\sigma}_\parallel \cdot \bm{k}_\parallel)^2}{(m+\omega)^2 +2i(m+\omega)k_1 -  ({k_1}^2 - {\sigma_1}^2 ( \bm{\sigma}_\parallel \cdot \bm{k}_\parallel)^2)}\varphi_{-}^{(+)}e^{-2ik_1 d}
    \\
    &= -\frac{m^2 + 2m\omega + \omega^2 + {k_1}^2 - 2\sigma_1 k_1  \bm{\sigma}_\parallel \cdot \bm{k}_\parallel + (\bm{\sigma}_\parallel \cdot \bm{k}_\parallel)^2 }{m^2 + 2m\omega + \omega^2 + 2i(m+\omega)k_1 - {k_1}^2 + (\bm{\sigma}_\parallel \cdot \bm{k}_\parallel)^2)}\varphi_{-}^{(+)}e^{-2ik_1 d}
    \\
    &= -\frac{2(m^2 + m\omega + {k_1}^2 - \sigma_1 k_1  \bm{\sigma}_\parallel \cdot \bm{k}_\parallel)}{2(m^2 + m\omega + ik_1(m+\omega))}\varphi_{-}^{(+)}e^{-2ik_1 d}
    \\
    &= -\frac{m(m+\omega) + {k_1}^2 - \sigma_1 k_1 \bm{\sigma}_\parallel \cdot \bm{k}_\parallel}{(m + ik_1)(m+\omega)}\varphi_{-}^{(+)}e^{-2ik_1 d}
\end{align}
Obtaining an additional relation for $\varphi_{+}^{(-)} $ and $\varphi_{-}^{(-)}$ using similar procedure, and combining the two results, we get :
\begin{equation}
    \varphi_{+}^{(\pm)}= -\frac{m(m+\omega) + {k_1}^2 - \sigma_1 k_1 \bm{\sigma}_\parallel \cdot \bm{k}_\parallel}{(m \pm ik_1)(m+\omega)}\varphi_{-}^{(\pm)}e^{-2ik_1 d}.
\end{equation}

\chapter{Derivations using Abel-Plana Formulae}
 
\section{Using MIT Bag boundary conditions} 
\subsection{For Naive fermion}
\label{Naivemitbag}
For the massless case ($am_f=0$), the dispersion relation obtained is :
\begin{equation}
    aE(ap)= a\sqrt{D_{nf}^\dagger D_{nf}}  = \sqrt{\sin^2 ap_1(n)}
\end{equation}
We shall now calculate the Casimir Energy for the massless Naive Fermion in $(1+1)$-dimensional spacetime using the  MITBag (B) boundary conditions in the contracted direction.
the function obtained can be easily reduced as :
\begin{align}
    \label{fnf}
     F(z) &=  \sqrt{\sin^2\left(\frac{\pi z}{N}\right)}\\
 \Rightarrow F(x,y) &= \sqrt{
    \begin{aligned}
    &\sin^2\left(\frac{ \pi x}{N} \right)\cosh^2\left(\frac{\pi y}{N}\right) - \cos^2\left(\frac{\pi x}{N}\right)\sinh^2\left(\frac{\pi y}{N} \right)\\&+2 i \sin(\frac{\pi x}{N})\cos(\frac{\pi x}{N})\sinh(\frac{\pi y}{N})\cosh(\frac{\pi y}{N})
    \end{aligned}
    }
\end{align}
Now note that the complex function $F(z)$ has branch cuts only at $x=0 \text{ and } x=N$ along the $y$-direction. Let us consider a path shifted by a small parameter $\epsilon\rightarrow +0 $. Function $F(z)$ on the paths $ x=0 +\epsilon \text{ and } x=N- \epsilon$ can be computed as :
\begin{align}
\sqrt{\sin^2\left(\frac{\pi}{N}(\epsilon \pm iy)\right)}
&=
\sqrt{\left(\pm i \sinh\left(\frac{\pi y}{N} + \epsilon\right)\right)^2}
\nonumber\\
&\simeq
\pm i \sinh\left(\frac{\pi y}{N}\right)
\end{align}
   \begin{align}
    \sqrt{\sin^2\left(\frac{\pi}{N}\left(N-\epsilon \pm iy \right)\right)} &= \sqrt{\sin^2\left(\pi -(\epsilon \mp iy)\frac{\pi}{N}\right)}\nonumber\\&= \sqrt{(\mp i \sinh(\frac{\pi y}{N} )+\epsilon)^2}\nonumber\\&\simeq \mp i \sinh(\frac{\pi y}{N})
\end{align} 
The above functional values and the generalized Abel Plana Formulae will be used to evaluate the Casimir Energy in Periodic and Antiperiodic boundary conditions. 
With the analytic continuation $z\rightarrow (n+\frac{1}{2})$  we have
 the function $F(z) = \sqrt{\sin^2(\pi z/N)}$ into the Abel Plana formulae in finite range for Antiperiodic boundary (\ref{nonint}).
 Similar to previous case substituting range $(a,b) = (0+\epsilon,N -\epsilon)$ , we get Casimir energy as:
\begin{equation}
 aE_{\text{Cas}}^{\text{1+1D,}nf} = \begin{aligned}
 \Bigg[i\int_0^\infty dy \frac{F(\epsilon +iy)}{e^{2\pi y}+1} - i&\int_0^\infty dy \frac{F(\epsilon -iy)}{e^{2\pi y}+1}\\& - i\int_0^\infty dy \frac{F(N-\epsilon +iy)}{e^{2\pi  (y-iN)}+1} +i\int_0^\infty dy \frac{F(N-\epsilon -iy)}{e^{2\pi (y+iN)}+1} \Bigg]
 \end{aligned} \nonumber
\end{equation}
\begin{equation}
  = \begin{aligned}
  \Bigg[ i\int_0^\infty dy \frac{i\sinh(\pi y/N)}{e^{2\pi y}+1} - i&\int_0^\infty dy \frac{-i\sinh(\pi y/N)}{e^{2\pi y}+1} \\& - i\int_0^\infty dy \frac{-i\sinh(\pi y/N)}{e^{2\pi y}+1} +i\int_0^\infty dy \frac{ i \sinh(\pi y/N)}{e^{2\pi y}+1} \Bigg]\end{aligned}\nonumber
\end{equation}
The constituent integrals can be computed to be:
\begin{align}
    \int_0^\infty \frac{dy\sinh(2\pi y/N)}{e^{2\pi y}-1} & = +\frac{N}{4\pi}-\frac{1}{4}\cot(\frac{\pi}{N})\label{minus}\\\int_0^\infty \frac{dy\sinh(2\pi y/N)}{e^{2\pi y}+1} & = -\frac{N}{4\pi}+\frac{1}{4}\csc(\frac{\pi}{N})\label{plus}
\end{align}
\begin{align}
   & =-4\int_0^\infty \frac{dy\sinh(\pi y/N)}{e^{2\pi y}+1}\nonumber \\
   & =-4\left[-\frac{N}{2\pi} +\frac{1}{4}\csc(\frac{\pi}{2N})\right] =\frac{2N}{\pi} -\frac{1}{4}\csc(\frac{\pi}{2N})
 \end{align} 
  
\subsection{For Wilson fermion}
\label{Wilsonmitbag}
Calculating for the massless Wilson fermion :
\begin{align}
  aE_{\text{Cas}}^{\text{1+1D,W}} = a\sqrt{D_W^\dagger D_W} &= \sqrt{\sin^2(ap_1) +(1-\cos(ap_1))^2 } \nonumber\\&=2\sqrt{\sin^2\left(\frac{ap_1}{2}\right)}
\end{align}
 We shall now calculate the Casimir Energy for the massless Wilson fermion in $(1+1)$-dimensional spacetime for the allowed modes in the contracted direction.
the function obtained can be easily reduced as :
\begin{equation}
\label{fw}
     F(z) = 2 \sqrt{\sin^2\left(\frac{\pi z}{2N}\right)}\\
\end{equation}
Now note that the complex function $F(z)$ has only one branch cut at $x=0 \text{ and } x=N$ along the $y$-direction. Let us consider a path shifted by a small parameter $\epsilon\rightarrow +0 $ at $x=0$. Function $F(z)$ on the paths $ x=0 +\epsilon$ can be computed as :
\begin{align}
    2\sqrt{\sin^2\left(\frac{\pi}{2N}\left(0+\epsilon \pm iy\right)\right)} & = 2\sqrt{\left(\pm i\sinh\left(\frac{\pi y}{2N}\right)+ \epsilon\right)^2}\nonumber\\&\simeq \pm 2i \sinh\left(\frac{\pi y}{2N}\right)
    \end{align}
    \begin{align}
    2\sqrt{\sin^2\left(\frac{\pi}{2N}\left(N \pm iy \right)\right)} &= 2\sqrt{\sin^2\left(\frac{\pi}{2} \pm \frac{i\pi y}{2N}\right)}\nonumber\\&= 2\sqrt{\cosh^2\left(\frac{\pi y}{2N}\right)}= 2\Big\lvert \cosh(\frac{\pi y}{2N})\Big\rvert
\end{align} 
The above functional values and the generalized Abel Plana Formulae will be used to evaluate the Casimir Energy in Periodic and Antiperiodic boundary conditions.  \\ 
With the analytic continuation $z\rightarrow (n+\frac{1}{2})$  we have the function $F(z) = 2 \sqrt{\sin^2\left(\frac{\pi z}{2N}\right)}$ into the Abel Plana formulae in finite range for Antiperiodic boundary (\ref{nonint}). Similar to previous case substituting range $(a,b) = (0+\epsilon,N)$ , we get Casimir energy as:
 
 \begin{equation}
 aE_{\text{Cas}}^{\text{1+1D,W}} = 
 \begin{aligned}
\Bigg[ i\int_0^\infty dy \frac{F(\epsilon +iy)}{e^{2\pi y}+1} - i&\int_0^\infty dy \frac{F(\epsilon -iy)}{e^{2\pi y}+1}\\& - i\int_0^\infty dy \frac{F(N +iy)}{e^{2\pi  (y-iN)}+1}
 +i\int_0^\infty dy \frac{F(N -iy)}{e^{2\pi (y+iN)}+1}\Bigg]
 \end{aligned}\nonumber
 \end{equation}
  \begin{equation}
 = \begin{aligned}2\Bigg[ i\int_0^\infty dy \frac{i\sinh(\pi y/2N)}{e^{2\pi y}+1} - i&\int_0^\infty dy \frac{-i\sinh(\pi y/2N)}{e^{2\pi y}+1}\\& - i\int_0^\infty dy \frac{\cosh(\pi y/2N)}{e^{2\pi y}+1} +i\int_0^\infty dy \frac{\cosh(\pi y/2N)}{e^{2\pi y}+1} \Bigg]
 \end{aligned}\nonumber\\\nonumber
\end{equation}
\begin{align}
   & =-4\int_0^\infty \frac{dy\sinh(\pi y/2N)}{e^{2\pi y}+1} \nonumber\\
   & =-4\left[-\frac{N}{\pi} +\frac{1}{4}\csc(\frac{\pi}{4N})\right] = \frac{4N}{\pi} -\csc(\frac{\pi}{4N})\nonumber
 \end{align}
The validity of the functions $F(z)$ chosen in (\ref{fnf}) and (\ref{fw}) is verified in accordance with the validity conditions for which the Abel-Plana Formulae in finite range \cite{math_casimir}.


\bibliographystyle{unsrturl}
\bibliography{bibliography}

\end{document}